\pdfoutput=1

\documentclass[apj]{emulateapj}
\usepackage{apjfonts}

\slugcomment{}

\shorttitle{The Stellar Halos and Thick Disks of the Milky
Way}
\shortauthors{Carollo et al.}

\begin{document}


\title{Structure and Kinematics of the Stellar Halos and Thick Disks  \\
of the Milky Way Based on Calibration Stars from SDSS DR7}

\author{Daniela Carollo\altaffilmark{1,2}}

\author{Timothy C. Beers\altaffilmark{3}}

\author{Masashi Chiba\altaffilmark{4}}

\author{John E. Norris\altaffilmark{1}}
\author{Ken C. Freeman\altaffilmark{1}}

\author{Young Sun Lee\altaffilmark{3}}

\author{Zeljko Ivezic\altaffilmark{5}}

\author{Constance M. Rockosi\altaffilmark{6}}

\author{Brian Yanny\altaffilmark{7}}


\altaffiltext{1} {Research School of Astronomy and Astrophysics, Australian National University, Cotter Road, Weston, ACT 2611, Australia;
carollo@mso.anu.edu.au;jen@mso.anu.edu.au; kcf@mso.anu.edu.au}
\altaffiltext{2} {INAF-Osservatorio Astronomico di Torino, Italy}
\altaffiltext{3} {Department of Physics \& Astronomy and JINA: Joint Institute for Nuclear Astrophysics, Michigan State University, E. Lansing, MI 48824, USA; beers@pa.msu.edu;lee@pa.msu.edu}
\altaffiltext{4}{Astronomical Institute, Tohoku University, Sendai 980-8578, Japan; chiba@astr.tohoku.ac.jp}
\altaffiltext{5}{Department of Astronomy, University of Washington, Box 351580, Seattle, WA 98195, USA; ivezic@astro.washington.edu}
\altaffiltext{6}{Astronomy and Astrophysics Department, University of California, Santa Cruz, CA 95064, USA; crockosi@ucolick.org}
\altaffiltext{7}{Fermi National Accelerator Laboratory, P.O. Box 500, Batavia, IL 60510, USA; yanny@fnal.gov}

\begin{abstract}

The structure and kinematics of the recognized stellar components of the Milky
Way are explored, based on well-determined atmospheric parameters and kinematic
quantities for 32360 ``calibration stars'' from the Sloan Digital Sky Survey
(SDSS) and its first extension, (SDSS-II), which included the sub-survey SEGUE:
Sloan Extension for Galactic Understanding and Exploration. Full space motions
for a sub-sample of 16920 stars, exploring a local volume within 4 kpc of the
Sun, are used to derive velocity ellipsoids for the inner- and outer-halo
components of the Galaxy, as well as for the canonical thick-disk and proposed
metal-weak thick-disk populations. This new sample of calibration stars
represents an increase of 60\% relative to the numbers used in a previous
analysis. We first examine the question of whether the data require the presence
of at least a two-component halo in order to account for the rotational behavior
of likely halo stars in the local volume, and whether more than two components
are needed. We also address the question of whether the proposed metal-weak
thick disk is kinematically and chemically distinct from the canonical thick
disk, and point out that the Galactocentric rotational velocity inferred for the
metal-weak thick disk, as well as its mean metallicity, appear quite similar to
the values derived previously for the Monoceros stream, suggesting a possible
association between these structures. In addition, we consider the fractions of
each component required to understand the nature of the observed kinematic
behavior of the stellar populations of the Galaxy as a function of distance from
the plane. Scale lengths and scale heights for the thick-disk and metal-weak
thick-disk components are determined. Spatial density profiles for the inner-
and outer-halo populations are inferred from a Jeans Theorem analysis. The full
set of calibration stars (including those outside the local volume) is used to
test for the expected changes in the observed stellar metallicity distribution
function with distance above the Galactic plane {\it in-situ}, due to the
changing contributions from the underlying stellar populations. The above issues
are considered, in concert with theoretical and observational constraints from
other Milky-Way-like galaxies, in light of modern cold dark matter galaxy
formation models.

\end{abstract}

\keywords{Galaxy: Evolution, Galaxy: Formation, Galaxy: Halo, Galaxy: Disks,
Galaxy: Kinematics, Galaxy: Structure, Methods: Data Analysis, Stars: Abundances, Surveys}

\section{Introduction}

The Milky Way is a unique laboratory, as it is the one large galaxy where it is
possible to analyze the full space motions and chemical compositions of its
stellar populations based on individual stars, and to use this information to
disentangle its structural properties in great detail. Knowledge of the full
six-dimensional location and velocity phase space and the chemical properties of
stellar populations in the Galaxy provides invaluable information, not only on
its recognized structures (and sub-structures, such as tidal streams), but by
extension, on the nature of its formation and evolution. These concepts are
well-understood, and have been widely used in previous studies, from the seminal
work of Eggen, Lynden-Bell, \& Sandage (1962) and Searle \& Zinn (1978), to more
recent research based on much larger samples of tracer objects (e.g., globular
clusters: Zinn 1993; field stars: Hartwick 1987; Sandage \& Fouts 1987;
Sommer-Larsen \& Zhen 1990; Allen, Schuster, \& Poveda 1991; Ryan \& Norris
1991; Kinman et al. 1994; Norris 1994; Carney et al. 1996; Chiba \& Beers 2000;
Ivezi{\'c} et al. 2008; Xue et al. 2008; Morrison et al. 2009), all working
toward drawing a much more detailed picture of our host galaxy.

The primary recognized stellar populations in the solar neighborhood are the
thin disk (sometimes separated into the old and young thin disks), the thick
disk, and the halo; the properties of these components have been reasonably
well-determined by numerous previous authors. However, some issues remain
unresolved. Examples include the origin of the thick disk, and the lower
metallicity limit of its member stars. While many authors argue that the thick
disk is likely to have formed in response to ancient accretion events (e.g.,
Quinn \& Goodman (1986), Freeman (1987); Abadi et al. 2003, and references therein),
recent suggestions have been made that some properties of the thick disk could
be understood as the result of the radial migration of stars within the disk
itself (e.g., Schoenrich \& Binney 2009). Furthermore, despite numerous efforts, the
issue of whether or not the so-called metal-weak thick disk (MWTD; see Chiba \&
Beers 2000, and references therein) is a distinct structure from the canonical
thick disk has yet to be settled. Details concerning the origin of the Galactic
halo are of particular interest for understanding the formation and evolution of
our Galaxy. Indeed, the halo contains most of the metal-poor stars of the Milky
Way, objects that encode the nature of the first stars to form in the Universe,
which can be used to constrain models of large galaxy formation and evolution in
general.

The Galactic halo has long been considered a single component. However, evidence
has accumulated over the past few decades that it may be more complex (
Carney 1984; Hartwick (1987); Allen et al. 1991; Preston, Shectman, \& Beers 1991; Majewski 1992;
Zinn 1993; Kinman et al. 1994; Norris 1994; Carney et al. 1996; Chiba \& Beers
2000; Kinman et al. 2007; Lee, Gim, \& Casetti-Dinescu 2007; Miceli et al.
2008). Recently, based on an analysis of a large sample of calibration stars from the
Sloan Digital Sky Survey (SDSS) DR5 (Adelman-McCarthy et al. 2007), Carollo et
al. (2007) argued for the existence of {\it at least} a two-component halo. In
their view the Galactic halo comprises two broadly overlapping structural
components, an inner and an outer halo. These components exhibit different
spatial density profiles, stellar orbits, and stellar metallicities. It was
found that the inner-halo component dominates the population of halo stars found
at distances up to 10-15 kpc from the Galactic center, while the outer-halo
component dominates in the region beyond 15-20 kpc. The inner halo was shown to
comprise a population of stars exhibiting a flattened spatial density
distribution, with an inferred axial ratio on the order of $\sim$ 0.6. According
to Carollo et al. (2007), inner-halo stars possess generally high orbital
eccentricities, and exhibit a small (or zero) net prograde rotation around the
center of the Galaxy. The metallicity distribution function (MDF) of the inner
halo peaks at [Fe/H] = $-$1.6, with tails exending to higher and lower
metallicities. By comparison, the outer halo comprises stars that exhibit a more
spherical spatial density distribution, with an axial ratio $\sim$ 0.9.
Outer-halo stars possess a wide range of orbital eccentricities, exhibit a clear
retrograde net rotation, and are drawn from an MDF that peaks at [Fe/H] =
$-$2.2, a factor of four lower than that of the inner-halo population.

In this paper we present further details on the structural and kinematic
parameters for the inner and outer halos, as well as for the thick disk and MWTD,
based on analysis of a large sample of calibration stars from SDSS DR7
(Abazajian et al. 2009). The DR7 sample includes stars obtained during the Sloan
Extension for Galactic Understanding and Exploration (SEGUE; see Yanny et al.
2009), one of three sub-surveys obtained during the first extension of SDSS,
known as SDSS-II.

This paper is outlined as follows. For the convenience of the
reader the basic assumptions of our analytic approach
and our main results are briefly summarized in \S 2. Section 3 describes the
calibration stars from SDSS DR7, the methodology used to derive their
stellar atmospheric parameter estimates (T$_{\rm eff}$, log g, [Fe/H]), and the
criteria used to identify a local sample for our kinematic analysis. The
techniques used to derive estimates of the kinematic and orbital parameters for
stars in our sample are described in \S 4. This section also provides a
comparison of the changes in the sampling of the spatial distribution and in the
derived local kinematic properties for the calibration-star samples due to the
expansion from DR5 to DR7. Section 5 considers the optimal variables for
extracting knowledge of the primary stellar-population components revealed in
the solar neighborhood, and addresses the question of whether a dual-halo
population is in fact required to understand the observed local kinematics of
the halo. This section also describes the general properties of the maximum
likelihood technique we employ, and its application for de-coupling the
thick-disk, inner-, and outer-halo components. In this section we also evaluate
the fractions of each component as a function of distance from the Galactic
plane inferred from the present data, consider whether the MWTD might be an
independent component from the thick disk, and derive the rotational lag and
dispersion of the MWTD, as well as the range of metallicities covered by the
MWTD. Derived velocity ellipsoids for the thick-disk, MWTD, and the halo
components are reported in \S 6. In \S 7 the scale length and scale height of
the thick disk are derived; the MWTD scale length is also constrained. Section 8
derives the inferred spatial density profiles for the two halo components, based
on a Jeans Equation analysis. In \S 9 we consider the observed tilts of the
velocity ellipsoids as a function of vertical distance and metallicity. The {\it
in-situ} ``as-observed'' MDF, based on cuts in distance from the Galactic plane
for the full sample of calibration stars, is described in \S 10. Section 11
presents a summary and discussion of our results.

\section{Basic Assumptions and Brief Synopsis \\of Main Results}

In order to carry out our analysis we make the central assumption that the Galaxy
comprises a number of principal stellar components that are well-described
by Gaussian kinematic distribution functions and individual metallicity
distribution functions of (for now) unknown shape.  We then model these
components using Maximum Likelihood techniques, and compare these models
with the observed properties of the SDSS DR7 calibration stars. We make the
additional assumption that the number of stellar components should be taken as
the minimum necessary to adequately describe the observations, and no additional
components are added to the models if they do not result in statistically
significant improvements in the ability of our models to fit the data.

With the above set of assumptions, we find: (a) A single-population halo is
incompatible with the observed kinematics of local stars with [Fe/H] $< -2.0$,
while a dual (inner/outer) halo model is necessary to describe the observations.
(b) The net rotation of the outer halo is slightly more retrograde than
previously reported, $\langle$V$_{\phi}$$\rangle$ $ = -80 \pm 13 $ km~s$^{-1}$.
(c) The net rotation of the inner halo is consistent with zero (at the two-sigma
level), $\langle$V$_{\phi}$$\rangle$ $ = 7 \pm 4 $ km~s$^{-1}$. (d) There exists
a gradient in the mean rotational velocity of likely inner-halo stars,
$\triangle$$\langle$V$_{\phi}$$\rangle$/$\triangle |$Z$| = -28$ $\pm$ 9
km~s$^{-1}$ kpc$^{-1}$, for stars located within 2 kpc of the Galactic plane.
(e) The dispersions in the rotational-velocity component for both the inner- and
outer-halo populations increase with distance from the plane. This may apply to
the MWTD as well. (f) The net rotation of the canonical thick-disk population,
for stars close to the plane, is $\langle$V$_{\phi}$$\rangle$ = $182$ $\pm$ 2
km~s$^{-1}$. (g) The asymmetric drift of the thick disk varies with distance
from the plane, as shown by previous work,
$\triangle$$\langle$V$_{\phi}$$\rangle$/$\triangle$ $|$Z$|$ = $-$36 $\pm$ 1
km~s$^{-1}$ kpc$^{-1}$. (h) The thick-disk and MWTD systems only contribute to
our sample within 5 kpc of the plane, the inner-halo population dominates
between 5 and 10 kpc, and the outer-halo population dominates beyond 20 kpc. The
inversion point in the dominance of the inner/outer halo is located in the
distance range 15-20 kpc. (i) A MWTD, kinematically independent of the canonical
thick disk (with net rotation $\langle$V$_{\phi}$$\rangle$ = 100-150
km~s$^{-1}$), is required to account for the rotational properties of
low-metallicity stars close to the Galactic plane. (j) The metallicity range for
stars that are likely members of the MWTD is $-1.8 \lesssim$ [Fe/H] $\lesssim
-0.8$, possibly up to [Fe/H] $\sim -0.7$. (k) The derived velocity ellipsoids, ($\sigma_{V_{R}}$, $\sigma_{V_{\phi}}$, $\sigma_{V_{Z}}$),
for the four components we have considered are as follows -- Thick Disk
 = (53 $\pm$ 2, 51
$\pm$ 1, 35 $\pm$ 1) km~s$^{-1}$, MWTD  = (59 $\pm$ 3, 40 $\pm$ 3, 44 $\pm$ 3) km~s$^{-1}$, Inner Halo
= (150 $\pm$ 2, 95
$\pm$ 2, 85 $\pm$ 1) km~s$^{-1}$, Outer Halo = (159 $\pm$ 4, 165 $\pm$ 9, 116 $\pm$ 3)
km~s$^{-1}$. (l) The kinematically derived scale length and scale height for the thick disk
are h$_{R}$ = 2.20 $\pm$ 0.35 kpc and h$_{Z}$ = 0.51 $\pm$ 0.04 kpc,
respectively. The radial scale length for the MWTD is on the order of h$_{R}$
$\sim$ 2 kpc, while the scale height is 1.36 $\pm$ 0.13 kpc. (m) The slopes of
the inferred power-law spatial-density profiles for the inner- and outer-halo
populations are $-3.17$ $\pm$ 0.20 and $-1.79$ $\pm$ 0.29, respectively. (n) The
tilt angle, $\alpha$, between the long axis of the Schwarzschild velocity ellipsoid and the
sun-center line is, in Galactocentric
cylindrical coordinates, for stars of relatively high metallicity ($-$0.8 $<$
[Fe/H] $<$ $-$0.6), $\alpha$ = 7$^{\circ}$.1 $\pm$ 1$^{\circ}$.5 for 1 $<$
$|$Z$|$ $<$ 2 kpc, and $\alpha$ = 5$^{\circ}$.2 $\pm$ 1$^{\circ}$.2 for 2 $<$
$|$Z$|$ $<$ 4 kpc, consistent with one another. For stars of intermediate
metallicity ($-1.5 <$ [Fe/H] $<$ $-$0.8), the tilt is $\alpha$ = 10$^{\circ}$.3
$\pm$ 0$^{\circ}$.4 close to the plane, and $\alpha$ = 15$^{\circ}$.1 $\pm$
0$^{\circ}$.3 farther from the plane, which differ significantly from one
another. For stars of lower metallicity ([Fe/H] $<$ $-$1.5), the tilt is
$\alpha$ = 8$^{\circ}$.6 $\pm$ 0$^{\circ}$.5 close to the plane, and $\alpha$ =
13$^{\circ}$.1 $\pm$ 0$^{\circ}$.4 farther from the plane, again significantly
different. (o) The ``as observed'' metallicity distribution function of stars in
our sample exhibits clear changes with distance from the Galactic plane,
exhibiting the anticipated shift from dominance by thick-disk and MWTD
populations, with peaks at [Fe/H] = $-0.6$ and $-1.3$, respectively, to
dominance by the inner-halo ([Fe/H] $= -1.6$) and outer-halo ([Fe/H] $= -2.2$)
populations.

\section{The SDSS Calibration-Star Sample}

\subsection{SDSS and SEGUE Data}

SDSS-I was an imaging and spectroscopic survey that began routine operations in
April 2000, and continued through June 2005. The SDSS and its extensions used a
dedicated 2.5m telescope (Gunn et al. 2006) located at the Apache Point
Observatory in New Mexico. The telescope is equipped with an imaging camera and
a pair of spectrographs, each of which are capable of simultaneously collecting
320 medium-resolution ($R \sim 2000$) spectra over its seven square degree field
of view, so that on the order of 600 individual target spectra and roughly 40
calibration-star and sky spectra are obtained on a given spectroscopic
``plug-plate'' (York et al. 2000).

The SEGUE sub-survey, carried out as part of SDSS-II, ran from July 2005 to July
2008. SEGUE obtained some 240000 medium-resolution spectra of stars in the
Galaxy, selected to explore the nature of stellar populations from 0.5 kpc to
100 kpc (Yanny et al. 2009). In the present paper we make use of the
spectroscopy and photometry obtained by SDSS/SEGUE for the small number (16) of
calibration stars obtained for each spectroscopic plug-plate, chosen for two
primary reasons. The first set of these objects, the spectrophotometric
calibration stars, are stars that are selected to approximately remove the
distortions of the observed flux of stars and galaxies arising from the
wavelength response of the Astrophysical Research Consortium (ARC) 2.5m
telescope and the SDSS spectrographs, as well as the distortions imposed on the
observed spectra by the Earth's atmosphere. The spectrophotometric calibration
stars cover the apparent magnitude range 15.5 $< g_{0} <$ 17.0, and satisfy the
color ranges 0.6 $< (u-g) _{0} <$ 1.2 ; 0.0 $< (g-r) _{0} <$ 0.6.\footnote{The
subscript 0 in the magnitudes and colors indicates that they are corrected for
the effects of interstellar absorption and reddening, following standard
procedures, based on the dust map of Schlegel, Finkbeiner, \& Davis (1998).} The
second set of stars, the telluric calibration stars, is used to calibrate and
remove night-sky emission and absorption features from SDSS spectra. The
telluric calibration stars cover the same color ranges as the spectrophotometric
calibration stars, but at fainter apparent magnitudes, in the range 17.0 $<
g_{0} <$ 18.5.

The SEGUE Stellar Parameter Pipeline (SSPP) processes the wavelength- and
flux-calibrated spectra generated by the standard SDSS spectroscopic reduction
pipeline, obtains equivalent widths and/or line indices for about 80 atomic or
molecular absorption lines, and estimates the effective temperature, T$_{\rm
eff}$, surface gravity, log g, and metallicity, [Fe/H], for a given star through
the application of a number of approaches. A given method is usually optimal
over specific ranges of color and signal-to-noise ($S/N$) ratio. The SSPP
employs 8 primary methods for the estimation of T$_{\rm eff}$, 10 for the
estimation of log g, and 12 for the estimation of [Fe/H]. The final estimates of
the atmospheric parameters are obtained by robust averages of the methods that
are expected to perform well for the color and $S/N$ of the spectrum obtained
for each star. The use of multiple methods allows for empirical determinations
of the internal errors for each parameter, based on the range of reported values
from each method -- typical internal errors for stars in the temperature range
that applies to the calibration stars are $\sigma_{\rm Teff} \sim$ 100 K to
$\sim$ 125 K, $\sigma_{\rm log g} \sim $ 0.25 dex, and $\sigma_{\rm [Fe/H]}
\sim$ 0.20 dex. The external errors in these determinations are of similar size. See
Lee et al. (2008a,b) and Allende Prieto et al. (2008) for more details.

Over the past several years, large-aperture telescopes have been used to obtain
high-resolution spectroscopy for over 300 of the brighter (14.0 $< g_0 <$ 17.0)
SDSS stars (see Allende Prieto et al. 2008; W. Aoki et al., in preparation; D.
Lai et al., in preparation). W. Aoki et al. (in preparation) and D. Lai et al.
(in preparation) suggest that the current SSPP is actually somewhat conservative
in the assignment of metallicities for stars of the lowest [Fe/H], in the sense
that high-resolution estimates of [Fe/H] are on the order of 0.2-0.3 dex lower than
those reported by the SSPP. For the purpose of our present analysis, we have
used these preliminary results to apply a quadratic correction to the
SSPP-derived metallicities, of the form:

\begin{equation}
{\rm [Fe/H]}_C = -0.186 + 0.765*{\rm [Fe/H]}_A - 0.068*{{\rm [Fe/H]}_A}^2
\end{equation}

\noindent where [Fe/H]$_A$ is the adopted metallicity from the SSPP, and
[Fe/H]$_C$ is the corrected metallicity.  This polynomial has little effect on
stars with metallicity greater than about [Fe/H] = $-2.5$, but lowers the estimated
metallicities for stars below this abundance by 0.1 to 0.2 dex.

Once estimates of the atmospheric parameters for each of the calibration stars
are obtained, one can use the derived surface gravity of each star to infer
whether it is a likely dwarf, main-sequence turnoff star, subgiant, or giant.
Photometric estimates of the distance to each star (accurate to an estimated
10\%-20\%) are then obtained by comparison of its observed apparent magnitude
(corrected for interstellar absorption) with its expected absolute magnitude
based on calibrated open-cluster and globular-cluster sequences, using the
techniques described by Beers et al. (2000). See the discussion by Ivezi{\'c} et
al. (2008) for a comparison between the Beers et al. (2000) distances
with those obtained by use of cluster fiducials in the native SDSS $ugriz$
system.

Radial velocities for stars in our sample are derived from matches to an
external library of high-resolution spectral templates with accurately known
velocities, degraded in resolution to match the SDSS spectra (see Yanny et al.
2009). The typical precision of the resulting radial velocities is on the order
of 5-20 km~s$^{-1}$ (depending on the $S/N$ of the spectra), based on multiple
repeat observations of individual objects with different plug-plates. Zero-point
errors are negligible (after correction for a global offset of 7.3 km~s$^{-1}$
of unknown origin), and exhibit a dispersion of no more than 2 km~s$^{-1}$,
based on a comparison of the subset of stars in our sample with radial
velocities obtained from the high-resolution spectra taken for testing and
validation of the SSPP (Allende Prieto et al. 2008; Yanny et al 2009).

The full sample considered in the present paper consists of 32360
unique stars with acceptable derived atmospheric parameters. This represents a
60\% increase with respect to the starting sample of 20236 calibration stars
considered by Carollo et al. (2007). Although there is clearly a bias towards
the identification of metal-poor stars arising from the color selections (in
particular for the spectrophotometric calibration stars), it should be kept in
mind that it is not possible to discriminate the lowest metallicity stars (e.g.,
those with [Fe/H] $< -$2.5) from those with [Fe/H] $\sim -$2.0, since the effect
of declining metallicity on broadband stellar colors is minimal in this regime
(see Ivezi{\'c} et al. 2008). Hence, we expect that the distribution of
metallicities for the calibration stars should reflect the true shape of the
low-metallicity tail of the MDF for stars outside the disk populations. The same
does not apply for stars with [Fe/H] $> -$2.0; the observed distributions of
metallicity mis-represents the true MDFs at these abundances. It is important to
note that no selection on the kinematics, e.g., by making use of measured proper
motions, is made in the choice of the calibration objects. This enables our
kinematic studies to be carried out without the need to apply explicit
corrections or modeling of any selection bias for the motions of the stars in
the sample.

\subsection{Selection of Local Sample of Stars}

We begin with a total sample of 32360 unique calibration stars with available
spectroscopy and average $S/N \geq 10/1$ over the spectral range 3800~{\AA} to
8000~{\AA}. We then consider all stars with effective temperatures in the range
4500~K $\leq$ T$_{\rm eff} \leq $ 7000~K, over which the SSPP is expected to
provide the highest accuracy for the derived atmospheric parameters. This
reduces the sample to 32329 stars. In addition to this restriction, stars in our
primary sample for kinematic analysis must satisfy the following criteria:

\begin{itemize}

\item Stars must have derived distances d $<$ 4 kpc from the Sun, in order to
restrict the kinematic and orbital analyses to a local volume (where the
assumptions going into their calculation are best satisfied). This selection
also mitigates against the increase in the errors in the derived transverse
velocities, which scale with distance from the Sun (e.g., for a typical star in
our sample, at d = 1 kpc, errors of 3.5 mas/yr in proper motions and 15\% in
distance result in errors in the derived transverse velocities of 22
km~s$^{-1}$; these rise to 37 km~s$^{-1}$ and 46 km~s$^{-1}$ for stars at d = 2
kpc and 4 kpc, respectively).  This cut reduces the numbers of stars to 24824.

\item Stars must have a measured radial velocity, accurate to better than
20 km~s$^{-1}$ (which eliminates only a handful of stars from consideration), as
well as an ``acceptable'' proper motion available (which means that the star
satisfies additional criteria designed to eliminate spurious reported motions;
see Munn et al. 2004). Note that all proper motions have been corrected for the
systematic error described by Munn et al. (2008). \footnote {We call
attention to the extensive testing of the SDSS-recalibrated proper-motion errors
carried out by Bond et al. (2009). By comparison with the derived (non-)motions
of distant quasars, they demonstrated that systematic errors on proper motions
are no more than 1 milli-arcsecond per year (with an rms of 0.6 milli-arcseconds
per year).} After these cuts are applied, the total numbers of stars remaining
is 23643.

\item Stars must have a present Galactocentric distance, projected onto the
plane, in the range 7 $<$ R $<$ 10 kpc. This restriction also serves to
improve the applicability of the simple models for the adopted form of the
Galactic potential we apply during the kinematic analysis.  This cut reduces the
numbers of stars in our local sample to 16920.

\end{itemize}

Figure 1 compares the (projected) spatial distributions of the SDSS/SEGUE
calibration stars used by Carollo et al. (2007), based on DR5 (Adelman-McCarthy
et al. 2007), with that of our present analysis, based on DR7 (Abazajian et al.
2009). In addition to the overall increase in the numbers of stars, one can also
notice the impact of the SEGUE fields, which included substantial numbers of
stars at lower Galactic latitudes. The upper panel of Figure 2 shows the
``as-observed'' MDF for the DR7 calibration stars in the full sample (black
histogram), as well as for stars selected for the local sample, as described
above. The MDFs of the two samples are clearly very similar, and comprise stars
with metallicities that sample all of the primary stellar components of the Galaxy.
The lower panel of Figure 2 shows the distribution of derived surface gravity,
$\log$ g, versus effective temperature, T$_{\rm eff}$, obtained by the SSPP.
Stars in the full sample are indicated by black dots, while those in the local
sample are represented by red dots. Note that the full sample contains
substantial numbers of objects with gravities consistent with main-sequence
turnoff stars, subgiants, and giants, while the local sample primarily comprises
main-sequence turnoff stars and dwarfs. We emphasize that both of the panels in
Figure 2 are influenced by the selection functions used for targeting of the
calibration stars, and should not be taken as representative of the distribution
of either [Fe/H] or log g in the explored volume.

\section{Derivation of Full Space Motions and Orbital Parameters}

Proper motions, used in combination with distance estimates and radial
velocities, provide the information required to calculate the full space motions
(the components of which are referred to as U,V,W) of our program stars with
respect to the Local Standard of Rest (LSR; defined as a frame in which the mean
space motions of the stars in the solar neighborhood average to zero)
\footnote{The velocity component U is taken to be positive in the direction
toward the Galactic anticenter, the V component is positive in the direction of
Galactic rotation, and the W component is positive toward the North Galactic
Pole.}. Corrections for the motion of the Sun with respect to the LSR are applied
during the course of the calculation of the full space motions; here we adopt
the values (U,V,W) = ($-9$,12,7) km~s$^{-1}$ (Mihalas \& Binney 1981). For the
purpose of analysis it is also convenient to obtain the rotational component of
a star's motion about the Galactic center in a cylindrical frame; this is
denoted as V$_{\phi}$, and is calculated assuming that the LSR is on a circular
orbit with a value of 220 km~s$^{-1}$ (Kerr \& Lynden-Bell 1986).  It is worth noting
that our assumed values of R$_{\sun}$ (8.5 kpc) and the circular velocity of the LSR
are both consistent with two recent independent determinations of these
quantities by Ghez et al. (2008) and Koposov, Rix, \& Hogg (2009).

The orbital parameters of the stars, such as the perigalactic distance (the
closest approach of an orbit to the Galactic center), r$_{peri}$, and
the apogalactic distance (the farthest extent of an orbit from the Galactic center),
r$_{apo}$, of each stellar orbit, the orbital eccentricity, $e$, defined as $e$
= (r$_{apo}$ $-$ r$_{peri}$)/(r$_{apo}$ + r$_{peri}$), as well as Z$_{max}$ (the
maximum distance of a stellar orbit above or below the Galactic plane), are
derived by adopting an analytic St\"ackel-type gravitational potential (which
consists of a flattened, oblate disk, and a nearly spherical massive dark-matter
halo; see the description given by Chiba \& Beers 2000, Appendix A) and
integrating their orbital paths based on the starting point obtained from the
observations.

We have considered the possible systematic uncertainty in Z$_{max}$ when one
adopts a different gravitational potential, such as the model originally
constructed by Dejonghe \& de Zeeuw (1988), and later elaborated upon by Batsleer
\& Dejonghe (1994) and Chiba \& Beers (2001), which takes a Kuzmin-Kutuzov
potential for both a highly flattened disk and a nearly spherical halo.
Experiments with our data indicate that, with the adoption of this alternative
potential, the systematic change in this orbital parameter is on the order of
10\% for Z$_{max}$ $<$ 50 kpc. A larger systematic applies when Z$_{max}$ $>$ 50
kpc, where this parameter is more sensitive to the potential. However, the range
of Z$_{max}$ explored in this paper is always below 50 kpc, even though a
handful of stars have larger values. Errors on the derived orbital parameters
due to the observational errors have been estimated through a Monte Carlo
simulation (100 realizations for each star). Although the sizes of these errors
depend on sample selection, for the St\"ackel potential used in Chiba \& Beers
(2000), together with constraints Z$_{max}$ $<$ 50 kpc and R$_{apo}$ $<$ 50 kpc
(apogalactic distance projected onto the Galactic plane), the estimated
one-sigma errors are $\sigma_{r_{peri}}$ = 1.1 kpc, $\sigma_{r_{apo}}$ = 2.2
kpc, $\sigma_{ecc}$ = 0.12, and $\sigma_{Z_{max}}$ = 1.3 kpc. The errors in
these parameters show little dependence on Z$_{max}$ over the range we consider
in this analysis.

Figure 3 shows the derived U,V,W velocity components, as a function of
metallicity, for both the DR5 and DR7 calibration-star samples. A comparison of
the eccentricity versus [Fe/H] diagrams from DR5 and DR7 is shown in Figure 4.
As can be appreciated from inspection of these figures, the increased sample of
stars in DR7, as well as refinements in the estimates of metallicities made
recently in the SSPP, have better delineated stars in the three primary stellar
populations represented in the local volume -- the thin disk, the thick disk,
and the halo. It should be kept in mind that the numbers of stars shown in
Figures 3 and 4 {\it do not reflect the actual relative proportions} of these
populations in our sample volume, due to the selection criteria used and the
bright limit of $g \sim 14.5$ imposed by the SDSS imaging.

Figure 5 shows the distribution of derived orbital eccentricities for stars
covering different regions of metallicity, chosen to isolate the contributions
of the various stellar components we describe in more detail below, for two
different intervals on distance $|$Z$|$ from the Galactic plane. In the upper
panels, for stars with $-1.0 < $ [Fe/H] $< -0.6$, the dominance of the canonical
thick disk is clear in both distance intervals. In the second row of panels, for
stars in the metallicity region $-1.5 < $ [Fe/H] $< -1.0$, the feature extending
to $e \sim 0.5$ indicates the likely presence of a MWTD population close to the
Galactic plane. Farther from the Galactic plane the observed distribution of
eccentricity appears more consistent with inner-halo stars, with some residual
contribution from the MWTD. The eccenticity distribution for the metal-poor halo
stars with $-2.0 <$ [Fe/H] $< -1.5$, shown in the third row of panels, is close
to the linear distribution $f(e) \propto e$ for both intervals on distance from
the plane. This form of $f(e)$ is expected for a well-mixed steady-state system
of test particles for which the distribution function is a function of the
energy only, moving in the potential of a strongly centrally concentrated
density distribution (see Binney \& Tremaine, 2008, p. 387). We see below,
however, that the distribution function for the metal-poor halo stars cannot
simply depend on energy only, because the three components of their velocity
dispersion are significantly different. In the lower row of panels, for stars
with [Fe/H] $< -2.0$, the distribution of orbital eccentricies appears
strikingly different than in the third row of panels. As discussed below, we
expect that this distribution reflects the contribution from an outer-halo
population with a less prominent fraction of stars on highly eccentric orbits,
superposed on the inner-halo population that dominates the third row of panels.

\section{Extracting the Major Stellar-Population Components
in the Solar Neighborhood}

\subsection{Basic Parameters}

Carollo et al. (2007) explored how to combine the derived kinematic parameters
and metallicity estimates for the calibration stars in DR5 to reveal the
presence of the primary stellar populations of the Galactic halo. With the much
larger sample of calibration stars available from SDSS/SEGUE DR7, we now refine
our investigation in order to disentangle the main components in the solar
neighborhood and extract their spatial and kinematic parameters. Although the
Galactic disk system comprises both a thin disk and a thick disk (and likely
MWTD), the thin disk is not well-represented in our present sample due to the
color and apparent magnitude selection criteria used for the SDSS calibration
stars. Thus, the main populations we expect to contribute to our sample of local
stars are the thick disk (and possibly a MWTD), as well as the inner and outer
halos.

Given the strong overlap in the spatial and metallicity distributions of the
components we consider, it is necessary to define a suitable set of parameters
to best identify the presence of each population. Based on our tests of
alternatives, the best parameters for this purpose are the distance from the
Galactic plane, the rotational velocity of a star in a cylindrical frame with
respect to the Galactic center, and the stellar metallicity. As shown in Figure
1, the local sample within 4 kpc of the Sun well-covers the region where the
thick-disk (and MWTD) components are expected to be prominently represented.
Thus, for exploration of these components we employ $|$Z$|$, the present
distance of a star above or below the Galactic plane. In contrast, for
exploration of the halo components we use Z$_{max}$, which depends on the
adopted gravitational potential, as an indicator of their vertical extension,
The choice of Z$_{max}$ for the Galactic halo is necessary because of the much
larger spatial extent of its two primary components, relative to that of the
thick-disk components. Below we consider possible correlations of Z$_{max}$ with
other kinematic or spatial parameters used in this analysis.

Velocity ellipsoids are evaluated in a Galactocentric cylindrical reference
frame, (R,$\phi$,Z), where R is the distance of the star from the Galactic
center, projected onto the Galactic plane. The angular coordinate $\phi$ is
taken to be positive in the direction of Galactic rotation, and Z is the height
above or below the Galactic plane. In this system, the mean velocities and their
dispersions are denoted by ($\langle$V$_{R}$$\rangle$, $\sigma_{V_{R}}$),
($\langle$V$_{\phi}$$\rangle$, $\sigma_{V_{\phi}}$), and
($\langle$V$_{Z}$$\rangle$, $\sigma_{V_{Z}}$). It is also explicity assumed, for
the present analysis, that the velocity ellipsoids of the disk and halo systems
are aligned on the coordinate axes. We explicitly test this assumption in \S 9
below.

The first step is to understand the {\it global behavior} of the derived
kinematic parameters as a function of metallicity and $|$Z$|$ or Z$_{max}$.
Figure 6 shows the derived velocity distributions for V$_{R}$, V$_{\phi}$, and
V$_{Z}$ for different metallicity ranges. As expected, the rotational velocity
exhibits rather different behavior as the range in [Fe/H] changes. In the two
upper panels of this figure the thick disk is dominant (V$_{\phi}$ $\sim$ 200
km~s$^{-1}$), while in the lower panels, as metallicity decreases, the halo
becomes more clearly present (V$_{\phi}$ $\sim$ 0 km~s$^{-1}$). The long tail of
the rotational velocity distribution towards retrograde velocities seen in the
lower panels arises from stars that likely belong to the outer-halo population.
In contrast to this behavior, the distributions of V$_{R}$ and V$_{Z}$ remain
almost symmetric over all ranges of metallicity, and the mean velocities are
always close to zero, as expected. Notice that the dispersions of V$_{R}$ and
V$_{Z}$ increase as [Fe/H] decreases, reflecting the dominance of the halo
components at lower metallicity. This figure suggests that, among these
parameters, V$_{\phi}$ most clearly distinguishes the influence of the different
stellar populations in the Galaxy represented in our local sample.

It is important to check for possible correlations that might exist between
Z$_{max}$ and (V$_{R}$, V$_{\phi}$, V$_{Z}$) before their use in the kinematic
analysis of the inner- and outer-halo populations. Figure 7 shows the
cylindrical velocity components, as a function of Z$_{max}$, for the stars in
our local sample. The upper panel indicates a correlation between V$_{R}$ and
Z$_{max}$, in particular for extreme values of V$_{R}$. An even stronger
correlation is seen in the lower panel of the figure, which represents the
vertical velocity component, V$_{Z}$. These correlations arise because large
values of V$_{R}$ and V$_{Z}$ correspond to large orbital energy in the
meridional plane, (R,Z), and thus also to large values of Z$_{max}$. By
contrast, the rotational velocity V$_{\phi}$ (middle panel) does not exhibit a
strong correlation with Z$_{max}$, other than that expected from the presence of
the thick-disk population at high positive rotation velocity and the halo at
lower rotational velocity. Note, in the middle panel, the clear excess of stars with
retrograde motions for Z$_{max}$ $>$ 15 kpc, which we associate with the
outer-halo component, as discussed below. We conclude that the Galactocentric
rotational velocity, V$_{\phi}$, and the vertical distance, Z$_{max}$ (or
$|$Z$|$), when combined with metallicity, can be used to obtain useful
information on the different stellar populations present in the local volume.

\subsection{How Many Components in the Stellar Halo?}

If the halo of the Galaxy is spatially and kinematically complex, as argued by
Carollo et al. (2007) and other previous authors, a natural question that arises
is just how many components must be invoked to accommodate the available data.
Note that here we are considering the definition of a ``component'' to mean a
{\it bona-fide} stellar population, in the traditional sense, which is to say
that the member stars share a common set of locations, kinematics, compositions
(MDFs), and possibly ages (or distributions of ages), indicative of a likely
common astrophysical origin. This definition would not extend to include the
presence of {\it individual} substructures, such as debris streams (e.g., Yanny
et al. 2003; Belokurov et al. 2006; Grillmair 2006; Bell et al. 2008; Grillmair
et al. 2008; Klement et al. 2009; Schlaufman et al. 2009), and other
overdensities such as the Virgo Overdensity (An et al. 2009, and references
therein). Rather, our analysis seeks to model the kinematics of the halo using a
minimum number of components, which themselves may or may not comprise the
superposition of numerous individual substructures. The distinction is
irrelevant for our approach.

Our most powerful tool for distinguishing various halo components is
consideration of the rotational behavior of likely halo stars. A sub-sample of
the local calibration stars, satisfying rather strict limits on metallicity,
[Fe/H] $ < -$2.0, and orbital distances above the plane, Z$_{max} >$ 5 kpc,
should be dominated by the halo components, with little or no thick-disk or MWTD
contamination. We now consider such a sample for addressing this issue.

The question of how many components may comprise a complex halo can be subtle,
since in principle individual components should be free to take on wide ranges
of mean rotational velocity and dispersion, and it may or may not be possible to
reliably separate them if there exists too large an overlap in their kinematic
behaviors. Simple subjective guesses for the likely input parameters needed to
carry out mixture-model analyses are not suitable for obtaining statistically
sound answers to the question of the need (or not) for multiple components.
Fortunately, alternatives exist. Kauffman \& Rousseeuw (1990) describe the use
of a clustering algorithm ``Partitioning Around Medoids'' (PAM), and its large-N
extension ``Clustering LARge Applications'' (CLARA). These algorithms seek
clusters of objects that have a high degree of similarity, while maximizing the
degree of dissimilarity between different clusters. The original data set is
initially partitioned into clusters around $k$ so-called ``representative
objects,'' referred to as the medoids, then an iterative scheme is applied to
locate the medoids that best achieve the similarity/dissimilarity goal. The
algorithm employed by PAM/CLARA is similar to the well-known {\it k-means}
clustering algorithm, but it is more robust to outliers, and enjoys the
computational advantage that it obtains a unique partitioning of the data
without the need for explicit multiple starting points for the proposed
clusters.

\begin{deluxetable*}{ccrrccrrcc}[!t]
\tablewidth{0pt}
\tablenum{1}
\tablecaption{Cluster Analysis Results: Partitioning Around Medoids}
\tablehead{
\colhead{$k$} &
\colhead{Objects} &
\colhead{Coordinate} &
\colhead{IQR} &
\colhead{$\overline{s}(i)$} &
\colhead{$\overline{s}(k)$} &
\colhead{$\langle$V$_{\phi}$$\rangle$ } &
\colhead{ $\sigma_{V_{\phi}}$} &
\colhead{Fraction}  &
\colhead{p-value} \\
\colhead{} &
\colhead{} &
\colhead{(km~s$^{-1}$)} &
\colhead{(km~s$^{-1}$)} &
\colhead{} &
\colhead{} &
\colhead{(km~s$^{-1}$)} &
\colhead{(km~s$^{-1}$)} &
\colhead{}&
\colhead{} \\
\colhead{(1)} &
\colhead{(2)} &
\colhead{(3)} &
\colhead{(4)} &
\colhead{(5)} &
\colhead{(6)} &
\colhead{(7)} &
\colhead{(8)} &
\colhead{(9)} &
\colhead{(10)}
}
\startdata
     &     &       &     &         &          &         &      &          &         \\
 1   & 710 &  $-$31& 178 & \nodata &  \nodata &  $-$71  &  142 &  1.00    & 0.002    \\
     &     &       &     &         &          &         &      &          &          \\
 2   & 363 &   60  &  40 & 0.65    &  0.61    &  $-$18  &   99 &  0.51    & 0.350    \\
     & 347 & $-$142& 142 & 0.57    & \nodata  & $-$128  &  159 &  0.49    & \nodata  \\
     &     &       &     &         &          &         &      &          &          \\
 3   & 310 &  $-$23&  74 & 0.66    &  0.53    &  $-$55  &   94 &  0.66    & 0.321    \\
     & 230 & $-$175& 132 & 0.50    & \nodata  & $-$277  &  104 &  0.17    & \nodata  \\
     & 170 &  135  &  77 & 0.43    & \nodata  &     75  &   95 &  0.17    & \nodata  \\
\enddata
\tablecomments{The interquartile range (IQR) is a measure of the scale
(dispersion) of the data. In the case of a normal distribution the IQR is
related to the dispersion by $\sigma$ = IQR/1.349}
\end{deluxetable*}

The quality of the resulting partitions selected by PAM/CLARA can be assessed by
inspection of the dimensionless parameter $s(i)$, which takes on values in the
range $-1 \le s(i) \le +1$. This parameter measures the similarity of a given
datum $i$ to its closest represesentative object, relative to its next nearest
representative object. If $s(i)$ is close to $+1$, we can reasonably conclude
that the datum is properly assigned. If not, and $s(i)$ takes on values closer
to 0 or $-1$, we can conclude that either it is not clear to which
representative object the datum should be assigned ($s(i) \sim 0$) or that it
has been misassigned ($s(i) \sim -1$). Once a successful partition of the data
has been made, we can inspect the following summary values: (1) $\overline{s}(i)
$, the average of $s(i)$ within a given cluster, and (2) $\overline{s}(k)$, the
average of $s(i)$ for the entire data set. This last number provides some
guidance on what value of $k$ is most appropriate for a given data batch, i.e.,
the number of clusters is chosen such that $\overline{s}(k)$ is a
maximum\footnote{The partitioning experiment described here only considers
individual velocity components that have larger dispersions than could be
accounted for by expected errors in the derived rotation velocities.}.

Table 1 summarizes the results of the partitioning exercise (using CLARA) for
different input numbers of clusters, $1 \le k \le 3$. Column (1) lists the
numbers of clusters. Column (2) lists the numbers of stars assigned to each
cluster. Column (3) lists the coordinate (V$_\phi$) for the representative
object nearest the center of each proposed cluster. The inter-quartile range
(IQR; see definition in Table 1) in V$_\phi$ covered by the proposed cluster is
listed in column (4). The summary statistics $\overline{s}(i)$ and
$\overline{s}(k)$ are listed in columns (5) and (6), respectively. Note that the
results are ordered in Table 1 from high to low numbers of the proposed objects
in each cluster.

As seen in Table 1, the first (non-trivial) split suggested by CLARA ($k = 2$)
results in essentially equally populated clusters with positive (60 km~s$^{-1}$)
and negative ($-142$ km~s$^{-1}$) coordinates, albeit with rather different
IQRs. Each of the two clusters has moderately high values of $\overline{s}(i)$,
and the average over both clusters, $\overline{s}(k)$, indicates a significant
split. As one progresses to the case $k = 3$, CLARA splits the negative
coordinate cluster into two pieces (at coordinates $-23$ km~s$^{-1}$ and $-175$
km~s$^{-1}$), and moves the positive cluster to higher velocity (135
km~s$^{-1}$). Note that the significance for two of the individual splits drops
to below $\overline{s}(i) = 0.60$, as does the average $\overline{s}(k)$. This
provides an indication that there is likely to be no need for three or more
components in order to fit a reasonable mixture model to these data.

The partitions suggested by CLARA can be used to guide choices of input
parameters for mixture-modeling of the proposed clusters. The R statistical
software package (\texttt{http://www.r-project.org/}), and in particular its
assocated package R-Mix
(\texttt{http://www.math.mcmaster.ca/peter/mix/mix.html}), provide convenient
tools for simple analyses of mixtures.

Here we take the coordinates of the proposed medoids as input starting points
for $\langle$V$_{\phi}$$\rangle$, and $\sigma_{V_{\phi}}$ = IQR/1.349 as
estimates of the input starting points for the dispersions of the (assumed
Gaussian) individual components. R-Mix quickly converges, and provides the final
estimates of the means, dispersons, and fractions listed in columns (7)-(9) of
Table 1, respectively. The final column of Table 1 lists the so-called p-value,
which represents the probability of obtaining a value of reduced $\chi^2$ for
the fit as large (or larger) by chance. The results of the fits are shown in
Figure 8. It is clear, even from casual inspection, that the case of a single
cluster ($k = 1$) is a poor description of the data. This impression is
confirmed from the p-value of the chi-square statistic for this fit (0.002),
which strongly rejects the null hypothesis of a single cluster.

As can be seen, the $k = 2$ component fit appears much improved, and indeed the
p-value cannot reject the null hypothesis that two components are sufficient to
describe these data. The p-value for the $k = 3$ case is also unable to reject
the null hypotheses, but we emphasize that the quality of the initial partitions
from CLARA strongly suggest that $k = 2$ is an appropriate choice for the number
of components. A likelihood ratio test, which compares the statistical
improvement of the fit obtained from $k = 3$ components over that with $k = 2$
components, indicates only minimal improvement is obtained from the four
additional degrees of freedom (the two additional means and dispersions, and the
two additional mixture fractions). In other words, the value of the maximum
likelihood is not increased sufficiently to justify the need for an additional
component in the fit. The conclusion of this exercise is that, while larger
numbers of halo components beyond two can be {\it accommodated} by the available
kinematical information, they are not {\it required}, nor do they provide
superior fits to the data.

\subsection{The Maximum Likelihood Technique}

For more flexibility in our further analyses of mixture models, and to specify the
kinematic parameters of the various components in the local volume, it is
convenient to implement a maximum likelihood approach. This has been carried
out, making use the routine AMOEBA (Press et al. 1992) for performing searches
for the maximum in the likelihood function, as described below.

Maximum Likelihood (ML) analysis is a well-known general method for statistical
estimation. The basic theory of this technique assumes that the data can be
described by a probability density function, $f(X;\alpha)$, where X is a
variable, and $\alpha$ represents the parameter (or vector of parameters)
characterizing the known form of $f$. The aim of the method is then to estimate
$\alpha$. If X$_{1}$, X$_{1}$...X$_{n}$ are the individual data, and assuming
that they are independent and drawn from $f$, then the likelihood function can
be written as:

\bigskip

\begin{eqnarray}
L(X_{1},X_{2},...,X_{n}) &=& f(X_{1},X_{2},...,X_{n}|\alpha) \nonumber \\
                         &=& f(X_{1}|\alpha)f(X_{2}|\alpha)...f(X_{n}|\alpha) \nonumber \\
                         &=& {\displaystyle\prod_{1\leq i\leq n} f(X_{i}|\alpha)}
\end{eqnarray}

\noindent In its classical form, this equation gives the likelihood of obtaining the data,
given $\alpha$ (Wall \& Jenkins 2003).

The basic parameters we consider in this analysis, V$_{\phi}$ and Z$_{max}$, are
used to define a likelihood function such that all the main structural
components in the solar neighborhood (with the exception of the thin disk, which
contains very few stars in our sample) are included in $\alpha$.

It has been argued that the thick-disk component may include stars with
substantially lower abundance than its peak value (around [Fe/H] = $-$0.6).
Several authors have claimed a low-metallicity tail for the thick disk (the
MWTD) extending to stars as metal deficient as [Fe/H] $\sim -1.6$, or even lower
(Norris, Bessell, \& Pickles 1985; Morrison, Flynn, \& Freeman 1990; Beers \&
Sommer-Larson 1995; Layden 1995; Martin \& Morrison 1998; Chiba, Yoshii, \&
Beers 1999; Katz et al. 1999; Chiba \& Beers 2000; Beers et al. 2002). The
question of the limiting abundance of the MWTD has been considered several times
in the past (Rodgers \& Roberts 1993; Layden 1995; Beers \& Sommer-Larsen 1995;
Ryan \& Lambert 1995; Twarog \& Anthony-Twarog 1996; Chiba \& Beers 2000), and
in more recent work as well (Reddy \& Lambert 2008). In particular, Chiba \&
Beers (2000) claimed that the MWTD contributes about 30\% of the metal-poor
stars in the abundance range $-1.7 < $[Fe/H]$ \leq -1.0$ in the solar
neighborhood. Note that, in the majority of these efforts, the kinematics of the
MWTD were assumed to be similar (or identical) to those of the canonical thick
disk. The small samples of stars available previously for investigation of this
question limited the available options.

Ivezi\'{c} et al. (2008) adopted a different approach, modeling the observed
photometric metallicities and rotational velocities of main-sequence turnoff stars
from SDSS DR6 (Adelman-McCarthy et al. 2008) as a function of distance from the
Galactic plane, for a narrow cone at the North Galactic Pole. The marginal
distributions of metallicity and rotation velocities for these data were fit
{\it independent of one another} to non-Gaussian functions for the disk (and
single Gaussians in metallicity and rotation velocity for the halo), from which
an adequate description of the global behavior of the disk+halo was obtained.
We remind the reader that in our approach we seek to model the observed
properties of the stars in our local sample by assuming that stars are
associated with identifiable individual stellar populations, with differing
kinematics and metallicity distributions, that comprise the disk and halo
populations.

In the present paper we explore the possibility that the thick-disk system may
comprise two separate stellar populations, with independent kinematics, that we
associate with the canonical thick-disk and MWTD components. The Galactic halo is
assumed to be a two-component structure, comprising the inner and outer halos, as
discussed by Carollo et al. (2007). Thus, the general expression for the likelihood
function takes the form:

\begin{eqnarray}
\log f = \displaystyle\sum_{i=1}^n \log [F_{td}\cdot f^{i}_{td} + F_{mwtd}\cdot f^{i}_{mwtd} + F_{in}\cdot f^{i}_{in} +
F_{out}\cdot f^{i}_{out}]
\end{eqnarray}


\noindent where F$_{td}$, F$_{mwtd}$, F$_{in}$, and F$_{out}$ are the stellar
fractions of the thick disk, MWTD, inner halo, and outer halo, respectively.
On the left side of Eqn. 3, the function $f$ depends on the mean velocity,
dispersion, and stellar fraction of each component. On the right side of
the equation, the functions $f^{i}$ denote the (Gaussian) velocity distributions
adopted to represent each component.






Thus, the likelihood function has twelve output parameters that must be determined
(the fractions, and the mean rotation velocities and dispersions of each component).
However, the problem can be greatly simplified by working in different ranges of
metallicity, combined with different values of Z$_{max}$ (or $|$Z$|$), where one
component is dominant with respect to the others, and then fitting the distribution of
the rotational velocity in each range of metallicity separately. Adopting this
approach, we first consider the extremes of the MDF for the local sample ([Fe/H] $<
-2.0$ for the metal-poor sub-sample, and $-0.8 <$ [Fe/H] $< -0.6$ for the metal-rich
sub-sample). The first range is completely dominated by the two overlapping halo
components, the inner and outer halo; we expect that many of the stars in this
sub-sample belong to the outer halo, which has a peak in its MDF at [Fe/H] $\simeq
-2.2$ (Carollo et al. 2007). The second range contains stars belonging primarily to
the thick-disk population, whose MDF has a peak at [Fe/H] = $ -0.6$. The ML analysis
proceeds by specifying several input parameters {\it a-priori}, then holding them
fixed during subsequent analysis (e.g., by derivation of the global behavior of
V$_{\phi}$ at the extremes of the metallicity ranges). A similar approach was adopted
by Chiba \& Beers (2000) and other previous work.

In \S 5.4 we first derive the trends of $\langle$V$_{\phi}$$\rangle$ and
$\sigma_{V_{\phi}}$ for the low-metallicity sub-sample (the thick disk and MWTD
are assumed to not be present in this region of [Fe/H]) as a function of
Z$_{max}$. This analysis fixes the values of the mean rotational velocity and
dispersion for the outer-halo population by consideration of the extrema of
these trends. Once these values are specified, ML fits are obtained for the mean
rotation and dispersion of the inner-halo population.

In \S 5.5 we examine the trends of $\langle$V$_{\phi}$$\rangle$ and
$\sigma_{V_{\phi}}$ for the high-metallicity sub-sample, as a function of
$|$Z$|$, to derive the mean rotation velocity and dispersion for the thick-disk
population. We also derive the gradient in the rotation velocity for the thick
disk as a function of $|$Z$|$. The ML analysis is then extended to include the
full range of metallicity for the local sample, in order to derive the values of
the stellar fractions ($F_{td}$, $F_{in}$, $F_{out}$) as a function of
Z$_{max}$.

In \S 5.6 we consider stars in the intermediate-metallicity sub-sample ($-2.0 <$
[Fe/H] $< -0.6$), split into three distinct metallicity intervals, to assess the
requirement for a MWTD as an independent component. We then employ a
mixture-model analysis to test for the presence of a MWTD component in each of
these intervals as a function of $|$Z$|$. Finally, we explore the metallicity
range of the MWTD component.

\subsection{Extracting the Inner/Outer Halo Components in the Low-Metallicity Range}

\subsubsection{Characteristic Properties of the Outer-Halo Component}

Here we examine the properties of the mean Galactocentric rotational velocity,
$\langle$V$_{\phi}$$\rangle$, and its dispersion, $\sigma_{V_{\phi}}$, as a
function of the vertical distance, Z$_{max}$, for stars in the local sample
falling into the low-metallicity range ([Fe/H] $< -2.0$). At high Z$_{max}$ this
sample is expected to be dominated by halo stars with large retrograde motions,
suggesting membership in the outer-halo population.

Figure 9 shows the result of this exercise. Here, the values of
$\langle$V$_{\phi}$$\rangle$ and $\sigma_{V_{\phi}}$ are evaluated by passing a
box of 100 stars, with an overlap of 70 stars per bin, through the data. The
blue curves represent the sub-sample of stars in the range of metallicity $-1.5
<$ [Fe/H] $< -1.2$, shown for comparison purposes, while the red curves indicate
stars in the low-metallicity sub-sample. In the upper panel, for Z$_{max} <$ 5
kpc, the rotational velocities of the intermediate-metallicity stars exhibit a
strong gradient, declining from a value of $\sim$ 100 km~s$^{-1}$ down to an
average velocity of $\sim$ 0 km~s$^{-1}$. This behavior is likely due to the
overlap of two stellar populations, the MWTD and the inner halo, which dominates
the sample starting from Z$_{max}$ $\sim$ 5 kpc. Above this value of Z$_{max}$
the average rotational velocity is $\sim$ 0 km~s$^{-1}$, or slightly prograde,
over the range 5 kpc $<$ Z$_{max}$ $<$ 15 kpc, then trends toward retrograde
values as Z$_{max}$ increases.

The low-metallicity sub-sample exhibits an abrupt change of
$\langle$V$_{\phi}$$\rangle$ in the range of Z$_{max}$ between 5 and 10 kpc,
which indicates a transition between the inner- and outer-halo populations. The
$\langle$V$_{\phi}$$\rangle$ for this sub-sample then slowly trends to a
constant value $\langle$V$_{\phi}$$\rangle$ $\sim$ $-100$ km~s$^{-1}$ at large
Z$_{max}$, after undergoing a few oscillations that are likely due to
correlations between contiguous overlapping bins.

For comparison with previous work we have also considered the rotational
behavior of the low-metallicity sub-sample, as a function of $|$Z$|$, up to
$|$Z$|$ = 2 kpc. In this metallicity range we expect that most of the stars are
associated with the inner-halo component, with little contamination from the
MWTD (which we show below ceases contributing significant numbers of stars below
[Fe/H] $\sim -1.8$) or from outer-halo stars due to the proximity to the disk
plane. We confirm the existence of a gradient in the mean rotational velocity
over this range of $\triangle$$\langle$V$_{\phi}$$\rangle$/$\triangle |$Z$| =
-28$ $\pm$ 9 km~s$^{-1}$ kpc$^{-1}$, as also reported in Chiba \& Beers (2000)
(see discussion in \S 11.1).

The lower panel of Figure 9 shows the dispersion of the Galactocentric
rotational velocity, as a function of Z$_{max}$, for the two sub-samples. Note
that the velocity dispersions discussed here still include the effects of
observational errors, and are thus somewhat higher than their ``true'' values.
Below we make use of the ``as observed'' dispersions for the low-metallicity
sub-sample for ML modeling of the observed distribution of
V$_{\phi}$.\footnote{Below we correct, as appropriate, derived velocity
dispersions for the effects of measurement errors following the formalism of
Jones \& Walker (1988). In order to remove the effects of observational errors,
the term $\frac{1}{n}\displaystyle\sum_{i=1}^n\xi_{i}^2$, where $\xi_{i}$ is the
error on measurement, is subtracted from the observed dispersion.}

The velocity dispersion exhibits a mean value around 100 km~s$^{-1}$ for
Z$_{max} <$ 5 kpc for both sub-samples (slightly lower for
intermediate-metallicity stars, and slightly higher for low-metallicity stars).
At larger values of Z$_{max}$ the dispersion increases for both sub-samples,
with intermediate-metallicity stars trending to $\sigma_{V_{\phi}}$ $\sim$ 125
km s$^{-1}$, and low-metallicity stars trending to $\sigma_{V_{\phi}}$ $\sim$
190 km s$^{-1}$ for Z$_{max}$ $>$ 20 kpc.

Figure 9 clearly indicates that above Z$_{max}$ between 10 and 20 kpc, the
low-metallicity sub-sample is dominated by a highly retrograde halo component
with a large dispersion (outer halo), while the slightly prograde and lower
dispersion component (inner halo) is dominant for Z$_{max} <$ 10 kpc. For the
purpose of the ML analysis described in the following subsection the input
values of the mean rotational velocity and dispersion for the outer-halo
component are evaluated by choosing stars from the low-metallicity sub-sample
with Z$_{max} >$ 15 kpc (where contamination from the inner-halo population
should be minimal). For the outer-halo component this analysis leads to a value
for $\langle$V$_{\phi}$$\rangle$ = $-$ 80 $\pm$ 13 km~s$^{-1}$, slightly more
retrograde than that obtained by Carollo et al. (2007) ($\sim -70$ km~s$^{-1}$)
from the DR5 sample of local calibration stars. For the dispersion we obtain
$\sigma_{V_{\phi}}$ = 180 $\pm$ 9 km~s$^{-1}$ (uncorrected for observational
errors).

\subsubsection{Maximum Likelihood Results}

We adopt a likelihood function for V$_{\phi}$ for the low-metallicity sub-sample,
based on a mixture of two Gaussian distributions representing the two halo components,
with a normalization forcing their sum to unity (F$_{in}$ + F$_{out}$ = 1). The input
parameters for the likelihood function are the values of $\langle$V$_{\phi}$$\rangle$
and $\sigma_{V_{\phi}}$ for the outer halo, as obtained in the previous section; the
output parameters are the mean rotational velocity and its dispersion for the inner
halo and the stellar fractions for the inner- and outer-halo components. The fit is
then performed selecting different intervals of the vertical distance, Z$_{max}$.

As a first application, we have performed a fit for the low-metallicity
sub-sample of stars, selected with 1 kpc $<$ Z$_{max}$ $<$ 10 kpc, and Z$_{max}$
$>$ 10 kpc, respectively.\footnote{It should be noted that this sample differs
from that used in \S 5.2 above, which was employed to assess the appropriate number of
halo components.} While a contribution from a MWTD population
may still be present, the lower limit of Z$_{max}$ for the sub-sample close to the
Galactic plane has been chosen to reduce its possible impact.

Figure 10 shows the result of this exercise. The histograms in each panel are
the distribution of the observed V$_{\phi}$ in the two selected ranges of
Z$_{max}$, while the green (inner halo) and red (outer halo) curves represent
the results of the ML analysis for these two components; the blue curves are the
mixture models obtained by their sum. Visual inspection suggests that the fits
well-represent the observed distribution of the data for both cuts on Z$_{max}$.
It is clear that the inner-halo component dominates the sub-sample of stars at 1
kpc $<$ Z$_{max}$ $<$ 10 kpc, while the net-retrograde, high-dispersion
outer-halo population dominates for Z$_{max}$ $>$ 10 kpc.

Prior experience suggests that visual inspection of multiple-component fits are
often insufficient for interpreting results of mixture-model analyses. Thus, we
have also evaluated the so-called Double Root Residuals (DRRs) for these fits.
The DRR plot is a variation of a hanging histogram, which is just a
(non-normalized) difference plot between binned data and a proposed fitting
function. One limitation of the hanging histogram is that it gives equal weight
to residuals from bins near the extrema of a distribution (which are usually
poorly populated) as to residuals near the middle of a distribution (which are
usually well-populated, and hence subject to greater $\sqrt{N}/N$ variations). A
square-root transformation applied to the residuals provides a desirable
variance stabilization. The DRR plot is a refinement to the canonical
root-residual plot (Velleman \& Hoaglin 1981), which avoids some difficulties with
small data sets. The set of DRRs is obtained using:

\begin{mathletters}
\begin{eqnarray}
DRR &=& \sqrt{2 + 4(N_{obs})} - \sqrt{1 + 4(N_{fit})} \quad  {\rm if\, N_{obs}} \ge 1 \\
    &=& 1 - \sqrt{1 + 4(N_{fit})} \qquad \qquad \qquad \,\, {\rm if\, N_{obs}} =  0
\end{eqnarray}
\end{mathletters}

\noindent In the above, $N_{obs}$ and $N_{fit}$ are the number of observed and predicted stars in each bin, respectively.

There are several advantages of a DRR plot over a simple difference plot
for illustrating the residuals. First, it graphically emphasizes {\it where} in a
distribution lack of fit between data and model exists -- the square-root
transformation puts the residuals throughout the fit on an equal footing.
Secondly, if the model is close to an adequate fit to the data, then the DRRs
are roughly equivalent to normal deviates. Thus, a DRR with numerical value $>
\pm$ 2 is significant at the 95\% (2$\sigma$) level, whereas DRRs less than this
value indicate a reasonable agreement between data and model in a given bin. As
can be appreciated from inspection of the right-hand panels in Figure 10, the
DRRs for the inner/outer halo fits obtained from the ML analysis are acceptable
for most bins, with the exception of bins near the high positive and high
negative velocity regions for the 1 kpc $<$ Z$_{max}$ $<$ 10 kpc cut. Indeed,
the p-value of the reduced $\chi^2$ for this cut strongly rejects the null
hypothesis of a good fit ($p = 0.0003$).This suggests that complexity beyond the
simple description of an inner-halo population we have adopted may remain to be
explored.

By contrast, the DRRs for the Z$_{max} >$ 10 kpc cut, and the p-value of the
reduced $\chi^{2}$ for the global fit ($p = 0.950$), both indicate that the
proposed fit is acceptable. In this first application the utility of the DRRs
becomes clear. Visual inspection of the fit shown in the upper-left panel of
Figure 10 might have suggested that the proposed mixture model was adequate
(although the chi-square analysis rejected this hypothesis), while the DRRs
clearly revealed the location in the distribution where the lack of fit was most
obvious. In addition, the possible deviations from the proposed model fit in the
lower-left panel of Figure 10 might have suggested lack of fit, but the DRRs
indicated that the deviations are in fact not significant (as confirmed by the
chi-square analysis). It is quite difficult for the eye to separate out the
effects of root-N variations from real deviations for populations of different
sizes. Indeed, it is the smaller size of the population for the fit in the
lower-left panel, as well as the smaller individual bin populations, which
account for the fact that the proposed model cannot be rejected, while it {\it
is} rejected for the larger population shown in the upper-left panel.

To evaluate the mean rotational velocity, its dispersion, and the fractions of
stars for the inner-halo components, a second set of Z$_{max}$ cuts was
examined. In this case we fit sub-samples of stars selected with different upper
limits on Z$_{max}$ (cumulative Z$_{max}$), and keeping the lower limit set at 1
kpc. The complete set of the derived inner-halo parameters and their $1~\sigma$
errors are summarized in Table 2.

\begin{deluxetable}{crcc}[!t]
\tablewidth{0pt}
\tablenum{2}
\tablecaption{Maximum Likelihood Results for the Low-Metallicity Sub-sample --
Inner Halo Parameters}
\tablehead{
\colhead{Z$_{max}$} &
\colhead{$\langle$V$_{\phi,in}$$\rangle$} &
\colhead{$\sigma_{V_{\phi,in}}$} &
\colhead{F$_{in}$} \\
\colhead{(kpc)} &
\colhead{(km~s$^{-1})$} &
\colhead{(km~s$^{-1})$} &
\colhead{ }
}
\startdata
        &               &             &                 \\
1$-$5   &    11 $\pm$ 4 & 113 $\pm$ 3 &  1.00 $\pm$ 0.01\\
1$-$10  &     7 $\pm$ 4 & 110 $\pm$ 3 &  0.93 $\pm$ 0.01\\
1$-$15  &    10 $\pm$ 4 & 106 $\pm$ 3 &  0.83 $\pm$ 0.02\\
1$-$20  &    13 $\pm$ 4 & 105 $\pm$ 4 &  0.78 $\pm$ 0.02\\
1$-$25  &    20 $\pm$ 4 & 100 $\pm$ 3 &  0.70 $\pm$ 0.02\\
1$-$30  &    23 $\pm$ 4 &  97 $\pm$ 3 &  0.66 $\pm$ 0.02\\
\enddata
\tablecomments{The input parameters for the outer halo have been fixed at
($\langle$V$_{\phi,out}$$\rangle$, $\sigma_{V_{\phi,out}})$ = ($-80$, 180)
km~s$^{-1}$. The listed dispersions are not corrected for observational errors.
The normalization F$_{in}$ + F$_{out}$ = 1 is assumed. }
\end{deluxetable}

Estimates of the errors on the output parameters were obtained following a
well-known procedure, based on the assumption that the maximum of the
likelihood, $L_{max}$, is the best estimator for the parameters of the model,
and that the logarithm of the likelihood ratio, $-2\ln(\frac{L'}{L_{max}})$, is
asymptotically distributed like the $\chi^2$ distribution (e.g., Lupton 1993).

\subsection{Extracting the Thick-Disk Component and the Fractions of Each Galactic Component}

\subsubsection{Characteristic Properties of the Thick-Disk Component}

To extract information on the canonical thick-disk component we select a sub-sample
of stars in the metallicity range $-0.8 <$ [Fe/H] $< -0.6$, and derive its mean
rotational velocity and dispersion as a function of vertical distance, $|$Z$|$,
up to 4 kpc from the Galactic plane. It should be kept in mind that there may
remain some contamination from MWTD stars even in this high-metallicity range,
which might have an influence on the observed behavior.

Figure 11 shows the result of this exercise. As seen in the upper panel of this
figure, for $|$Z$|$ $<$ 1 kpc, the mean rotational velocity is $\sim$ 200
km~s$^{-1}$; as has been previously noted by several authors, it decreases at
larger heights above the plane. A least-squares fit to the data yields
$\triangle$$\langle$V$_{\phi}$$\rangle$/$\triangle |$Z$| = -36$ $\pm$ 1
km~s$^{-1}$ kpc$^{-1}$. Inspection of the lower panel of Figure 11 shows that,
for $|$Z$|$ $> 2$ kpc, $\sigma_{V_{\phi}}$ increases from the canonical value of
around 40-50 km~s$^{-1}$ to on the order of 90 km~s$^{-1}$. In the same panel we
note the presence of some oscillations that are larger than the errors on the
dispersion for each bin of $|$Z$|$. We suspect that this jagged structure of
$\sigma_{V_{\phi}}$ could be due to the presence of substructures, which will be
investigated in a future paper. The fiducial parameters we adopt for the thick
disk are obtained by selecting the sample at 1 $<$ $|$Z$|$ $<$ 2 kpc, in the
high-metallicity range mentioned above. This selection is intended to avoid
contamination by thin-disk stars (including those with possibly mis-estimated
metallicities).

We obtain a mean velocity and dispersion of $\langle$V$_{\phi}$$\rangle$ = 182
$\pm$ 2 km s$^{-1}$ and $\sigma_{V_{\phi}}$ = 57 $\pm$ 1 km~s$^{-1}$
(uncorrected for the observational errors, which are typically on the order of
5-6 km s$^{-1}$ for large sub-samples, and 10-15 km~s$^{-1}$ for smaller
sub-samples), respectively.

\subsubsection{Maximum Likelihood Results}

We now evaluate the stellar fractions of the thick-disk, inner-halo, and
outer-halo components with the ML analysis, as a function of Z$_{max}$,
including stars over the full metallicity range of our local sample. The
analysis was performed by selecting sub-samples of stars located in {\it strips}
of Z$_{max}$ (differential Z$_{max}$). Recall that previously, for the
low-metallicity sub-sample, cuts were selected for different values of the upper
limit on Z$_{max}$ (cumulative Z$_{max}$). The differential distribution is
better able to respond to changes in the dominance of the overlapping
structures, and thus is suitable for examination of the trend of the stellar
fractions as a function of Z$_{max}$.

The mean rotational velocities and dispersions for each component are fixed to the
values obtained above (dispersions uncorrected for the observational errors):
($\langle$V$_{\phi,td}$$\rangle$, $\sigma_{V_{\phi,td}})$ = (182, 57) km~s$^{-1}$;
($\langle$V$_{\phi,in}$$\rangle$, $\sigma_{V_{\phi, in}})$ = (7, 110) km~s$^{-1}$;
($\langle$V$_{\phi,out}$$\rangle$, $\sigma_{V_{\phi,out}})$ = ($-80$, 180)
km~s$^{-1}$. Note that for the inner-halo parameters we have adopted the values
listed in Table 2 for 1 $<$ Z$_{max} < 10$ kpc. Values for these three components are taken
as input parameters for the ML analysis, while the output parameters are the stellar
fractions, F$_{td}$, F$_{in}$, and F$_{out}$, respectively. Note that it is not
strictly necessary that we fix the input parameters of the various components; we do
this in order to obtain robust estimates of their fractions, which would change if the
inputs were allowed to vary.

The likelihood function comprises three components, F$_{td}$$f_{td}$, F$_{in}$$f_{in}$
and F$_{out}$$f_{out}$; as before, the $f$s are assumed to be Gaussian
distributions representing the corresponding V$_{\phi}$ components. The likelihood
function is normalized to F$_{td}$ + F$_{in}$ + F$_{out}$ = 1. The results of the
analysis are shown in Figure 12. It is clear from inspection of the left-hand
panels of Figure 12 that, for Z$_{max} <$ 5 kpc, only two components contribute
significantly -- the thick disk and the inner halo. At 5 $<$ Z$_{max}$ $<$ 10
kpc, only the inner-halo component contributes (the green line representing this
component exactly overlays the blue mixture-model line). At 10 $<$ Z$_{max}$ $<$
15 kpc the outer halo begins to contribute, although the inner halo is still
dominant. At 15 $<$ Z$_{max} <$ 20 kpc, the inner halo and outer halo make
comparable contributions. Finally, when Z$_{max} >$ 20 kpc, the outer-halo
component is completely dominant, with very little contribution from the
inner-halo component.

The right-hand panels of Figure 12 are the DRRs of the corresponding fits. The
individual bin residuals for the subsets of the data above 15 kpc are consistent
with adequate fits (as are the p-values of their reduced $\chi^2$ determined for
the global fits), but those for the cuts below 15 kpc exhibit significant
deviations that may be of interest. We consider these in turn.

We originally assumed that the extreme negative residual seen for Z$_{max} <$ 5
kpc at highly prograde V$_{\phi}$ arose primarily from the absence of a
thin-disk component in our model. However, additional experiments were performed
over the distance range 1 $<$ Z$_{max}$ $<$ 5 kpc, to explicitly avoid
substantial thin-disk contamination; the lack of fit remained.

We also considered a possible difference in the thick-disk fiducial parameters
when evaluated using Z$_{max}$ instead of $|$Z$|$. The estimated values obtained
in the range 1 $<$ Z$_{max}$ $<$ 2 kpc are $\langle$V$_{\phi}$$\rangle$ = 190
$\pm$ 2 km s$^{-1}$ and $\sigma_{V_{\phi}}$ = 46 $\pm$ 1 km~s$^{-1}$ (not
corrected for observational errors). When the ML analysis is performed with
these thick-disk parameters, the lack of fit is slightly reduced, but the
residuals in the velocity range 0 $\lesssim $ V$_{\phi} \lesssim$ 150
km~s$^{-1}$ increase. In this velocity range the lack of fit may indicate the
presence of a MWTD component. To explore this possibility, a further experiment
has been performed, adding a fourth component in the model to fit the sub-sample
of stars in the upper panel of Figure 12, using the R-Mix modeling approach. We
find that a four-component model provides a better fit (p-value = 0.11)
.\footnote{With derived velocity dispersions $\sigma_{V_{\phi,I}}$ = 97 $\pm$ 4
km~s$^{-1}$, $\sigma_{V_{\phi,II}}$ = 59 $\pm$ 19 km~s$^{-1}$, $\sigma_{V_{\phi,
III}}$ = 37 $\pm$ 19 km~s$^{-1}$, and $\sigma_{V_{\phi,IV}}$ = 31 $\pm$ 8
kms$^{-1}$, and fractions for each component of F$_{I}$ = 0.26, F$_{II}$ =
0.20, F$_{III}$ = 0.30 and F$_{IV}$ = 0.24, respectively.}

With no restrictions on metallicity, there is considerable latitude allowed for
the kinematic parameters derived for each component. This is, unfortunately,
rather difficult to overcome on the basis of the simple mixture-model analysis
we have employed. More detailed modeling, taking metallicity into account, is
described below.

This experiment shows that, close to the Galactic plane, the structure of the
Galaxy is complex, and a fourth component is required to obtain a better fit.
However, for the purpose of obtaining the stellar fractions of the inner- and
outer-halo components as a function of the vertical distance Z$_{max}$, which is
our primary objective in this section, we have chosen to employ the simpler
three-component model, even though the fit is not optimal over all distance
cuts. Issues related to the MWTD are discussed in the following sections.

Figure 13 shows the derived fractions of the three components in our basic
model, the canonical thick disk, the inner halo, and the outer halo, as a
function of Z$_{max}$, as derived from the ML analysis. The thick disk is
denoted by a black dot, and is only present, as expected, in the first bin of
vertical distance (0 $<$ Z$_{max}$ $<$ 5 kpc). In this range the fraction of the
thick-disk and inner-halo (green dots) components are roughly equal. Above
Z$_{max}$ = 5 kpc, only the inner-halo and outer-halo (red dots) components are
present. The inner-halo stellar fraction decreases with increasing Z$_{max}$,
while the outer-halo stellar fraction increases. The {\it inversion point} in
the dominance of the inner/outer halo occurs in the range of Z$_{max}$ between
15 and 20 kpc.

Table 3 reports the values of the stellar fractions for the three components, and
their errors, evaluated as before.

\begin{deluxetable}{cccc}[!t]
\tablewidth{0pt}
\tablenum{3}
\tablecaption{Stellar Fractions for the Thick Disk, Inner Halo, \\
 and Outer Halo}
\tablehead{
\colhead{Z$_{max}$} &
\colhead{F$_{td}$} &
\colhead{F$_{in}$} &
\colhead{F$_{out}$} \\
\colhead{(kpc)} &
\colhead{} &
\colhead{} &
\colhead{}
}
\startdata
          &                   &                 &                  \\
  $<$ 5   &  0.55 $\pm$  0.03 & 0.45 $\pm$ 0.02 &  0.00  $\pm$ 0.02\\
 5$-$10   &  0.00 $\pm$  0.03 & 1.00 $\pm$ 0.02 &  0.00  $\pm$ 0.02\\
 10$-$15  &  0.00 $\pm$  0.04 & 0.80 $\pm$ 0.03 &  0.20  $\pm$ 0.02\\
 15$-$20  &  0.00 $\pm$  0.05 & 0.55 $\pm$ 0.04 &  0.45  $\pm$ 0.03\\
 $>$ 20   &  0.00 $\pm$  0.02 & 0.08 $\pm$ 0.01 &  0.92  $\pm$ 0.02\\
\enddata
\tablecomments{The input parameters for all three components have been
fixed at the following values:  ($\langle$V$_{\phi,td}$$\rangle$, $\sigma_{V_{\phi,td}})$
= (182, 57) km~s$^{-1}$; ($\langle$V$_{\phi,in}$$\rangle$, $\sigma_{V_{\phi,
in}})$ = (7, 110) km~s$^{-1}$; ($\langle$V$_{\phi,out}$$\rangle$,
$\sigma_{V_{\phi,out}})$ = ($-80$, 180) km~s$^{-1}$, respectively.
The normalization F$_{td}$ + F$_{in}$ + F$_{out}$ = 1 is assumed.}
\end{deluxetable}

\subsection{How Many Components in the Stellar Disk ?}

We now consider the nature of the disk populations, and in particular explore
the question of whether an independent MWTD must be present in order to account
for the rotational properties of the stars in our local sample close to the
Galactic plane. The approach used here is similar to that employed for modeling
of the Galactic halo components. We seek to model the kinematics of the
thick-disk system using a minimum number of components, assuming as before that
the presence (or not) of individual substructures in each component is not
relevant for our approach. We also assume that Schwarzschild velocity ellipsoids
apply, and that the ellipsoids are aligned to the cylindrical coordinate system
we employ.

In an initial set of experiments, we used the ML method to explore stars in an
intermediate-metallicity range ($-2.0 <$ [Fe/H] $< -0.6$) over different
intervals in vertical distance from the plane, $|$Z$|$, but the results were
inconclusive. The overlaps in the rotational behavior of the various components
present over such a wide metallicity range made it difficult to isolate the
kinematic properties of the individual components. Instead, we hope to simplify
the problem by dividing our local sample of stars into three primary metallicity
regions, which are intended to include as few populations as possible. These
regions are ({\it i}) $-2.0 <$ [Fe/H] $< -1.5$, ({\it ii}) $-1.5 <$ [Fe/H] $<
-1.0$, and ({\it iii}) $-1.0 <$ [Fe/H] $< -0.6$. For the purpose of this
discussion, we consider the contribution from a given component to be
significant if its fraction, as assigned by the mixture-modeling analysis, is
$>$ 10\%.

We proceed by carrying out objective initial partitions for each metallicity
region in two sets of intervals on distance from the plane, 1 $<$ $|$Z$|$ $< 2$
kpc, and 2 $<$ $|$Z$|$ $< 4$ kpc. The lower limit of the first interval is set
to 1 kpc to avoid contamination from stars of the thin-disk population, which,
although they might not be expected to be present at metallicities [Fe/H] $<
-0.6$, may have had their metallicities slightly mis-estimated by the current
SSPP, and entered the sample unintentionally. As previously, we used CLARA to
obtain splits of the V$_{\phi}$ parameter for cases of $k = 1$, $k = 2$, and $k
= 3$ clusters. The initial estimates for $\langle$V$_{\phi}$$\rangle$, and its
dispersion, $\sigma_{V_{\phi}}$, were then used as inputs for the R-Mix routine,
which produced final estimates of the best-fit mixture models for each
metallicity region and distance interval. Note that for this set of experiments
all parameters are allowed to be ``free''; we do not hold any of their
properties fixed during the modeling procedure.

The results of the above experiment are listed in Table 4 and shown in Figures
14 and 15. Table 4 is divided into two sections corresponding to the intervals
on vertical distance; within each section the results for the one-, two-, and
three-cluster splits are considered separately. Column (1) of this table lists
the metallicity region while column (2) lists the numbers of stars in the cut.
Columns (3)-(5) list the resulting $\langle$V$_{\phi}$$\rangle$, its dispersion,
$\sigma_{V_{\phi}}$, and the fractions obtained from the R-Mix analysis for
component I, respectively. Columns (6)-(8) list the same quantities, where
applicable, for component II. Columns (9)-(11) list the same quantities, where
applicable, for component III. The final column lists the p-value for the
chi-square of the residuals to the listed fit.

\begin{deluxetable*}{cccccccccccc}[!t]
\tablewidth{0pt}
\tabletypesize{\scriptsize}
\tablenum{4}
\tablecolumns{12}
\tablecaption{R-Mix Results for Multiple-Component Fits to the Local Sample}
\tablehead{
\colhead{[Fe/H]} &
\colhead{N$_{Stars}$} &
\colhead{$\langle$V$_{\phi, I}$$\rangle$} &
\colhead{$\sigma_{V_{\phi, I}}$} &
\colhead{F$_{I}$} &
\colhead{$\langle$V$_{\phi, II}$$\rangle$} &
\colhead{$\sigma_{V_{\phi, II}}$} &
\colhead{F$_{II}$} &
\colhead{$\langle$V$_{\phi, III}$$\rangle$} &
\colhead{$\sigma_{V_{\phi, III}}$} &
\colhead{F$_{III}$} &
\colhead{p-value} \\
\colhead{} &
\colhead{} &
\colhead{(km~s$^{-1}$)} &
\colhead{(km~s$^{-1}$)} &
\colhead{} &
\colhead{(km~s$^{-1}$)} &
\colhead{(km~s$^{-1}$)} &
\colhead{} &
\colhead{(km~s$^{-1}$)} &
\colhead{(km~s$^{-1}$)} &
\colhead{} &
\colhead{}\\
\colhead{(1)} &
\colhead{(2)} &
\colhead{(3)} &
\colhead{(4)} &
\colhead{(5)} &
\colhead{(6)} &
\colhead{(7)} &
\colhead{(8)} &
\colhead{(9)} &
\colhead{(10)} &
\colhead{(11)} &
\colhead{(12)}
}
\startdata
\cutinhead{Close to the Galactic Plane ($1 < $ $|$Z$|$ $ < 2$ kpc)}\\
                 &      &      &     &      &         &         &         &         &         &         &            \\
\multicolumn{12}{c}{\it Single Component}\\
                 &      &      &     &      &         &         &         &         &         &         &            \\
$-$2.0 to $-$1.5 & 1851 & $-$10& 103 & 1.00 & \nodata & \nodata & \nodata & \nodata & \nodata & \nodata & $< 0.001$  \\
$-$1.5 to $-$1.0 & 2277 &  46  & 104 & 1.00 & \nodata & \nodata & \nodata & \nodata & \nodata & \nodata & $< 0.001$  \\
$-$1.0 to $-$0.6 & 2217 & 146  &  64 & 1.00 & \nodata & \nodata & \nodata & \nodata & \nodata & \nodata & $< 0.001$  \\
                 &      &      &     &      &         &         &         &         &         &         &            \\
\multicolumn{12}{c}{\it Two Components}\\
                 &      &      &     &      &         &         &         &         &         &         &            \\
$-$2.0 to $-$1.5 & 1851 &$-$104& 136 & 0.13 &    4    &  89     & 0.86    & \nodata & \nodata & \nodata & $<$ 0.01   \\
$-$1.5 to $-$1.0 & 2277 &    1 &  96 & 0.67 &  135    &  44     & 0.33    & \nodata & \nodata & \nodata & $<$ 0.001  \\
$-$1.0 to $-$0.6 & 2217 &   64 &  77 & 0.19 &  165    &  42     & 0.81    & \nodata & \nodata & \nodata & $<$ 0.001  \\
                 &      &      &     &      &         &         &         &         &         &         &            \\
\multicolumn{12}{c}{\it Three Components}\\
                 &      &      &     &      &         &         &         &         &         &         &            \\
$-$2.0 to $-$1.5 & 1851 & $-$95& 152 & 0.11 &  $-$24  &  83     & 0.71    &  93     & 48      & 0.19    & 0.272      \\
$-$1.5 to $-$1.0 & 2277 &$-$193& 142 & 0.01 &  $-$3   &  88     & 0.63    & 135     & 45      & 0.36    & 0.175      \\
$-$1.0 to $-$0.6 & 2217 &   54 &  79 & 0.15 &  146    &  40     & 0.58    & 198     & 25      & 0.27    & 0.137      \\
                 &      &      &     &      &         &         &         &         &         &         &            \\
\cutinhead{Farther from the Galactic Plane ($2 < $ $|$Z$|$ $< 4$ kpc)}\\
                 &      &      &     &      &         &         &         &         &         &         &            \\
\multicolumn{12}{c}{\it Single Component}\\
                 &      &      &     &      &         &         &         &         &         &         &            \\
$-$2.0 to $-$1.5 & 2115 & $-$30& 117 & 1.00 & \nodata & \nodata & \nodata & \nodata & \nodata & \nodata & $< 0.001$  \\
$-$1.5 to $-$1.0 & 2710 &   14 & 104 & 1.00 & \nodata & \nodata & \nodata & \nodata & \nodata & \nodata & $< 0.001$  \\
$-$1.0 to $-$0.6 & 1963 &  108 &  81 & 1.00 & \nodata & \nodata & \nodata & \nodata & \nodata & \nodata & $< 0.001$  \\
                 &      &      &     &      &         &         &         &         &         &         &            \\
\multicolumn{12}{c}{\it Two Components}\\
                 &      &      &     &      &         &         &         &         &         &         &            \\
$-$2.0 to $-$1.5 & 2115 &$-$162& 156 & 0.14 &  $-$9   &  93     & 0.86    & \nodata & \nodata & \nodata & 0.037      \\
$-$1.5 to $-$1.0 & 2710 &  $-$7&  98 & 0.85 &  133    &  45     & 0.15    & \nodata & \nodata & \nodata & 0.007      \\
$-$1.0 to $-$0.6 & 1963 &   32 &  88 & 0.29 &  139    &  52     & 0.71    & \nodata & \nodata & \nodata & 0.058      \\
                 &      &      &     &      &         &         &         &         &         &         &            \\
\multicolumn{12}{c}{\it Three Components}\\
                 &      &      &     &      &         &         &         &         &         &         &            \\
$-$2.0 to $-$1.5 & 2115 &$-$311& 200 & 0.02 & $-$168  &  95     & 0.14    &   0     & 88      & 0.84    &  0.857     \\
$-$1.5 to $-$1.0 & 2710 &$-$121& 134 & 0.03 &   $-$8  &  91     & 0.80    & 134     & 48      & 0.17    &  0.816     \\
$-$1.0 to $-$0.6 & 1963 &   38 &  88 & 0.33 &  133    &  45     & 0.60    & 220     & 11      & 0.07    &  0.096     \\
\enddata
\tablecomments{In the labels: I = Comp. I, II = Comp. II, and III = Comp. III}
\end{deluxetable*}

\subsubsection{Close to the Galactic Plane (1 $<$ $|$Z$|$ $ <$ 2 kpc) }

We first consider the behavior for the distance interval close to the Galactic plane.

The p-values for the one-cluster and two-cluster splits clearly indicate that there is
no metallicity region over which acceptable mixture models exist. Acceptable fits are
only obtained for the three-cluster splits, as indicated by the listed p-values. This
is also reflected in the appearance of the components shown in Figure 14, and their
associated DRRs. Note that in the case of metallicity region {\it iii} (the higher
metallicity range shown in the lower panel of Figure 14), the DRRs do suggest some
lack of fit in the region near 220 km~s$^{-1}$, which might suggest contamination from
thin-disk stars. However, recall that our distance interval was set to avoid such a
contamination. We return to this point below.

Examination of the mean rotational velocities and dispersions for each of the
metallicity regions for the three-cluster split is informative. Notice that for
metallicity region {\it i} (the lowest metallicity considered), the three components
are all considered significant (fractions $>$ 10\%) contributors, and that their means
and dispersions might be plausibly assigned to (I) the outer halo, (II) the inner
halo, and (III) the MWTD. There is no component with the properties of the canonical
thick disk present in this metallicity region, as expected. Inspection of the upper
panel of Figure 14 is also revealing. The ``flat-topped'' (more properly, platykurtic)
appearance of the distribution of V$_{\phi}$ seen here is a classic signature of a
mixture of populations. Furthermore, the long tail in V$_{\phi}$ extending to large
negative velocities requires an additional component in order to obtain an acceptable
fit.

For metallicity region {\it ii} (intermediate-metallicity stars), there are only
two significant contributors, components II and III. Examination of their means
and dispersions suggest their identification with the inner-halo and MWTD
components, respectively. The appearance of the middle panel of Figure 14
supports this interpretation as well. Note that the dominance of components II
and III is so great that one might have expected that a two-cluster split would
have been acceptable. However, some contamination from likely outer-halo stars
at high negative velocities pushes the formal chi-square analysis into the
region where the two-cluster hypothesis can be rejected. The derived means and
dispersions of the dominant components in either the two-cluster or
three-cluster splits are similar, in any case. The component we identify with
the inner halo accounts for roughly 60\% of the stars, while that identified
with the MWTD accounts for roughly 40\% of the stars in this metallicity region.

For metallicity region {\it iii} (the highest metallicity considered), all three
components are significant contributors. Inspection of the means and dispersions
suggests identification of the individual components with (I) the inner halo, (II) the
MWTD, and (III) the canonical thick disk. It is of interest that the component we
identify with the MWTD is the dominant contributor in this metallicity region,
accounting for roughly 60\% of the stars, while the component we identify with the
inner halo only contributes 15\%, the remaining 25\% of the stars being accounted for
by the canonical thick disk.

The possible lack of fit at high rotation velocities revealed by the DRRs
is not likely to be simply the resulthe 1 kpc from the Galactic plane. As shown
below, this component appears even more strongly present farther from the plane.

For all three metallicity regions we conclude that an independent MWTD
component, with $\langle$V$_{\phi}$$\rangle$ in the range 100-150 km~s$^{-1}$,
and $\sigma_{V_{\phi}}$ in the range 40-50 km s$^{-1}$ ($\sim$ 35-45
km~s$^{-1}$, when corrected for observational errors), is required in order to
account for the rotational behavior of the stars in our local sample located
relatively close to the Galactic plane.

\subsubsection{Additional Splits of the Nearby Sample on $|$Z$|$ and R}

At the suggestion of the referee, we have carried out additional checks on the
nature of the stellar populations close to the disk plane, and on the robustness
of our results. We first split the local sample of stars with 1 $<$ $|$Z$|$ $ <$
2 kpc into two pieces: sub-sample A corresponding to stars above the Galactic
plane (1 $<$ Z $<$ 2 kpc), and sub-sample B corresponding to stars below the
plane ($-2 <$ Z $< -1$ kpc). The CLARA analysis with one-, two-, and three-cluster
splits, for all three metallicity regions, was carried out within each
sub-sample, followed by an R-mix mixture-model analysis using the starting
points derived from CLARA. It is worth noting that, due to the much higher
sampling frequency of SDSS for stars in the North Galactic Hemispere, the total
numbers of stars in each of these splits was typically 80\% in sub-sample A,
and 20\% in sub-sample B.

The results of this exercise are as follows. For sub-sample A (the 5749 stars
above the plane), in all metallicity regions, one- and two-cluster splits were
strongly rejected (as before), while three-cluster splits produced acceptable fits.
As found for the full-sample analysis, the intermediate-metallicity component in
the three-cluster split provided evidence for only two significant contributors
that we identify as the inner halo and the MWTD (a highly retrograde component
is represented by a fractional contribution of 0.5\%, compared with a
fractional contribution of 1\% in the full-sample analysis).

For sub-sample B (the 1669 stars below the plane), in all metallicity regions, a
one-cluster split was strongly rejected. Two-cluster splits were rejected for
metallicity regions {\it i} and {\it iii}, but an acceptable fit was found for
the intermediate-metallicity stars of region {\it ii}. The derived parameters
for this split are consistent with the association of an inner-halo component
($\langle$V$_{\phi}$$\rangle$ = $-6$ km~s$^{-1}$, $\sigma_{V_{\phi}}$ = 104
km~s$^{-1}$) and a MWTD component ($\langle$V$_{\phi}$$\rangle$ = 150
km~s$^{-1}$, $\sigma_{V_{\phi}}$ = 45 km~s$^{-1}$). Once again, as found for the
full-sample analysis, the intermediate-metallicity region in the three-cluster split
provided evidence for only two significant contributors, identified as the inner
halo and MWTD (a highly retrograde component is represented by a fractional
contribution of 0.8\%). The three-cluster split for the higher-metallicity region,
while acceptable, had only two significant contributors, associated with the inner
halo and MWTD (a highly retrograde component contributed a 7\% fraction). We
conclude that, in spite of the rather large difference in the numbers of stars
contained in the two sub-samples, individual analysis of these do not differ
significantly from the analysis of the entire sample in this distance cut.

Similar to the above, we then split the local sample of stars with 1 $<$ $|$Z$|$
$ <$ 2 kpc into two pieces: sub-sample C corresponding to stars located in the
inner solar cylinder (7 $<$ R $<$ 8.5 kpc), and sub-sample D corresponding to
stars located in the outer solar cylinder (8.5 $<$ R $<$ 10.0 kpc). The CLARA analysis
with one-, two-, and three-cluster split was carried out within each sub-sample,
and for all metallicity regions, followed by an R-mix mixture-model analysis
using the starting points derived from CLARA. In the case of the splits on
R, similar numbers of stars were included in each sub-sample (C: 3287 stars; D: 4129
stars).

The results of this exercise are as follows. For sub-sample C, in all
metallicity regions, one-cluster splits were strongly rejected (as before).
Two-cluster splits were rejected for metallicity regions {\it i} and {\it ii},
while an acceptable two-cluster split was obtained for the higher-metallicity
stars (metallicity region {\it iii}). The derived parameters for this split are
roughly consistent with association of an inner-halo component
($\langle$V$_{\phi}$$\rangle$ = 61 km~s$^{-1}$, $\sigma_{V_{\phi}}$ = 82
km~s$^{-1}$) and a MWTD component ($\langle$V$_{\phi}$$\rangle$ = 154
km~s$^{-1}$, $\sigma_{V_{\phi}}$ = 43 km~s$^{-1}$). Three-cluster splits were
found to be acceptable for metallicity regions {\it i} and {\it iii}, but were
marginally rejected (p = 0.04) in the case of the intermediate-metallicity stars
of region {\it ii}.

For sub-sample D, in all metallicity regions, 1-cluster splits were strongly
rejected (as before). Two-cluster splits were rejected for metallicity regions
{\it ii} and {\it iii}, while an acceptable 2-cluster split was obtained for the
lower-metallicity stars in metallicity region {\it i}. The derived parameters
for this split are consistent with the association with an outer-halo component
($\langle$V$_{\phi}$$\rangle$ = $-97$ km~s$^{-1}$, $\sigma_{V_{\phi}}$ = 166
km~s$^{-1}$) and an inner-halo component ($\langle$V$_{\phi}$$\rangle$ = $-5$
km~s$^{-1}$, $\sigma_{V_{\phi}}$ = 87 km~s$^{-1}$). Three-cluster splits were
found to be acceptable in all three metallicity regions.

We conclude that the results obtained from these additional splits are fully
consistent, within the expected variation induced by the different sub-sample
sizes, with the previous results obtained of the full local samples in this
distance interval.

\subsubsection{Farther from the Galactic Plane (2 $<$ $|$Z$|$ $ <$ 4 kpc)}

In this distance interval we expect that the contribution from the outer-halo
component will increase in metallicity region {\it i}, while in this same region
the contribution of the MWTD should be minimal. For metallicity region {\it ii}
we expect that the MWTD, although present, should be of less importance than the
contribution from the inner-halo component. For metallicity region {\it iii} we
might expect that, as in the case of the distance interval close to the Galactic
plane, the MWTD should dominate over the inner-halo component. As described
below, all of these expectations are indeed met.

Inspection of Table 4 indicates that, as before, the p-values for the
one-cluster and, to a somewhat lesser extent, the two-cluster splits, indicate
that there is no metallicity region over which acceptable mixture-model fits
exist. The p-value for the two-cluster split for metallicity region {\it iii} is
marginal (p = 0.058), indicating that a two-cluster split is almost an
acceptable fit. If one were to accept this split as reasonable, the mean
velocities and dispersions suggest identification of component I with the inner
halo, and component II with the MWTD. Statistically acceptable fits are only
obtained for the three-cluster splits, as indicated by the listed p-values. This
is also reflected in the appearance of the components shown in Figure 15, and
their associated DRRs (with the exception of the lower panel for metallicity
region {\it iii}, which suggests an even greater lack of fit for high rotation
velocities than for the distance interval close to the plane).

We now examine the three-cluster splits in more detail.

For metallicity region {\it i} only components II and III are significant
contributors. As can be seen in Table 4, the R-Mix procedure splits off a highly
retrograde component, which may well be spurious, and in any case is only a
minor (2\%) contributor. 	

For metallicity region {\it ii} we identify component I with the outer halo,
component II with the inner halo, and component III with the MWTD. Note,
however, that component I again makes only a minor contribution, although its
inclusion is apparently necessary to account for the asymmetric lower tail of
the rotational velocity distribution.

For metallicity region {\it iii} we identify component I with the inner halo,
and component II with the MWTD. Component III (which is clearly required, as
seen from examination of the DRRs in the lower panel of Figure 15) makes only a
minor (7\%) contribution. This high-velocity feature is of interest, since it is
stronger than the similar feature identified in the DRRs of this metallicity
region in the distance interval closer to the plane. Further investigation of
the nature of the stars in this region is clearly necessary, and is already
underway. Stars with ``thin-disk-like" metallicities and roughly solar (rather
than thick-disk-like) [$\alpha$-element/Fe] ratios located several kpc above the
Galactic plane have been previously reported by Lee \& Beers (2009), who
speculated that such stars may have been originally formed in the thin disk, but
were perturbed to large distances from the plane by past minor mergers. We defer
a full discussion to a future paper.

We conclude that an independent MWTD component must exist in the region farther
from the Galactic plane as well, in order to accommodate the observed
rotational behavior of our local sample of stars. It is interesting to note how
dominant the component we associate with the MWTD is (60\%), relative to the
component we identify with the inner halo (33\%), for the metal-rich metallicity
region. These roles quickly reverse in the intermediate-metallicity region, with
the component associated with the inner halo accounting for 80\% of the stars in
this sample, and the component identified with the MWTD only 17\%.

\subsubsection{The Metallicity Range of the MWTD}

Having established that a kinematically independent MWTD is indeed required in
order to understand the rotational behavior of our local sample of stars, we now
use a similar approach to the above to determine over what metallicity interval
the MWTD is present. Ideally, we would like to have information on the shape of
the MDF for this component, but our present data (due to the selection biases
involved) is not suitable for such an analysis. We can, however, determine the
upper and lower limits of its MDF. We accomplish this by considering metallicity
regions {\it i} and {\it iii} (the lower- and higher-metallicity regions,
respectively) for the distance interval close to the Galactic plane, where the
MWTD component is more dominant and less overlapped with the inner-halo
component than is the case for the interval farther from the plane.

For metallicity region {\it i} we repeat the $k = 2$ and $k =3$ cluster-split
experiments, lowering the upper limit on metallicity from [Fe/H] $< -1.5$, in
steps of 0.1 dex, until a three-cluster split is no longer required, and a
two-cluster split is adequate. That is, we seek the limiting metallicity where
only the inner- and outer-halo components make significant ($>$ 10\%)
contributions. We find that this occurs for the limit [Fe/H] $< -1.8$.
Similarly, for metallicity region {\it iii}, we repeat the $k = 2$ and $k =3$
cluster-split experiments, but this time raising the lower limit on metallicity
from [Fe/H] $> -1.0$, in steps of 0.1 dex, until a three-cluster split is no
longer required, and a two-cluster split is adequate. A successful two-cluster
split would then only require the presence of an inner-halo and a canonical
thick-disk component. We find that a two-cluster split is adequate for [Fe/H] $>
-0.8$, although there is weak evidence for the presence of the MWTD even for
$-0.7 <$ [Fe/H] $< -0.6$.

We conclude from this exercise that the MWTD is a significant contributor to the
numbers of stars in the local volume over the metallicity range $-$1.8 $<
$ [Fe/H] $< -$0.8, and possibly up to [Fe/H] $\sim$ $-$0.7.

\bigskip
\bigskip

\section{Velocity Ellipsoids}

We now derive estimates of the velocity ellipsoids for the thick disk, inner
halo, and outer halo. A similar derivation has been attempted for the MWTD, as
discussed at the end of this section. Note that all values of the velocity
dispersions reported in this section are corrected for observational errors.

As already shown in Figure 6, the mean velocity in the (cylindrical coordinate)
R and Z directions does not change with metallicity, remaining close to zero
over all ranges of [Fe/H] considered. The behavior of the velocity dispersions
is quite different, revealing a strong dependence on metallicity, as shown in
Figure 16. In this figure the observed velocity dispersions for the V$_{R}$ and
V$_{Z}$ components are plotted versus [Fe/H], with each bin representing an
interval in metallicity of 0.1 dex. The observed trends are certainly related to
the differing contributions of the various components in space and metallicity.
In both panels the change in the behavior of the velocity dispersion versus
metallicity (change in slope, and the increasing values) is particularly clear
at $-1.0 <$ [Fe/H] $< -0.5$, $-2.0 <$ [Fe/H] $< -1.0$, and [Fe/H] $< -2.0$. The
first metallicity interval is dominated by the thick-disk population, with a
mean dispersion of $\sigma_{V_{R}} \sim$ 60 km~s$^{-1}$ and $\sigma_{V_{Z}}
\sim$ 40 km~s$^{-1}$. In the second metallicity range the sample is dominated by the
inner-halo population, with some overlap from the MWTD. Finally, the most
metal-deficient stars, which are dominated by members of the outer-halo population,
exhibit much higher dispersions. The dependence (or not) of the shape of the halo
velocity ellipsoid with metallicity was a matter of considerable discussion in past
work (e.g., Norris 1994; Carney et al. 1996; Chiba \& Beers 2000). It was
understandably difficult to resolve, due to the small numbers of stars in these
previous samples. It is now clear that both the radial ($\sigma_{V_{R}}$) and
vertical ($\sigma_{V_{Z}}$) velocity dispersions increase at lower abundances.
Similar results have been reported by Bond et al. (2009).

For the thick disk, using the same metallicity ranges ($-0.8 <$ [Fe/H] $< -0.6$)
and distance ranges (1 $<$ $|$Z$|$ $<$ 2 kpc) as chosen above for determination
of the V$_{\phi}$ component, we obtain $\langle$V$_{R}$$\rangle$ = 2.5 $\pm$ 2
km~s$^{-1}$ and $\langle$V$_{Z}$$\rangle$ = 0.4 $\pm$ 1 km~s$^{-1}$, and
$\sigma_{V_{R}}$ = 53 $\pm$ 1 km~s$^{-1}$ and $\sigma_{V_{Z}}$ = 35 $\pm$ 1
km~s$^{-1}$, for the radial and vertical velocity components, respectively.

To evaluate these quantities for the outer halo, we have explored a finer grid
of metallicity for stars in the low-metallicity range ([Fe/H] $< -2.0$), as
shown in Figure 17. In each panel, the points represent the values of the
dispersions for a metallicity below [Fe/H]$_{max}$, the maximum metallicity for
stars in each sub-sample. As in Figure 16, the dispersions are corrected for the
observational errors.

As shown in the upper panel of Figure 17, the dispersion of the V$_{R}$
component is constant (or slightly increasing) in the range of metallicity $-2.4
<$ [Fe/H]$_{max}$ $< -2.0$, reaching a value of $\sigma_{V_{R}}$ $\sim$ 180
km~s$^{-1}$ at the lowest metallicity. Similar behavior can be noticed for the
velocity dispersion in the vertical direction. A constant value of about
$\sigma_{V_{Z}}$ $\sim$ 120 km~s$^{-1}$ is obtained for $-2.4 <$ [Fe/H]$_{max}$
$< -2.0$, falling slightly for the more metal-poor stars (albeit with large
error bars). We adopt the mean velocities and dispersions for the outer halo
corresponding to the metallicity upper limit [Fe/H]$_{max} = -2.2$ (i.e., [Fe/H]
$< -2.2$), where contamination from the inner halo should be minimal. These are
$\langle$V$_{R}$$\rangle$ = $-$8.6 $\pm$ 6.1 km~s$^{-1}$ and
$\langle$V$_{Z}$$\rangle$ = 1.9 $\pm$ 4.4 km~s$^{-1}$, and $\sigma_{V_{R}}$ =
159 $\pm$ 4 km~s$^{-1}$ and $\sigma_{V_{Z}}$ = 116 $\pm$ 3 km~s$^{-1}$, for the
radial and vertical components, respectively.

Determination of the velocity ellipsoid for the inner-halo population proved
more challenging. A series of experiments in the range of metallicity $-2.0 <$
[Fe/H] $< -1.5$, where contamination from the outer halo should be minimized,
were been carried out to obtain this information from a mixture-model
analysis. However, the symmetry of the distribution for both the V$_R$ and V$_Z$
components precluded obtaining accurate estimates. Recall that, due to the
existence of clear correlations between Z$_{max}$ and other kinematic
parameters, we cannot use Z$_{max}$ in order to reduce the contamination of
the inner halo by other components that are present in this
metallicity interval (the outer halo and MWTD). When a two-component model is
adopted, with the outer-halo parameters fixed, the R-Mix procedure provides
estimates, for the radial cmponent, of $\langle$V$_{R}$$\rangle$ = 3 $\pm$ 17
km~s$^{-1}$, and $\sigma_{V_{R}}$ = 143 $\pm$ 14 km~s$^{-1}$ (p-value $\sim$
0.07; i.e., marginally adequate). The same experiment for the vertical velocity
component does not provide an adequate fit.

\begin{deluxetable*}{ccccccc}[!t]
\tablewidth{0pt}
\tablenum{5}
\tablecaption{Velocity Ellipsoids}\\
\tablehead{
\colhead{Component} &
\colhead{$\langle$V$_{R}$$\rangle$} &
\colhead{$\langle$V$_{\phi}$$\rangle$} &
\colhead{$\langle$V$_{Z}$$\rangle$} &
\colhead{$\sigma_{V_{R}}$} &
\colhead{$\sigma_{V_{\phi}}$} &
\colhead{$\sigma_{V_{Z}}$} \\
\colhead{} &
\colhead{(km~s$^{-1}$)} &
\colhead{(km~s$^{-1}$)} &
\colhead{(km~s$^{-1}$)} &
\colhead{(km~s$^{-1}$)} &
\colhead{(km~s$^{-1}$)} &
\colhead{(km~s$^{-1}$)}
}
\startdata
           &              &                 &              &              &             &            \\
Thick Disk &   3 $\pm$  2 &   182 $\pm$ 2   &  0 $\pm$ 1   &   53 $\pm$ 2 &  51 $\pm$ 1 &  35 $\pm$ 1\\
MWTD       & $-13$ $\pm$  5 &   125 $\pm$ 4 &  $-14$ $\pm$ 5 &   59 $\pm$ 3 &  40 $\pm$ 3 & 44 $\pm$ 3\\
Inner Halo & 3 $\pm$  2 &  7 $\pm$ 4   & 3 $\pm$ 1  &  150 $\pm$ 2 &  95 $\pm$ 2 &  85 $\pm$ 1\\
Outer Halo & $-$9 $\pm$ 6 & $-$80 $\pm$ 13  & 2 $\pm$ 4    &  159 $\pm$ 4 & 165 $\pm$ 9 & 116 $\pm$ 3\\
\enddata
\tablecomments{The reported values of V$_{\phi}$ and $\sigma_{V_{\phi}}$ for the MWTD
are put in the center of the ranges estimated for this component reported in the
text. The derived velocity dispersions of the velocity ellipsoids are corrected
for the effects of observational errors, typically on the order of 5-6
km~s$^{-1}$ for large sub-samples, and 10-15 km~s$^{-1}$ for smaller
sub-samples. Note that the errors listed in this table and throughout the text
refer to random errors only, and do not take into account the existence of
systematic errors, primarily related to uncertainties in the photometric distances,
which are expected to be at least on the order of 10\% (and in some cases as
large as 15-20\%).}
\end{deluxetable*}

Frustrated by the mixture-model approach, we chose to obtain at least an
estimated velocity ellipsoid for the inner halo by simply adopting the mean
velocities and dispersions of the sub-sample of stars in this same metallicity
range, with the knowledge that contamination from other components may still be
present. This results in estimates of $\langle$V$_{R}$$\rangle$ = 3.0 $\pm$ 2.4
km~s$^{-1}$ and $\langle$V$_{Z}$$\rangle$ = 3.4 $\pm$ 1.4 km~s$^{-1}$, and
$\sigma_{V_{R}}$ = 150 $\pm$ 2 km~s$^{-1}$ and $\sigma_{V_{Z}}$ = 85 $\pm$ 1
km~s$^{-1}$, for the means and dispersions of the radial and vertical velocity
components, respectively. Note that these estimates agree, within the errors,
with the $\langle$V$_{R}$$\rangle$ and $\sigma_{V_{R}}$ estimated using the
mixture-model analysis above.

Finally, we have attempted to estimate the remaining components of the velocity
ellipsoid for the MWTD. Based on the results obtained in \S 5.6.1, we have
selected a sub-sample of stars with $-1.5 <$ [Fe/H] $< -1.0$ and distance
interval 1 $<$ $|$Z$|$ $<$ 2 kpc, in which the dominant components are the inner
halo and the MWTD (Table 4). When the R-Mix mixture-modeling analysis is
employed with the stellar fractions of the two components fixed at the values
reported in Table 4, a two-component model provides very good fits, with values
$\langle$V$_{R}$$\rangle$ = $-13$ $\pm$ 5 km~s$^{-1}$ and
$\langle$V$_{Z}$$\rangle$ = $-14$ $\pm$ 5 km~s$^{-1}$, and $\sigma_{V_{R}}$ = 64
$\pm$ 4 km~s$^{-1}$ and $\sigma_{V_{Z}}$ = 48 $\pm$ 3 km~s$^{-1}$. When
corrected for the observational errors, the dispersions are $\sigma_{V_{R}}$ =
59 $\pm$ 4 km~s$^{-1}$ and $\sigma_{V_{Z}}$ = 44 $\pm$ 3 km~s$^{-1}$,
respectively. One-component models are strongly rejected, with p-values $<
0.001$. The two-component fit reveals a slight asymmetry in both of the velocity
distributions. This asymmetry may be an indication that the orbits of stars in
the MWTD component are non-circular and/or non-coplanar with those of stars in
the thin disk. Figure 18 shows the fits obtained and the associated DRRs. Table
5 lists our final determinations of the velocity ellipsoids for the thick disk,
MWTD, inner halo, and outer halo.

It should be noted that, in addition to errors in our determination of the
parameters of the velocity ellipsoids that arise from observational errors
(affecting primarily the derived dispersions, which we attempt to correct for),
there remain systematic errors that result from the propagation of errors in the
distance estimates and proper-motion measurements. Monte Carlo experiments
indicate that such errors are typically on the order of 5 km~s$^{-1}$. These are
not reported as separate components of the errors listed in Table 5.

An additional and much more difficult to quantify error comes from our use of
the mixture-modeling approach itself. Finite mixture models, by definition,
force the observed distribution of kinematic quantities to be described by the
adopted number of components, and normalize the derived model to the sum of
these components. If a component is {\it unrecognized}, and not included in the
model, the recognized components are forced to take up its missing ``signal'',
and to adjust their derived parameters accordingly. For instance, if we were not
to include the MWTD component, the parameters of the thick disk, inner halo, and
outer halo would be altered in an attempt to accommodate the observed data. The
same would apply, to an extent, if a superfluous component is included. These
problems can be mitigated by statistical tests of the ability of the derived
model to fit the observed distribution, but they cannot be entirely eliminated.
This is particularly the case when one or more of the components are strongly
overlapping in the kinematic quantities being modeled. The best defense against
``mixture-model errors'' of this sort is the examination of different samples of
stars, ideally selected in different ways or over different ranges of distance,
such that they sample the underlying components in ways that break possible
degeneracies. In future analyses we plan to consider the kinematic
properties of F-turnoff stars, G dwarfs, and subdwarf M stars from SDSS/SEGUE,
which should provide tests of the necessity (or not) of the presence of our
adopted components in the mixture-model analysis.

\section{On the Thick-Disk Scale Length \\ and Scale Height}

With the velocity-ellipsoid parameters for the canonical thick disk in hand, we
can estimate its scale length and scale height by assuming that it is in
equilibrium and well-mixed. Chiba \& Beers (2000) estimated the scale length
of the thick disk from its stellar kinematics, using their Eqn. 1, which assumes
that the scale height of the thick disk is constant and that the ratio of
$\sigma_{V_{R}}$ to $\sigma_{V_{Z}}$ is independent of radius. With their
adopted values for the velocity dispersions and mean rotational velocity, Chiba
\& Beers derive a scale length of 4.5 kpc for the thick disk.

In this paper we do not need to make the above assumptions -- we have a
sufficient number of stars occupying a large enough region of space to measure
the gradients in $\sigma_{V_{R}}$ and $\sigma_{V_{Z}}$ directly. We assume that
the thick disk has an exponential density distribution in $R$ and $Z$, with
scale length $h_R$ and scale height $h_Z$, respectively. The cross term,
$\sigma_{V_{R,Z}}$, of the velocity dispersion is observed to be small for the
thick disk. It takes a value of $\sigma_{V_{R,Z}}$ = 0.096 $\pm$ 0.020 at 1 $<$
$|$Z$|$ $<$ 2 kpc, and $-$0.8 $<$ [Fe/H] $<$ $-$0.6 (\S 9), so we can write the
radial and $Z$-components of the Jeans Equation as

\begin{equation}
\frac{\partial \ln \sigma^2_{V_{R}}}{\partial R} - \frac{1}{h_R}
 + \frac {1}{R} \{1 - \frac{\sigma_{V_{\phi}}^2}{\sigma_{V_{R}}^2}
 + \frac{V_c^2 - \langle V_{\phi} \rangle^2}{\sigma_{V_{R}}^2}\} = 0
\end{equation}

\begin{equation}
\frac{\partial \ln \sigma_{V_{Z}}^2}{\partial Z} - \frac{1}
{h_Z} + \frac{K_Z}{\sigma_{V_{Z}}^2} = 0
\end{equation}

\noindent where $V_c$ and $K_Z$ are the circular velocity and
vertical acceleration.  We measure the radial gradient of $\sigma_{V{_R}}$
and the Z-gradient of $\sigma_{V_{Z}}$ from our sample, and use the Kuijken
\& Gilmore (1991) estimate of $K_Z = 2\pi {\rm G} \times 71~M_\odot
~{\rm pc} ^{-2}$ at $|Z| = 1.1$ kpc. We adopt the following parameters
for the velocity dispersion and mean motion of the canonical thick
disk at the solar radius, and at a Z-height of 1.1 kpc: $(\sigma_{V_{R}},
\sigma_{V_{\phi}}, \sigma_{V_{Z}}) = (53 \pm 2, 51 \pm 1, 35 \pm 1)$ km s$^{-1}$
and $\langle V_{\phi} \rangle = 182 \pm 2$ km s$^{-1}$.

Over the interval $6 < R < 12$ kpc and $1 < |Z| < 3$ kpc, respectively, the
variations in $\sigma_{V_{R}}$ and $\sigma_{V_{Z}}$ closely follow a linear
behavior, with $\partial \sigma_{V_{R}} / \partial R = -5.8 \pm 1.3$ km s$^{-1}$
kpc $^{-1}$ and $\partial \sigma_{V_{Z}} / \partial |Z| = 6.8 \pm 1.7$ km
s$^{-1}$ kpc$^{-1}$, respectively. With these parameters, the scale length for
the thick disk is $h_R = 2.20 \pm 0.35$ kpc, and the scale height is $h_Z = 0.51
\pm 0.04$ kpc, where the errors represent the statistical uncertainties only.

The small value of $h_R$ relative to the Chiba \& Beers value comes about almost
entirely because our mean motion for the thick disk ($\langle V_{\phi} \rangle$
= 182 km s$^{-1}$) is significantly smaller than their value of 200 km s$^{-1}$.
However, we note that several other authors also find a small value of $h_R$
from independent arguments based on star counts, e.g., Reyl\'e \& Robin (2001).

Our value for the scale height of the thick disk is at the lower
end of the range of previously reported estimates, so we must look critically at the
input data from which $h_{Z}$ was derived. The parameters which are
used in estimating $h_{Z}$ are $K_{Z}$ , $\sigma_{V_{Z}}$ and its gradient,
all evaluated at $|$Z$|$ = 1.1 kpc.  The value of $K_{Z}$ at $|$Z$|$ =
1.1 kpc comes from the Kuijken \& Gilmore (1991) sample of K dwarfs, and is
consistent with the known density of disk objects. It is also in excellent agreement
with the $K_Z$ value independently determined by Holmberg \& Flynn (2000). For
determining $\sigma_{V_{Z}}$ and its gradient, we chose stars from a narrow
range of metallicity, $-$0.8 $<$ [Fe/H] $< -$0.6. Although the values of
$\sigma_{V_{Z}}$ and its gradient are well determined by our data, it is
possible that our thick-disk sample at $|$Z$|$ = 1.1 kpc is contaminated by
thin-disk stars, which would reduce the apparent value of $\sigma_{V_Z}$ and lead
to an underestimate of $h_{Z}$. Contamination by inner-halo stars could
also lead to an incorrect estimate of $\partial \sigma_{V_Z} / \partial |Z|$.
However, the positive gradient of $\sigma_{V_{Z}}$ makes only a relatively minor
contribution to the derived value of $h_Z$. For example, if the thick disk
$\sigma_{V_{Z}}$ were in fact isothermal, the value of $h_{Z}$ would increase
from 0.51 to only 0.64 kpc.

We note that our kinematic estimate of the scale height of the thick disk is
obtained at $|$Z$|$ = 1.1 kpc, while the (usually larger, $h_Z$ $\sim$ 1 kpc)
estimates come from star counts at $|$Z$|$ $\sim$ 2 to 4 kpc (eg Gilmore \& Reid
1983). Although the thick disk is often represented as vertically exponential,
and indeed appears to be exponential at larger $|Z|$, in fact it is unlikely to
be strictly exponential. In the general version of the vertical Jeans Equation
the $1/h_Z$ term in Eqn 7 is replaced by $\partial ln \nu / \partial Z$, where
$\nu(Z)$ is the vertical density distribution. If we adopt the form of $K_Z(Z)$
from Holmberg \& Flynn (2000), and we assume that the thick disk is vertically
isothermal with a velocity dispersion $\sigma_{V_{Z}} = 35$ km s$^{-1}$, then
the Jeans Equation can be integrated numerically to derive $\nu(Z)$. Near the
Galactic plane, $\nu(Z)$ changes slowly with $|$Z$|$ and becomes steeper as $Z$
increases. At $|$Z$|$ = 1.1 kpc, the ``local scale height"
[$-1/(\partial ln \nu/\partial Z)$] agrees almost exactly with our value of 0.64
kpc, as it must. On the other hand, one could define an effective scale height
of the thick disk as the height above the Galactic plane at which $\nu(Z)$ has
decreased by a factor $e$ from its value at $Z = 0$. This height is 0.99 $\pm$
0.04 kpc, but it is not directly comparable with the star-count scale heights at
larger $|$Z$|$.

For the MWTD we have only the kinematic parameters given in Table 5, but these
are sufficient to place some useful constraints on the radial structure of this
component. As we have no information on the radial gradient of the velocity
dispersion for this component, we assume that it follows the often-adopted relation
$\sigma_{V_{R}}^2 \propto \exp (-R/h_\sigma)$, where $h_\sigma$ is a length
scale. Then it follows from the radial component of the Jeans Equation that

\begin{equation}
\frac{1}{h_\sigma} + \frac{1}{h_R} = 1.04~{\rm kpc}^{-1}
\end{equation}

If the two scale lengths are comparable, as they would be for a
disk of uniform scale height and constant anisotropy, then
the radial scale length of the MWTD is again short, $\sim 2$
kpc.

Regarding the scale height for the MWTD, we find at $|$Z$|$ = 1.1 kpc
(and adopting the velocity dispersion reported in Table 5, $\sigma_{V_{Z}}$ = 44
$\pm$ 3 km s$^{-1}$) a value of h$_{Z}$ = 1.36 $\pm$ 0.13 kpc.

\section{Density Profiles for the Stellar \\ Halo Components}

Assuming that the Galaxy is in dynamical steady state, the Jeans Equation for
a spherically symmetric stellar system, in spherical coordinates (r,$\theta$,$\phi$
), can be written as (Binney \& Tremaine 1988):

\begin{equation}
\frac{d(\rho\langle V_{r}^{2} \rangle)}{dr} + \frac{\rho}{r} [2\langle V_{r}^{2} \rangle-(\langle V_{\theta}^{2} \rangle+\langle V_{\phi}^{2} \rangle)] = -\rho\frac{d\Phi}{dr}
\end{equation}

\noindent where $\langle V_{i}^{2} \rangle$ ($i =r,\theta,\phi$) is related to
the velocity dispersion $\sigma_{i}^{2}$ as:

\begin{equation}
\sigma_{i}^{2} = \langle V_{i}^{2} \rangle - \langle V_{i} \rangle^{2}.
\end{equation}



Based on the well-supported assumption for both the inner and outer halos, $\langle
V_{r} \rangle$ = 0, and $\langle V_{\theta} \rangle$ = 0, the above equation can
written as:

\begin{equation}
\frac{1}{\rho}\frac{d(\rho \sigma_{V_{r}}^{2})}{dr} + \frac{2}{\rho}\sigma_{V_{r}}^{2}[1 - \frac{\sigma_{V_{\theta}}^2 + \sigma_{V_{\phi}}^2}{2\sigma_{V_{r}}^2} - \frac{\langle V_{\phi} \rangle^{2}}{2\sigma_{V_{r}}^2}] = -\frac{d\Phi}{dr}
\end{equation}

\noindent where $\sigma_{V_{r}}$, $\sigma_{V_{\theta}}$, $\sigma_{V_{\phi}}$,  are the velocity dispersions in the
directions of (r, $\theta$, $\phi$), respectively (in the
spherical system $V_{\phi}$ and $V_{\theta}$ are taken positive in the direction
of Galactic rotation and in the direction away from the North Galactic Pole,
respectively).

The anisotropy parameter, $\beta$, which describes the degree of anisotropy of
the velocity distributions, is defined as $\beta = 1 -
(\sigma_{V_{t}}/\sigma_{V_{r}})^{2}$, where $\sigma_{V_{r}}$ and $\sigma_{V_{t}}$ are
the radial and tangential velocity dispersions, respectively. The latter is
given by $\sigma_{V_{t}}^{2} = \frac{1}{2}(\sigma_{V_{\phi}}^{2}+\sigma_{V_{\theta}}^{2})$.
The values of the anisotropy parameter obtained with our dataset
are $\beta_{in}$ = 0.6 for the inner halo and $\beta_{out}$ = 0.2 for the
outer halo. The anisotropy parameters for the inner halo indicate that its
velocity ellipsoid is radially elongated, while the outer halo exhibits a less
radially-elongated shape for its velocity ellipsoid.

Assuming that the potential of the Galactic halo is approximately spherical and
logarithmic, $\Phi(r) = V_{c}^{2}$ ln r, which corresponds to a flat rotation
curve of the Galaxy in the solar neighborhood, and assuming also $V_{c}(r)$ =
$V_{c}$ = 220 km~s$^{-1}$, Eqn. 11 can be written as:

\begin{equation}
V_{c}^{2} = -\sigma_{V_{r}}^{2}(\frac{dln\rho}{dlnr} + \frac{dln\sigma_{V_{r}}^{2}}{dlnr} + 2\beta - \frac{\langle V_{\phi} \rangle^{2}}{\sigma_{V_{r}}^2})
\end{equation}

The density of stars in the Galactic halo components can be approximated by a
power-law relation $\rho(r) \propto r^{n}$ (where n is negative, due to the
finite extent of the Galaxy). Making this substitution, the previous equation
becomes:

\begin{equation}
V_{c}^{2} = -\sigma_{V_{r}}^{2}(n + \frac{dln\sigma_{V_{r}}^{2}}{dlnr} + 2\beta - \frac{\langle V_{\phi} \rangle^{2}}{\sigma_{V_{r}}^2})
\end{equation}

\noindent Note that the term $\langle V_{\phi} \rangle^{2}/\sigma_{V_{r}}^2$ is
negligible for the inner halo ($\langle V_{\phi} \rangle$ = 7 km s$^{-1}$),
while it is {\it not negligible} for the outer halo ($\langle V_{\phi} \rangle$ = $-$80 km
s$^{-1}$). The values obtained for this term are, for the inner halo, $\langle
V_{\phi} \rangle^{2}/\sigma_{V_{r}}^2$ = 2$\cdot$10$^{-3}$, and for the outer
halo, $\langle V_{\phi}\rangle^{2}/\sigma_{V_{r}}^2$ = 0.25.

In the solar neighborhood (R = R$_{\sun}$) we can use the velocity
ellipsoids obtained in the cylindrical reference system
($\sigma_{V_{R}}$, $\sigma_{V_{\phi}}$, $\sigma_{V_{Z}}$) to derive
the parameter $n$ from the previous equation for the two halo
components. Adopting the values of the velocity ellipsoid reported in
Table 5, we find the following power law for the two components:

$$ \rho_{in} \sim r^{-3.17 \pm 0.20} \qquad \rho_{out} \sim r^{-1.79 \pm 0.29} $$

The above estimate includes the term $dln(\sigma_{V_{r}}^{2})/dlnr$ in Eqn. 13,
which takes into account the variation of $\sigma_{V_{r}}$ as a function of the
radius $r$. In the solar neighborhood it becomes $dln(\sigma_{V_{r}}^{2})/dlnr$
$\sim$ $dln(\sigma_{V_{R}}^{2})/dlnR$. This term has been evaluated by examining
$\sigma_{V_{r}}$ in 1 kpc intervals, in the range 6 $<$ R $<$ 12 kpc,
using the low-metallicity sub-sample ([Fe/H] $< -2$). The minimum and
maximum values of the dispersion in the above ranges are $\sigma_{V_{R}}$ = 156
$\pm$ 9 km s$^{-1}$, and $\sigma_{V_{R}}$ = 172 $\pm$ 8 km s$^{-1}$,
respectively. By taking the logarithm and fitting the corresponding curve, we
obtain $dln\sigma_{V_{R}}^{2}$/$dlnr$ = $-$0.26.

Note that if the above term is neglected in Eqn. 13 we obtain:

$$ \rho_{in} \sim r^{-3.43 \pm 0.20} \qquad \rho_{out} \sim r^{-2.05 \pm 0.29} $$

\noindent Thus, this term has a considerable effect, and needs to be included.
Regardless, our analysis points to an outer-halo population with a much
shallower spatial density profile than that of the inner-halo population.

\section{Evidence for Tilts in the Derived \\ Velocity Ellipsoids}

The orientation of the velocity ellipsoids is of interest for understanding the
equilibrium and assembly of the various Galactic components, and has recently
been studied by Siebert et al. (2008), Bienayme et al. (2009), and Smith,
Evans, \& An (2009), as well as by Bond et al. (2009). We have
examined the $(R,Z)$ kinematics of our local sample in order to estimate the
tilts of the velocity ellipsoids as a function of the height above the Galactic
plane and metallicity.

We note that it is not appropriate to use the maximum vertical height of the
orbit, Z$_{max}$, as a parameter here, because selecting on Z$_{max}$ is
equivalent to a selection in the velocity components V$_{R}$ and V$_{Z}$. This
makes it difficult to separate the tilt for the inner and outer halos.

Figure 19 shows the velocity distribution in the (V$_{R}$, V$_{Z}$) plane as a
function of the vertical distance, $|$Z$|$, and metallicity. Close to the
Galactic plane (top panels), at 0 $<$ $|$Z$|$ $<$ 1 kpc, and at metallicity
$-$0.8 $<$ [Fe/H] $<$ $-$ 0.6, the sub-sample contains mostly thick-disk stars,
and the inclination of the velocity ellipsoid is small. At higher $|Z|$ values,
the inclination slightly increases, possibly due to contamination from the MWTD
and the metal-rich part of the inner halo. At intermediate metallicity (middle
panels), $-1.5 < $ [Fe/H] $< - 0.8$, the ellipsoid shows a strong inclination,
increasing with $|Z|$. We note that this intermediate sample comprises two
overlapping components, the MWTD and the inner halo, with contamination from the
canonical thick disk and the outer halo. The thick disk and MWTD cannot be
easily separated in the V$_{R}$ and V$_{Z}$ component, as already discussed in
\S 6. The low-metallicity sample (bottom panel), [Fe/H] $< -1.5$, which
contains mostly halo stars, again exhibits a strong inclination increasing with
$|$Z$|$.

The tilt of the velocity ellipsoid in cylindrical coordinates can
be described by the tilt angle and the correlation coefficient,
given by:

\begin{equation}
\tan2\alpha = \frac{2\sigma_{V_{R,Z}}^{2}}{(\sigma_{V_{R}}^{2} - \sigma_{V_{Z}}^{2})}
\end{equation}

\begin{equation}
C[V_{R},V_{Z}] = \frac{\sigma_{V_{R,Z}}^{2}}{(\sigma_{V_{R}}^{2}\sigma_{V_{Z}}^{2})^{1/2}}
\end{equation}

\noindent
where $\sigma_{V_{R,Z}}$ is the covariance. The values obtained for the tilt
angle and the correlation coefficient, in various ranges of the vertical
distance and metallicity, are listed in Table 6. The errors on $\alpha$ and $C$ are
evaluated through a Monte Carlo simulation (10000 realizations for each star),
using the velocity errors for individual stars and correcting for the effect of
the errors on the diagonal $\sigma$ terms. We find that the tilt angle is
statistically different from zero for all the selected samples, and the
principal axes of the velocity ellipsoids point towards a location slightly
above the Galactic center. For metal-rich stars close to the Galactic plane we
obtain $\alpha = 7^{\circ}.1 \pm 1^{\circ}.5$, which is consistent with the
inclination found by Siebert et al. 2008 ($7^{\circ}.3 \pm 1^{\circ}.8$) for a
sample of stars at $|Z| < 1$ kpc. At intermediate metallicity and $1 < |Z| < 2$
kpc, the inclination is $\alpha = 10^{\circ}.3 \pm 0^{\circ}.4$; it increases to
$15^{\circ}.1 \pm 0^{\circ}.3$ farther from the Galactic plane. These
intermediate-metallicity samples are dominated by the MWTD and inner halo.
Finally, at low metallicity, the sample comes mainly from the overlapping
inner-/outer-halo populations, and the inclination of the velocity ellipsoid is
$\alpha = 8^{\circ}.6 \pm 0^{\circ}.5$ and $13^{\circ}.1 \pm 0^{\circ}.4$ at $1
< |Z| < 2$ kpc, and $2 < |Z| < 4$ kpc, respectively.

\begin{deluxetable}{ccccc}[!t]
\tablewidth{0pt}
\tablenum{6}
\tablecaption{Velocity Ellipsoids: Tilt and Correlation}\\
\tablehead{
\colhead{[Fe/H]} &
\colhead{} &
\colhead{Tilt Angle} &
\colhead{} &
\colhead{$C[V_{R},V_{Z}]$}
}
\startdata
\cutinhead{Close to the Galactic Plane ($1 < $ $|$Z$|$ $ < 2$ kpc)}
                    &  &                                    &  &                   \\
$-$0.8  to $-$0.6   &  & 7$^{\circ}$.1 $\pm$ 1$^{\circ}$.5  &  &0.096$\pm$ 0.020  \\
$-$1.5  to $-$0.8   &  & 10$^{\circ}$.3 $\pm$ 0$^{\circ}$.4 &  &0.224$\pm$ 0.009  \\
$<-$1.5             &  & 8$^{\circ}$.6 $\pm$ 0$^{\circ}$.5  &  & 0.151$\pm$ 0.008 \\
\cutinhead{Farther from the Galactic Plane ($2 < $ $|$Z$|$ $ < 4$ kpc)}
                    &  &                                    &  &                  \\
$-$0.8  to $-$0.6   &  & 5$^{\circ}$.2 $\pm$ 1$^{\circ}$.2  & &0.106$\pm$ 0.024 \\
$-$1.5  to $-$0.8   &  & 15$^{\circ}$.1 $\pm$ 0$^{\circ}$.3 &  &0.367$\pm$ 0.007  \\
$<-$1.5             &  & 13$^{\circ}$.1 $\pm$ 0$^{\circ}$.4 &  & 0.248$\pm$ 0.008 \\
\enddata
\end{deluxetable}

Having demonstrated that the velocity ellipsoid for the inner halo is pointing
close to the Galactic center, we re-evaluated its velocity ellipsoid in
spherical coordinates, in order to evaluate whether the inner halo is
kinematically close to isotropic in $(\theta,\phi)$. If it were so, then the
inner-halo kinematics would be consistent with a system for which the
distribution function depends on energy and the {\it total} angular momentum,
which would in turn be an interesting clue to its assembly. The sample was
selected in the range of metallicity of $-2.0 <$ [Fe/H] $< -1.6$ and $2 < |Z| <
4$ kpc, where the dominant component is the inner halo, albeit with
contamination from the MWTD and the outer halo. A similar estimate for the
outer halo is more uncertain, due to the difficulty in separating the two halo
components, as mentioned above. However, we have attempted to select mostly
outer-halo stars using the metallicity interval [Fe/H] $< -2.4$ and $2 < |Z| <
4$ kpc, being aware that the sample remains contaminated by the inner halo. For
the inner halo, we find ($\sigma_{V_{r}}$, $\sigma_{V_{\phi}}$,
$\sigma_{V_{\theta}}$) = (160 $\pm$ 3, 102 $\pm$ 2, 83 $\pm$ 2) km s$^{-1}$ and
for the outer halo ($\sigma_{V_{r}}$, $\sigma_{V_{\phi}}$,
$\sigma_{V_{\theta}}$) = (178 $\pm$ 9, 149 $\pm$ 7 , 127 $\pm$ 6) km s$^{-1}$
respectively. We conclude that both halo components are significantly
anisotropic in $(\theta,\phi)$.

\section{The MDF of the Full Sample as a Function of Height Above the Galactic Plane}

There is a sufficient number of stars in the DR7 calibration-star sample to
construct ``as observed'' MDFs for the entire sample (regardless of whether the
star has an available proper motion, or is located within the cylinder we have
defined above for the local sample of stars), in order to look for confirmation
of the expected changes as a function of height above the Galactic plane.
The number of stars employed is 30956.

We examine the MDFs for different intervals in $|$Z$|$, with cuts chosen to ensure
there remain adequate numbers of stars in each interval.

Figure 20 shows the results of this exercise. Examination of the left-hand
column of panels in this figure shows how the MDF changes from the upper panel,
in which there are obvious contributions from the thick-disk, MWTD, and
inner-halo components in the cuts close to plane, to the lower panel, with an
MDF dominated primarily by inner-halo stars. In the right-hand column of panels,
with distances from the plane greater than 5 kpc, the transition from inner-halo
dominance to a much greater contribution from outer-halo stars is obvious.
This demonstration is, by design, independent of any errors that might arise
from derivation of the kinematic parameters, and provides confirmation
of the difference in the chemical properties of the inner- and outer-halo
populations originally suggested by Carollo et al. (2007).

Although it might appear possible to attempt mixture-model analysis to obtain
MDFs for each of the individual components, we recall that biases in the
selection of the stars in the calibration-star sample would confound such an
attempt.  Other samples of SDSS stars, which are more suitable for such an
analysis, are being studied for this purpose, and will be reported on in due
course.

\section{Summary and Discussion}

We have analyzed the SDSS/SEGUE sample of 32360 unique calibration stars from
DR7, amplifying the previous analysis carried out by Carollo et al. (2007) on a
smaller sample of similar stars from DR5. A Maximum Likelihood analysis of a
local sub-sample of 16920 calibration stars has been developed in order to
extract kinematic information for the major Galactic components (thick disk,
inner halo, and outer halo), as well as for the elusive metal-weak thick disk
(MWTD). We have measured velocity ellipsoids for the thick disk, the MWTD, the
inner halo, and the outer halo, demonstrated the need for a MWTD component that
is kinematically and chemically independent of the canonical thick disk (and put
limits on the metallicity range of the MWTD MDF), obtained estimates of the
scale length and scale height of the thick disk and MWTD, and derived the
inferred spatial density profiles of the inner/outer halo components by use of
the Jeans Equation. We have also presented evidence for tilts in the velocity
ellipsoids for stars in our sample as a function of height above the plane, for
several ranges in metallicity. Changes in the {\it in-situ} observed MDF for the
full calibration-star sample, as a function of height above the Galactic plane,
are consistent with those expected from kinematic analysis of the local sample.
Our results are summarized and compared with previous related analyses below,
followed by a discussion on their implication for the formation of the Milky Way
in the context of other recent results.

\subsection{Summary of Results and Comparison\\ with Previous Work}

We began by comparing the changes in the nature of the calibration-star sample
that have occurred in going from the SDSS DR5 to the DR7 data releases. Aside
from the 60\% increase in the total sample size, the addition of numerous
low-latitude observations from SEGUE, and recent improvements in the SSPP (in
particular for the determination of metallicities for metal-rich stars) have
enabled much clearer distinctions between the kinematic behavior of the various
Galactic components we consider in our analysis.

\subsubsection{Distribution of Orbital Eccentricity}

The distribution of observed orbital eccentricity for our local sample of stars
contains interesting information on the presence of various Galactic components
as a function of height above the Galactic plane and over a number of cuts in
metallicity. We found that, in the metallicity range $-$1.0 $<$ [Fe/H] $< -$0.6,
the canonical thick disk dominates close to the Galactic plane (1 $<$ $|$Z$|$
$<$ 2 kpc), as well as at 2 $<$ $|$Z$|$ $<$ 4 kpc. At intermediate metallicity,
$-$1.5 $<$ [Fe/H] $< -$1.0, the distribution exhibits an interesting feature
extending to $e$ $\sim$ 0.5, suggesting a likely MWTD population close to the
Galactic plane. At larger distances the distribution appears more consistent
with the inner-halo population. In the metallicity range $-$2.0 $<$ [Fe/H] $<
-$1.5 the distribution assumes the form $f(e) \sim e$, which is associated
entirely with the inner-halo population, and consistent with a well-mixed
steady-state system of test particles for which the distribution function
depends only on energy. In contrast to this behavior, we noted that the
distribution of orbital eccentricity for the most metal-poor stars in our local
sample ([Fe/H] $<-$2.0) deviates significantly from linearity. Such a
distribution reflects the contribution from an outer-halo population with a
smaller fraction of stars on high-eccentricity orbits.

\subsubsection{Identification of Halo Components}

We next explored which quantities can best be used to constrain the mixtures of
components present in the local calibration-star sample, concluding that
V$_{\phi}$ and Z$_{max}$ are superior to available alternatives. \footnote{After
the majority of our analysis was completed, we also investigated using the
meridional velocity, defined as V${_M}$ = (V$_{R}^2$ + {V$_{Z}^2)^{1/2}$, as a possible variable
for separation of the inner- and outer-halo components. This experiment proved
interesting (i.e., a retrograde V$_{\phi}$ was found for stars with high V${_M}$).
This variable is worthwhile to investigate further, since it does not require
adoption of a gravitational potential, as is the case for Z$_{max}$.}} The subtle, but
important, question of how many components are required to account for the
observed kinematics of halo stars in the local sample was then considered. On
the basis of an objective clustering approach, applied to the low-metallicity
sub-sample of local stars with [Fe/H] $< -2.0$, we demonstrated that (a) a
single-population halo is incompatible with the observed kinematics, that (b) a
dual-component halo, comprising populations we associate with the inner and
outer halo, is sufficient to accommodate the observed kinematics, and that (c)
although additional components (of unspecified physical meaning) can be added,
they are not required (and in any case, provide no statistically significant
improvement in the kinematic fits). We concluded that a dual-halo model is
preferred on the grounds of its simplicity. Previous claims for the existence of
a complex (often, dual) halo have been made over the past few decades, but these
were based on rather small samples of tracer stars or clusters, such that the
statistical significance of the dual signature was difficult to assess (e.g.,
Norris 1994; Carney et al. 1996). We believe the existence of an inner/outer
halo structure is now clearly confirmed.

\subsubsection{Maximum Likelihood Analysis \\ of Halo-Component Kinematics}

We developed a flexible Maximum Likelihood (ML) approach for further analysis of
the kinematics of Galactic components. We then used a low-metallicity sub-sample
of local stars in order to obtain estimates of the mean rotational velocity and
dispersion of the outer-halo population, obtaining values of
$\langle$V$_{\phi}$$\rangle$ = $-$80 $\pm$ 13 km~s$^{-1}$, slightly more
retrograde than that obtained by Carollo et al. (2007) ($\sim -70$ km~s$^{-1}$)
from the DR5 sample of local calibration stars. For the dispersion, we obtained
$\sigma_{V_{\phi}}$ = 180 $\pm$ 9 km~s$^{-1}$ ($\sigma_{V_{\phi}}$ = 165 $\pm$ 9
km~s$^{-1}$ when corrected for observational errors), a substantially larger
value than reported by previous authors who considered only a single-component
halo. The ML approach was then used in order to estimate the rotation and
dispersion of the inner-halo component, based on an examination of cuts on
Z$_{max}$ for the low-metallicity sub-sample. We obtained values for stars with
1 $<$ Z$_{max} < 10$ kpc (where the inner-halo component dominates) of
$\langle$V$_{\phi}$$\rangle$ = 7 $\pm$ 4 km~s$^{-1}$, i.e., an essentially
non-rotating inner halo, and $\sigma_{V_{\phi}}$ = 110 $\pm$ 3 km~s$^{-1}$
($\sigma_{V_{\phi}}$ = 95 $\pm$ 2 km~s$^{-1}$ when corrected for observational
errors). These values do not change substantially when considering cuts on
Z$_{max}$ up to 30 kpc from the plane, although the derived fraction of
inner-halo stars decreases with distance while the fraction of outer-halo stars
increases with distance.

We note that while candidate outer-halo stars with large Z$_{max}$ exhibit a
large retrograde mean rotation {\it in the solar neighborhood}, these stars
would have a smaller value of {\it in-situ} rotational velocity at greater
Galactocentric distances. This is because an orbit with large Z$_{max}$ tends to
have large R$_{apo}$, and thus smaller $| {\rm V}_\phi |$, at large
Galactocentric distance, due the conservation of angular momentum. The azimuthal
velocity at R$_{apo}$, V$_{\phi, apo}$ can be simply evaluated as V$_{\phi,
apo}$ = V$_{\phi}$$\cdot$(R/R$_{apo}$). Using the sub-sample with [Fe/H] $<$
$-2$ and Z$_{max}$ $>$ 10 kpc, we obtain $\langle {\rm V}_{\phi,apo}
\rangle = -14 \pm 2$ km~s$^{-1}$ as the most likely {\it in-situ} mean
rotation of outer-halo stars, given that stars are most probably observed when
they are near their apocentric distances. It is worth exploring this small
amount of {\it in-situ} retrograde rotation in numerical simulations of galaxy
formation and/or kinematic observations of extra-galactic halos, in order to
compare with the present results for the Milky Way. We also note that the
expected {\it in-situ} velocity dispersion of an outer-halo population also must
decrease with distance, as can be verified by inspection of the Jeans Equation.
Indeed, observations of distant samples of stars, which are expected to be
dominated by members of the outer-halo population, have been shown by a number
of authors to meet this expectation, at least for the radial-velocity component
(e.g, Sommer-Larsen et al. 1997; Xue et al. 2008; Brown et al. 2009). It should
be kept in mind that Sommer-Larsen et al. (1997) have also pointed out that the
{\it nature} of the orbits of the more distant stars must change from locally
radially anisotropic to tangentially anisotropic (with a transition point at
around 20 kpc), in order to be consistent with a flat rotation curve. This
latter point will prove challenging to verify observationally {\it in-situ}
until high-quality proper motions for more distant stars can be obtained, e.g.,
with the Gaia mission.

In the analysis of Chiba \& Beers (2000), which only considered a single halo
component, a slightly more prograde inner halo was obtained (based on stars with
[Fe/H] $< -2.0$), $\langle$V$_{\phi}$$\rangle$ $\sim 30$ km~s$^{-1}$, along with
a similar dispersion, $\sigma_{V_{\phi}}$ = 106 $\pm$ 9 km~s$^{-1}$. The
existence of a gradient in the mean rotational velocity for the inner-halo
component, as claimed previously by Chiba \& Beers (2000), is confirmed. We
obtained a value of $\triangle$$\langle$V$_{\phi}$$\rangle$/$\triangle |$Z$| =
-28$ $\pm$ 9 km~s$^{-1}$ kpc$^{-1}$, for stars located within 2 kpc of the
Galactic plane. This gradient is substantially lower than the value reported by
Chiba \& Beers ($\triangle$$\langle$ V$_{\phi}$$\rangle$/$\triangle |$Z$| = -52
\pm$ 6 km~s$^{-1}$ kpc$^{-1}$). Note that the gradient reported by Chiba \& Beers
is based on a sample of stars primarily located within 1 kpc of the plane;
beyond 1 kpc, their three bins are consistent with zero net rotation. Thus, the
apparent difference in the gradient of the inner-halo population obtained with
our present sample of local calibration stars may be related to the subtantially
larger number of stars included with 1 $< |$Z$| < 2$ kpc, lowering the derived
gradient. In any case, such a gradient may represent the signature of a
dissipatively-formed flattened inner halo. Clearly, in many respects our
inner-halo population is essentially kinematically identical to ``the halo''
population studied by Chiba \& Beers (2000), and many others.

\subsubsection{Maximum Likelihood Analysis \\ of Thick-Disk-Component Kinematics}

The mean rotational velocity and dispersion of the thick disk were then
considered, based on inspection of a metal-rich ($-0.8 < $ [Fe/H] $< -0.6$)
sub-sample of stars close to the Galactic plane. We obtained values of
$\langle$V$_{\phi}$$\rangle$ = $182$ $\pm$ 2 km~s$^{-1}$ and $\sigma_{V_{\phi}}$
= 57 $\pm$ 1 km~s$^{-1}$ ($\sigma_{V_{\phi}}$ = 51 $\pm$ 1 km~s$^{-1}$ when
corrected for observational errors) for stars in the range 1 $<$ $|$Z$|$ $<$ 2
kpc, where the thick disk is espected to dominate, and contamination from
thin-disk stars should be negligible. The gradient in asymmetric drift of the
thick-disk component as a function of height above the plane, noted by previous
authors, is very clear in our data as well; we derived
$\triangle$$\langle$V$_{\phi}$$\rangle$/$\triangle |$Z$| = -36$ $\pm$ 1
km~s$^{-1}$ kpc$^{-1}$, in excellent agreement with the rotational velocity
gradients for disk stars obtained by Chiba \& Beers (2000), Girard et al.
(2006), Sheffield et al. (2007), and Ivezi{\'c} et al. (2008). The disperson of
this population also increases with distance, although it is not clear how much
this might be influenced by overlap with metal-rich stars from the inner-halo
component. Several authors have argued that this gradient in the asymmetric
drift of the thick disk is consistent with heating of an early disk by satellite
merger(s) (Quinn, Hernquist, \& Fullagar 1993; Mihos et al. 1995; Robin et al.
1996; Walker, Mihos, \& Hernquist 1996; Vel\'{a}zquez \& White 1999; Aguerri,
Balcells, \& Peletier 2001; Chen et al. 2001; Hayashi \& Chiba 2006).

\subsubsection{Fractions of Galactic Components From the ML Analyses}

The ML approach was then applied to the full range of metallicities in our
sample of local calibration stars, in order to determine the fractional
contribution of the three primary components in our model as a function of
Z$_{max}$ (fixing as inputs the values of mean rotational velocities and
dispersions derived previously for each component). This exercise indicated
that, within 5 kpc from the plane, the thick-disk and inner-halo components
contribute roughly equally. Beyond 5 kpc the thick-disk component is absent, as
expected. The inner-halo population dominates between 5 and 10 kpc. Beyond 10
kpc the outer halo increases in importance, is present in equal proportion to
the inner halo between 15 and 20 kpc, and dominates beyond 20 kpc. The inversion
point in the dominance of the inner/outer halo is located in the range Z$_{max}$
= 15-20 kpc.

\subsubsection{Tests for an Independent MWTD Component}

We then used our extensive local dataset to examine the question of whether an
independent MWTD component is required in order to account for the rotational
properties of stars close to the Galactic plane. We selected sub-samples of
stars in three metallicity regions, ({\it i}) $-2.0 <$ [Fe/H] $< -1.5$, ({\it
ii}) $-1.5 <$ [Fe/H] $< -1.0$, and ({\it iii}) $-1.0 <$ [Fe/H] $< -0.6$, and two
intervals of distance from the plane, 1 $<$ $|$Z$|$ $<$ 2 kpc, and 2 $<$ $|$Z$|$
$<$ 4 kpc. The clustering analysis approach was applied to each sub-sample, and
used to demonstrate that, in the region closer to the Galactic plane, an
independent MWTD component with $\langle$V$_{\phi}$$\rangle$ = 100-150
km~s$^{-1}$ and $\sigma_{V_{\phi}}$ = 40-50 km~s$^{-1}$ (35-45, km~s$^{-1}$,
when corrected for the observational errors), is required in order to account
for the rotational behavior of the stars in our local sample. Similarly, an
independent MWTD component must also exist in the interval farther from the
Galactic plane, in order to accommodate the observed rotational behavior of the
local sample of stars in that region of $|$Z$|$. It is interesting to note that
the contribution of the MWTD component dominates over that of the inner-halo
component for stars in the most metal-rich metallicity interval considered, not
only close to the Galactic plane, but in the more distant interval as well. A
population of rapidly rotating stars was identified in the metal-rich
sub-sample, in both the nearby and more distant interval, which is deserving of
further study.

Although superposition of the inner-halo and thick-disk MDFs and selection
biases in the sample of local calibration stars preclude a determination of the
MDF for the MWTD based on these data, we were able to demonstrate that the
metallicity range covered by the MWTD is $-$1.8 $<$ [Fe/H] $<$ $-$0.8
($\langle$[Fe/H]$\rangle$ $\sim -1.3$), and possibly up to [Fe/H] $\sim -$0.7.

The net rotational lag for the stars we identify with the MWTD
($\langle$V$_{lag}$$\rangle$ $\sim 95$ km~s$^{-1}$) is quite different with
respect to the canonical thick-disk lag ($\sim 20$ km~s$^{-1}$: Chiba \& Beers
2000; 51 km~s$^{-1}$: Soubiran, Bienayme, \& Siebert 2003; $\sim 38$
km~s$^{-1}$: this paper). The results obtained with the mixture-modeling
analysis suggest that, if the MWTD is considered an independent stellar
component of the Galaxy, its most likely mean rotational velocity and dispersion
are $\langle$V$_{\phi}$$\rangle$ $\sim$ 100-150 km~s$^{-1}$
($\langle$V$_{lag}$$\rangle$ $\sim 70-120$ km~s$^{-1}$), and $\sigma_{V_{\phi}}$
$\sim$ 40-50 km~s$^{-1}$ ($\sigma_{V_{\phi}}$ = 35-45 km~s$^{-1}$ when corrected
for observational errors). It may be of interest that the value for the MWTD lag
obtained from our analysis is in good agreement with that derived for the
putative population of satellite debris found by Gilmore, Wyse, \& Norris
(2002), who obtained a lag in the mean azimuthal streaming motion of $\sim$ 100
km~s$^{-1}$ behind the Sun. Note that the reported Gilmore et al. dispersion, on
the order of 70 km~s$^{-1}$, is slightly higher than ours, but this might be
readily explained by contamination from inner-halo stars. These authors argued
that such parameters might well be associated with stellar debris from a past
merger event.

Of even greater interest is the possible connection between the MWTD stellar
component and the Monoceros stream. Indeed, as reported by Yanny et al. (2003),
the systemic Galactocentric rotational velocity of the stream is $\sim$ 110
$\pm$ 25 km~s${-1}$ , which is in very good agreement with the mean value of
V$_{\phi}$ derived for the MWTD component ($\sim$ 125 $\pm$ 4 km~s${-1}$). In
addition, the estimated mean value of the Monoceros stream metallicity is [Fe/H]
$= -1.3$ (Wilhelm et al. 2005), in agreement with the mean value of metallicity
of the MWTD reported above.\footnote {Note, however, that the mean photometric
metallicity estimate of the Monoceros stream reported by Ivezi{\'c} et al.
(2008) is somewhat higher, [Fe/H] $= -0.95$)}. The similarity in rotational
velocity and metallicity between the MWTD and the Monoceros stream requires
further investigation, and will be carried out in the near future. Our analysis
also indicated that the kinematic properties of the MWTD may change with height
above the plane, as was already demonstrated for the thick-disk component, but
superposition with inner-halo stars confounded a definitive determination. In
any event, the preponderance of evidence suggests that the MWTD population may
indeed be a kinematically and chemically distinct population from the canonical
thick-disk population. We plan to address this issue in more detail in future
investigations, through the use of stellar samples that are more suitable for
analysis of the complexity of the thick-disk structure.

\subsubsection{Velocity Ellipsoids}

We next derived estimates of the velocity ellipsoids for the thick disk, inner
halo, and outer halo; an approximate ellipsoid for the MWTD was also derived.

In the case of the thick-disk and outer-halo components, we examined the same
sub-samples of stars used for determination of the rotational properties of
these components. The inner-halo velocity ellipsoid component in the rotational
direction was obtained from the ML analysis by holding the values of the
rotational velocities and dispersions fixed for the thick-disk and outer-halo
components. In the radial and vertical directions (where the strong overlap with
multiple additional components complicates a mixture-model analysis), the
velocities and dispersions were obtained adopting the mean velocity and
dispersion of a sub-sample of stars in a more restricted range of metallicity
($-2.0 <$ [Fe/H] $< -1.5$), where the inner halo is expected to be dominant.

Our derived ellipsoid for the rapidly rotating canonical thick disk is
($\sigma_{V_{R}}$, $\sigma_{V_{\phi}}$, $\sigma_{V_{Z}}$) = (53 $\pm$ 2, 51
$\pm$ 1, 35 $\pm$ 1) km~s$^{-1}$, after correction of the derived dispersions
for measurement errors. These values closely match the thick-disk velocity
ellipsoid obtained by Chiba \& Beers (2000). For the outer halo, which exhibits
a large net retrograde rotation, we obtained ($\sigma_{V_{R}}$,
$\sigma_{V_{\phi}}$, $\sigma_{V_{Z}}$) = (159 $\pm$ 4, 165 $\pm$ 9, 116 $\pm$ 3)
km~s$^{-1}$, corrected for measurement errors. These values are consistent with
a more tangentially anisotropic velocity ellipsoid, which was previously
advocated by Sommer-Larsen et al. (1997), from an analysis of radial velocities
of distant horizontal-branch stars.

The inner halo is essentially non-rotating, with a derived mean value of
$\langle$V$_{\phi}$$\rangle$ = $7$ $\pm$ 4 km~s$^{-1}$, and a velocity ellipsoid
($\sigma_{V_{R}}$, $\sigma_{V_{\phi}}$, $\sigma_{V_{Z}}$) = (150 $\pm$ 2, 95
$\pm$ 2, 85 $\pm$ 1) km~s$^{-1}$, corrected for measurement errors. Although our
zero mean rotational velocity differs from that reported by Chiba \& Beers
(2000), $\langle$V$_{\phi}$$\rangle$ $\sim$ 30-50 km~s$^{-1}$, our velocity
ellipsoid values are consistent, within the reported errors, with the Chiba \&
Beers determinations of a radially anisotropic ellipsoid, ($\sigma_{V_{R}}$,
$\sigma_{V_{\phi}}$, $\sigma_{V_{Z}}$) = (141 $\pm$ 11, 106 $\pm$ 9, 94 $\pm$ 8)
km~s$^{-1}$). This agreement obtains even though the Chiba \& Beers analysis
adopted a one-component halo, from which we conclude that their sample (and
others) did not include significant numbers of outer-halo stars.

Our results can also be compared with the recent analysis of SDSS subdwarfs by
Smith et al. (2009). These authors obtained full space motions for a local
sample of some 1800 subdwarfs selected on the basis of a reduced proper motion
diagram, and used the same spectroscopic determinations of stellar atmospheric
parameters as the present work. Note that for the purpose of their analysis
they considered the halo as a single entity. They concur with our determination
that the inner halo exhibits essentially no net rotation (they obtained
$\langle$V$_{\phi}$$\rangle$ = 1 $\pm$ 2 km~s$^{-1}$), but reported a slightly
different derived velocity ellipsoid. Their values ($\sigma_{V_{R}}$,
$\sigma_{V_{\phi}}$, $\sigma_{V_{Z}}$) = (137 $\pm$ 2, 81 $\pm$ 1, 89 $\pm$ 1)
km~s$^{-1}$, agree with our determination for the vertical component of the
inner halo ellipsoid, but are 15-20 km~s$^{-1}$ lower than our determinations
for the radial and azimuthal components. Similarly, the values for the
(single-component) halo velocity ellipsoid obtained by Bond et al. (2009), from
an analysis of the space motions for some 100000 main-sequence stars from
SDSS/SEGUE with available spectroscopy, ($\sigma_{V_{R}}$, $\sigma_{V_{\phi}}$,
$\sigma_{V_{Z}}$) = (135 $\pm$ 5, 85 $\pm$ 5, 85 $\pm$ 5) km~s$^{-1}$, although
in agreement with the Smith et al. ellipsoid, also differ from our results for
the radial and azimuthal components.

We suspect that the resolution of these differences may be the presence of a
significant MWTD component. The lower velocity dispersion of such a component,
if unrecognized, may have an impact on the derived halo velocity ellipsoids. In
this regard, it may be of significance that Smith et al. noted that their
observed distribution of V$_{\phi}$ exhibited a significant asymmetry; their
attempted decomposition of the distribution with a two-component model yielded
estimates for the dispersions of 47 and 99 km~s$^{-1}$, values reminiscent of
what we would associate with the MWTD and inner-halo populations, respectively
(although the mean rotational velocity they obtained for the component with a
MWTD-like dispersion was only 17 km~s$^{-1}$). They also pointed out that the
asymmetry of their V$_{\phi}$ distribution increased for stars with [Fe/H] $>
-1.5$, consistent with this interpretation.

We have attempted to estimate the other two components (the radial and vertical
components) of the velocity ellipsoid for the MWTD. Application of a
mixture-model analysis, for the range of metallicity $-1.5 <$ [Fe/H] $< -1.0$
and distance interval 1 $<$ $|$Z$|$ $<$ 2 kpc, indicates slightly negative
velocities in the radial and vertical directions, $\langle$V$_{R}$$\rangle$ =
$-$13 $\pm$ 5 km~s$^{-1}$, and $\langle$V$_{Z}$$\rangle$ = $-$14 $\pm$ 5
km~s$^{-1}$, respectively. When combined with the previously estimated value for
the rotational velocity dispersion of the MWTD, the velocity ellipsoid for the
MWTD is ($\sigma_{V_{R}}$,$\sigma_{V_{\phi}}$, $\sigma_{V_{Z}}$) = (59 $\pm$ 5,
40 $\pm$ 3, 44 $\pm$ 2 km~s$^{-1}$). Note that, for the rotational direction, we
have adopted a value for the dispersion midway between the extremes obtained
from the mixture-model analysis of different metallicity regions and distance
intervals. The velocity asymmetries in the radial and vertical directions could
indicate that stars in the MWTD component may not be on circular orbits nor have
orbits that are co-planar with the orbits of stars associated with the thin-disk
component. This could be another sign that the MWTD might be assocated with
stars from the Monoceros stream, as mentioned above.

We also demonstrated that all three components of the velocity dispersions
increase with decreasing metallicity, in a manner suggesting discontinuous
transitions from the thick disk to the MWTD, and from the inner to the outer
halo. This is a fundamental new result (also reported by Bond et al. 2009),
which can be used to place constraints on possible formation scenarios for the
stellar components of the Milky Way by comparison with new-generation numerical
simulations.

\subsubsection{Thick-Disk System Scale Lengths and Heights}

Our extensive data set for the disk systems has also been used to obtain
kinematic estimates of the thick-disk scale length and scale height. In our form
of the Jeans Equation, separated into radial and vertical components, we have
included the gradients in $\sigma_{V_{R}}$ and $\sigma_{V_{Z}}$, which are
directly measured from the data. We found that the vertical acceleration,
K$_{Z}$, obtained by Kuijken \& Gilmore (1991) is in optimal agreement with the
value independently determined by Holmberg \& Flynn (2000). Adopting the
dispersion for the thick disk reported in Table 5, a mean rotational velocity,
$\langle$V$_{\phi}$$\rangle$ = 182 km s$^{-1}$, and K$_{Z}$ at 1.1 kpc, we found
a scale length, h$_{R}$ = 2.20 $\pm$ 0.35 kpc, and a scale height, h$_{Z}$ =
0.51 $\pm$ 0.04 kpc. The smaller value of our scale length with respect to that
of Chiba \& Beers (2000) (4.5 kpc) results from our mean rotational velocity of
the thick disk, which is significantly smaller than theirs. However, our
estimate of $h_{R}$ is in agreement with that of others, based on star counts
(Reyl$\acute{e}$ \& Robin 2001).

A large range of thick-disk scale heights has been reported in the literature --
from 0.6 kpc to 1 kpc (Gilmore \& Wyse 1985; Kuijken
\& Gilmore 1989; Robin et al. 1996; Norris 1996; Ng et al. 1997; Buser
et al. 1999; Chen et al. 2001; Kerber et al. 2001; Ojha 2001; Reyl$\acute{e}$ \&
Robin 2001; Du et al. 2003; Larsen \& Humphreys 2003; Du et al. 2006 ), and from
1.1 kpc to 1.6 kpc (Gilmore \& Reid 1983; Robin \& Crez$\acute{e}$ 1986; Spagna
et al. 1996; Buser et al. 1998). Our kinematic evaluation of h$_{Z}$ falls
at the lower end of these estimates. A possible explanation might be the derived
velocity dispersion in the vertical direction for the thick disk (35 km
s$^{-1}$), which could be reduced by the presence of thin-disk stars in the
selected sub-sample. The gradient of $\sigma_{V_{Z}}$ does not contribute
significantly in the Jeans Equation; even when an isothermal thick disk is
considered ($\partial \sigma_{V_{Z}} / \partial |Z|$ = 0), we obtain h$_{Z}$ =
0.64 kpc. Note that our estimate of the scale height is made at $|$Z$|$ = 1.1
kpc, while the ``usually larger" estimates come from star counts at $|$Z$|$
$\sim$ 2 to 4 kpc (see references above). In these works, the thick disk appears
as vertically exponential at larger $|$Z$|$, but in fact, it is unlikely to be
strictly exponential. If we consider the general version of the Jeans Equation,
and perform a numerical integration under the assumption that the thick disk is
vertically isothermal, with a velocity dispersion of $\sigma_{V_{Z}}$ = 35 km
s$^{-1}$, we find that at $|$Z$|$ = 1.1 kpc, the ``local scale height'' is in
perfect agreement with our value of 0.64 kpc. On the other hand, if we define an
effective scale height of the thick disk as the height above the Galactic plane
at which the density decreases by a factor $e$ from its value at $|$Z$|$ = 0, we
find that this height is 0.99 $\pm$ 0.04 kpc, which is not directly comparable
with the star count scale heights at larger $|$Z$|$.

In the case of the MWTD, we have only the kinematic parameters listed in Table 5
(i.e., no directly measured gradients could be obtained), which can be used to
derive important constraints on the radial structure of this component. Indeed,
using the radial component of the Jeans Equation, and assuming that
$\sigma_{V_{R}}^2 \propto exp(-R/h_{\sigma})$, where $h_{\sigma}$ is a scale
length, we find that the MWTD scale length is again short, $\sim$ 2 kpc. This
result comes from the assumption that the MWTD presents a uniform scale height
and constant anisotropy. The scale height has been derived adopting the same
value of the vertical acceleration reported above, and under the same
assumptions. We obtained $h_{Z}$ = 1.36 $\pm$ 0.13 kpc, in agreement with the
larger values of the scale height estimation of the canonical thick disk.

\subsubsection{Inferred Density Profiles of the Inner- \\ and Outer-Halo
Components}

The kinematics for our local sample of calibration stars were used to infer
density profiles for the inner-halo and outer-halo components. We obtained
$\rho_{in} \sim r^{-3.17 \pm 0.20}$ for the inner halo, consistent with the
derived density profile of ``the halo'' found by many previous authors (e.g.,
Harris 1976; Zinn 1985; Hartwick 1987; Carney, Latham, \& Laird 1990; Preston at
al. 1991; Chiba \& Beers 2000). In contrast, the density profile obtained for
the outer halo, $\rho_{out} \sim r^{-1.79 \pm 0.29}$, is substantially shallower
than that of the inner halo, as expected from the higher values of the velocity
dispersions for this component. It is thus of interest to note that Ivezi{\'c}
et al. (2000), based on a sample of RR Lyraes located between 10 and 50 kpc from
the Galactic center selected from SDSS commisioning data, obtained a slope of
$-2.7$ $\pm$ 0.2. From the QUEST RR Lyrae Data (which cover basically the same
distance range as the Ivezi{\'c} et al. sample), Vivas \& Zinn (2006) obtained
values for the derived slope between $-2.5$ $\pm$ 0.1 and $-3.1$ $\pm$ 0.1,
depending on the adopted level of flattening and local density. Our low value of
the slope for the outer-halo component is reminiscent to that obtained by Miceli
et al. (2008) for Oosterhoff Type~II RR Lyraes ($-2.88$ $\pm$ 0.11; the value
for Oosterhoff Type~I class is even smaller, $-2.26$ $\pm$ 0.07). Note that both
of these values assumed a spherical halo, and included stars only up to about 30
kpc away. When a flattened halo is considered, the Miceli et al. slope changes
from $-2.43$ $\pm$ 0.06 (all RR Lyrae types, spherical halo) to $-3.15$ $\pm$
0.07. Sesar et al. (2009) report a dramatic ``steepening'' of the halo density
profile beyond 30 kpc, based on an analysis of RR Lyraes and main-sequence stars
from the SDSS equatorial stripe, in the sense that the numbers of stars at large
Galactocentric distances is much lower than would be expected from extrapolation
of the (single halo component) density profile they adopt to describe the
density profile within 30 kpc. Alternatively, this result could be interpreted
as arising from the changeover from inner-halo to outer-halo dominance in the
star counts, with the outer-halo contributing substantially fewer stars beyond
30 kpc (we roughly estimate that the local relative normalization of
outer-halo to inner-halo spatial density laws is about 10\%). Thus, although
differences in the precise values of the slopes obtained from different data
sets and different analysis approaches remain, it appears that the inner/outer
halo model very naturally accounts for observed changes in the density profile
of halo tracers with distance reported by a number of previous groups.

\subsubsection{Derived Tilts of the Velocity Ellipsoids}

The kinematic parameters derived for stars in our local sample have been used to
examine tilts of the velocity ellipsoids, as a function of height above the
Galactic plane, over several intervals in metallicity.

Misalignment of the velocity ellipsoids with the adopted cylindrical coordinate
system has been found for all the selected sub-samples. At high metallicity the
tilt angle is 7$^{\circ}.1$ $\pm$ 1$^{\circ}$.5, when 1 $<$ $|$Z$|$ $<$ 2 kpc,
and 5$^{\circ}$.5 $\pm$ 1$^{\circ}$.2 at 2 $<$ $|$Z$|$ $<$ 4 kpc. A similar
value was reported by Siebert et al. (2008) for a sample of RAVE-survey stars at
$|$Z$|$ $<$ 1 kpc (7$^{\circ}.3$ $\pm$ 1$^{\circ}$.8). At intermediate
metallicity the tilt angle is 10$^{\circ}.3$ $\pm$ 0$^{\circ}$.4, and
15$^{\circ}.1$ $\pm$ 0$^{\circ}$.3, while for low-metallicity stars we found
8$^{\circ}.6$ $\pm$ 0$^{\circ}$.5, and 13$^{\circ}.1$ $\pm$ 0$^{\circ}$.4, at 1
$<$ $|$Z$|$ $<$ 2 kpc and  $<$ $|$Z$|$ $<$ 4 kpc, respectively. The velocity
ellipsoids point close to the Galactic center, as can be inferred from Figure
19. Bond et al. (2009) also report similar tilts of the velocity ellipsoid for
low-metallicity stars. The existence of these tilts indicates that motions of
stars in our local sample are aligned with respect to a spherical, rather than
cylindrical, coordinate system.

We considered the possibility that the inner halo and/or outer halo might exhibit
kinematics close to isotropic in ($\theta$,$\phi$), which would be consistent
with a system for which the distribution function depends on the energy and the
total angular momentum. For a sample of stars with $-$2.0 $<$ [Fe/H] $-$1.6, and
located at 2 $<$ $|$Z$|$ $<$ 4 kpc, where the dominant component is expected to
be the inner halo, we find ($\sigma_{V_{r}}$, $\sigma_{V_{\phi}}$,
$\sigma_{V_{\theta}}$) = (160 $\pm$ 3, 102 $\pm$ 2, 83 $\pm$ 2) km s$^{-1}$.
These values are somewhat higher than those recently obtained by Smith et al.
2009 (142, 81, 77) km s$^{-1}$, and Bond et al. (2009) (141, 85, 75)
km s$^{-1}$. In the case of the outer halo, a similar estimate is more uncertain,
due to the difficulty in separating the two halo components in V$_{\phi}$ and
V$_{Z}$. Nevertheless, an attempt was made, employing the interval of
metallicity [Fe/H] $< -2.4$ and 2 $<$ $|$Z$|$ $<$ 4 kpc, which should emphasize
outer-halo stars. We obtained ($\sigma_{V_{r}}$, $\sigma_{V_{\phi}}$,
$\sigma_{V_{\theta}}$) = (178 $\pm$ 9, 149 $\pm$ 7, 127 $\pm$ 6). The conclusion
of this exercise is that both the inner and outer halo are significantly
anisotropic in ($\theta$,$\phi$).

\subsubsection{Variation of the Observed Metallicity Distribution Function with
Distance from the Galactic Plane}

Finally, we examined the ``as observed'' MDF using the full sample of
calibration stars, to look for confirmation of the expected changes as
a function of the height above the plane (Fig. 20). We constructed MDFs at
different intervals of $|$Z$|$, close to the Galactic plane (0 $<$ $|$Z$|$ $<$ 5
kpc), and farther from the Galactic plane (5 $<$ $|$Z$|$ $<$ 9 kpc, and $|$Z$|$
$>$ 9 kpc). In the first case, the MDF exhibits a transition from a thick-disk
and MWTD-dominated population to an inner-halo dominated population, while in
the second case, there is clear evidence for a transition from inner-halo to
outer-halo dominance.

\subsubsection{Comparison with the Results of Bond et al. (2009)}

The recent analysis conducted by Bond et al. (2009) considered the kinematics of
a larger sample of SDSS stars than the present investigation, based on both
photometric and spectroscopic metallicity estimates.  Although the analysis
techniques and the adopted models differ somewhat from those used herein, the
basic conclusions of these two studies are very similar.  We point out that the
Bond et al. approach was not able to effectively split an assumed single-halo population
into inner- and outer-halo components, since the metallicities of outer-halo
stars are lower than can be usefully assigned abundances from photometric
techniques.  Nevertheless, one can see the presence of stars with retrograde
velocities in the lower-right panel of Figure 6 of Bond et al. (stars selected
with $|$Z$|$ in the range 5-7 kpc) as an ``excess" population relative to
their assumed single-halo population (note that in Bond et al., the rotational velocities
are assigned an opposite sign than in the present work; we are referring to
stars in their figure with V$_{\phi} > 250$ km~s$^{-1}$).  Similarly, again from
inspection of the same panel in their Figure 6, essentially all of their data
points with positive (meaning retrograde) rotational velocities fall {\it under}
their single-Gaussian fit, while those with negative (meaning prograde)
velocities in the range $-150 < $ V$_{\phi}$ $< 0$ km~s$^{-1}$ fall {\it above}
their single-Gaussian fit. This apparent lack of fit is almost certainly due to the lack
of an inner/outer split in their model of the halo population.  Nevertheless, due
to the relatively small number of outer halo stars at the distance limit of
their sample (7 kpc), the derived kinematics of their ``halo population" are
quite similar to our inner-halo population.

The results of these two studies for the thick-disk populations, although
approached in a different way, are also compatible.  In the Bond et al.
approach, the disk stars in their sample were modeled with a non-Gaussian
distribution for the rotational velocities, incorporating a component with a
fixed lag in velocity relative to what we would consider the canonical thick
disk.  We would identify this lower rotational velocity component with the MWTD
of our analysis.  It is interesting that they found that this model was able to
accommodate the observed behavior of their sample smoothly with increasing height
above the Galactic plane. Clearly, there remains more to be learned from future
analysis of these data.

\subsection{Discussion}

The structural and kinematic parameters obtained in this paper provide valuable
information for understanding the formation of the stellar halos and thick disks
of the Milky Way. We have confirmed that the inner- and the outer-halo
components exhibit different kinematic properties and inferred spatial
distributions, and (from the previous analysis of the DR5 sample of calibration
stars by Carollo et al. 2007, and confirmed by our {\it in-situ} analysis of the
the full calibration-star sample) that their metallicity distribution functions
differ significantly as well. We have also confirmed the presence of a
significant vertical gradient in $\langle$V$_{\phi}$$\rangle$ as a function of
$|$Z$|$ at low abundances, which suggests that the inner-halo component was most
likely formed by dissipational processes.

A number of recent theoretical and observational efforts provide supporting
evidence for the existence of an inner/outer halo dichotomy in the Milky Way,
and in other similar galaxies. We consider these in turn.

\subsubsection{The Halos:  Evidence from Theoretical Modeling}

De Lucia \& Helmi (2008) have adapted a high-resolution simulation of a Milky
Way-like halo, combined with a semi-analytic approach, to explore the formation
of the Galaxy. They found {\it no evidence} for a ``continuous" metallicity
gradient for halo stars, but rather, evidence for a strong concentration of
higher-metallicity halo stars toward the Galactic center, and the presence of
lower-metallicity stars at distances beyond 20 kpc from the Galactic center. The
results of this simulation bear a strong resemblance to the properties of the
dominant inner halo we find within 10-15 kpc from Sun, as well as with its
rather steep inferred density profile, and an outer halo that dominates beyond
15-20 kpc, with a much shallower inferred density profile.

In a recent paper, Zolotov et al. (2009) have investigated the formation of the
stellar halos in simulated disk galaxies, using high-resolution SPH + N-Body
simulations. They found that the inner regions (r $<$ 20 kpc) contain both
accreted and {\it in-situ} stellar populations (the latter includes stars formed
in the main halo). Contrasting with this, the outer regions of the simulated
halos were assembled through pure accretion and the disruption of satellites.

Other recent supporting evidence is found in a simulation by Salvadori et al
(2009). These authors studied the age and metallicity distribution function of
metal-poor stars in the Milky Way halo, as a function of Galactocentric radius,
from a study combining N-body simulations and semi-analytical methods. They
found that the stellar distributions, within 50 kpc from the center of their
simulated galaxy, follow a power-law in radius, $r^{n}$, where $n \sim$ $-$3.2
for stars with $-$2 $\leq$ [Fe/H] $\leq$ $-$1, and $n \sim$ $-$2.2 for stars
with [Fe/H] $\leq$ $-$2. In this context, the relative contribution of stars
with [Fe/H] $\leq$ $-$2 strongly increases with $r$ (16\% within 7 $< r <$ 20
kpc, up to $>$ 40\% for $r >$ 20 kpc ). Note that our spatial-density power laws
for the inner and outer halo are in reasonable agreement, within the errors,
with the values found by Salavadori et al (2009).

Cooper et al. (2009) have also conducted simulations of the formation of stellar
halos within the $\Lambda$CDM paradigm, offering a number of improvements over
previous efforts. Their results, which cover a range of possible halo formation
histories, produce predicted metallicity shifts at various radii, several of
which are reminiscent of what we have found in our own analysis of the Milky
Way's halo system. It is notable that, in all cases, the lowest metallicity
stars in their simulations are accreted (or stripped) from the very sort of
ultra-faint dwarf galaxies that Carollo et al. (2007) suggested may be related
to the origin of the outer-halo population.

\subsubsection{The Halos: Evidence from Observations\\ of Other Galaxies}

For at least a decade, the mystery of the relatively ``metal-rich" halo of M~31
has presented a challenge to our understanding of its formation history.
Photometric and spectroscopic observations by Kalirai et al. (2006b), and more
recently by Koch et al. (2008a), have demonstrated that a metal-poor halo of
M~31 does indeed exist, but is located beyond 40 kpc from its center
(substantially different than observed for the Milky Way, where the outer halo
becomes dominant beyond 20 kpc). Their description of a dual-halo model for the
Andromeda Galaxy parallels what we observe for the Milky Way, albeit on
different length scales.

Recent observations of the Milky Way analogue NGC~891 have provided additional
information on the structure and the stellar populations for this relatively
massive spiral galaxy in the NGC~1023 group (Ibata, Mouhcine, \& Rejkuba 2009;
Rejkuba, Houhcine, \& Ibata 2009). Analysis of their data revealed the presence
of a thick-disk structure with vertical scale height $h_{Z}$ = 1.4 kpc, and
scale length $h_{R}$ = 4.8 kpc, somewhat larger than values that have been
inferred for the Milky Way. The MDF of the thick-disk structure exhibits a peak
at [Fe/H] $\sim -0.9$ (as compared to [Fe/H] $\sim -0.6$ for the canonical thick
disk of the Milky Way) that does not change significantly in the vertical and
radial directions. When they considered the MDF for stars in regions resembling
the solar cylinder we have examined in the present paper (distance along the
major axis $\sim$ 8 kpc, vertical distance intervals of $\sim$ 3-4 kpc and 5-7
kpc), they detected a substantial population of intermediate-metallicity stars,
with [Fe/H] $\sim -1.0$. By way of contrast, at these same vertical distances
the MDF of the Milky Way is dominated by the inner halo ([Fe/H] $\sim -$1.6;
Ivezi{\'c} et al. 2008; this paper). The intermediate-metallicity population in
the solar cylinder of NGC~891 might be associated with a component similar to
the Milky Way's MWTD, but are present at larger distances from the plane in this
galaxy than in the Milky Way. The difference between the verticle extension and
MDF of the disk populations between NGC~891 and the Milky Way may simply reflect
different (stochastic) formation histories for the two galaxies.

The stellar halo of NGC~891 exhibits a shallow metallicity gradient, with
average metallicity decreasing from [Fe/H] $ = -1.15$ to $-1.27$ from the inner
regions to the outskirts. A similar trend was also found in the inner halo of
M~31 (20 kpc from the nucleus; Durrell, Harris, \& Pritchet 2001), and in the
halo of NGC~5128, which is the nearest giant elliptical galaxy (Rejkuba et al.
2005). In subsequent work, metal-poor halos were detected for these galaxies, at
distances larger than 12$r_{\rm eff}$ from the center (Chapman et al. 2006;
Kalirai et al. 2006a; Harris et al. 2007). However, in NGC~891, at the same
radius, a metal-poor halo has not (yet) been detected. In this regard we note
that in the right-hand panel of Figure 18 in Rejkuba et al. (2009) there is a
clear transition of the MDF, from intermediate ([Fe/H] $\sim -$1.0) metallicity
to a lower ([Fe/H] $\sim -$1.5) metallicity, which suggests the presence of a
metal-poor population like that in the inner halo of the Milky Way. We suspect
that when additional photometry at larger distances is carried out for NGC~891
its MDF will reveal metal-poor halo population(s) similar to those of the Milky
Way.

Recent spectroscopic studies of the metallicities and other elemental abundances
in the ultra-faint dSph galaxies (Kirby et al. 2008; Koch et al. 2008b; Norris
et al. 2008; Frebel et al. 2009) have clearly demonstrated that the frequently
cited result (e.g., Helmi et al 2006) that ``the progenitors of nearby dSph's
appear to have been different from the building blocks of the Milky Way,'' while
true for the more luminous members of the dSph population, does not necessarily
pertain to its fainter systems. While we refer the reader to the reviews of Koch
(2009) and Tolstoy, Hill, \& Tosi (2009) for more thorough discussions, we note
here two relevant points. First, it has been shown that the metallicities of
stars in even the more luminous dSphs (e.g., Sculptor; Battaglia et al. 2008)
are spatially stratified, with the more metal-poor objects being found
preferentially in their outer regions. These metal-poor stars are clearly the
most likely to be stripped from their parents during tidal interactions, with
the halo(s) of the Milky Way as the obvious recipients. Secondly, the works
cited above establish that extremely metal-poor stars ([Fe/H] $< -3.0$) exist in
the ultra-faint systems, and for [Fe/H] $< -1.6$, one finds [$\alpha$/Fe] $\sim$
0.3--0.5, similar to Galactic halo values. It is not difficult to imagine a
substantial population of low-luminosity dwarfs, most of which have already been
completely dissolved, accounting for the entire population of stars in the
presently observed outer-halo component of the Milky Way. We look forward to
further guidance for this idea to be obtained by additional high-resolution
spectroscopic observations of stars in the low-luminosity dwarfs, as well as
detailed elemental abundance determinations for stars with kinematics that
suggest membership in the outer-halo population (e.g., Ishigaki et al. 2009;
Roederer 2009; Zhang et al. 2009; Roederer et al., in preparation).

\subsubsection{The Thick Disk: Evidence from Theoretical Modeling}

Simulations that attempt to account for the formation of the thick-disk
component have suggested that the asymmetric drift we find for this structure
can be naturally accommodated. According to the simulations of Abadi et al.
(2003), it may also be consistent with the expected behavior if the thick disk
is comprised solely of debris from merging satellites. The formation of a MWTD
component may be the result of a stochastic event, such as assimilation of the
debris from a moderately metal-poor, low-mass satellite.

Modern cosmogonies predict the formation of the thick disk by predominantly
dissipationless processes. In this context, thick-disk stars were vertically
heated from a pre-existent thin disk during a significant minor merger or
perturbation by a cold dark matter sub-halo (Kazantzidis et al. 2009), or
directly deposited at large scale heights as tidally stripped debris during the
accretion of smaller satellite galaxies (e.g., Statler 1988; Abadi et al. 2003;
Yoachim \& Dalcanton 2008). Villalobos \& Helmi (2008) have investigated in
detail the heating process by a minor merger. They found that the trend of the
ratio $\sigma_{V_{Z}}$/$\sigma_{V_{R}}$ with radius in the final disks is a good
discriminant of the initial inclination of the merging satellite. For the Milky
Way, the observed $\sigma_{V_{Z}}$/$\sigma_{V_{R}}$ is $\sim$ 0.6 (Chiba \&
Beers 2000; this paper), which suggest that the thick disk of the Galaxy could
have been produced by a merger of intermediate inclination. The values found for
the MWTD velocity ellipsoid provide a ratio $\sigma_{V_{Z}}$/$\sigma_{V_{R}}$
$\sim$ 0.7, which could be explained through a merger of different inclination.

Hayashi \& Chiba (2006) have shown that disk thickening, quantified by the
change of its scale height (or the square of its vertical velocity dispersion),
strongly depends on the individual mass of an interacting sub-halo. In this
scenario of hierarchical formation, the MWTD could be formed through the early
merger (at least with respect to the merger responsible for the formation of the
canonical thick disk) of a satellite with a less massive, younger pre-existing
disk. Note, however, that the existence of stars with disk kinematics and older
than the last major merger event is difficult to accommodate in the $\Lambda$CDM
scenario (Abadi et al. 2003). The results of the simulations of Abadi et al.
suggest that the tidal debris from satellites can contribute not only to the
Galactic halo, but also to the disk system. The contribution to the halo or to
the disk depends on the orbit of the merging satellite, and on the degree to
which dynamical friction circularized the orbits before disruption (Statler
1988). An accreted satellite that contributes significantly to the thick disk
must be disrupted on an orbital plane very close to that of the disk, and be
massive enough to survive the disruption until its orbit is circularized within
the disk. In this scenario, the predicted location of the disrupted satellites
responsible for the formation of the thick disk and MWTD is within the
thick-disk system itself. We have noted in \S 11.1 that some properties of the
Monoceros stream (rotational velocity and metallicity) are in good agreement
with those of the MWTD. {\it The Monoceros stream could be part of the disrupted
satellite that was also responsible for the formation of the thick-disk system.}

\subsubsection{The Metal-Weak Thick Disk: Additional Evidence Needed
from Observations}

We have attempted to constrain the properties of the MWTD as an independent
kinematic component from the canonical thick disk. It was found that its
velocity lag is $\sim$ 60 km s$^{-1}$ lower than the canonical thick disk; its
velocity ellipsoid appears similar to that of the thick disk (although its
vertical velocity dispersion, and hence its inferred scale height, appear
somewhat larger). These values are, of course, still uncertain due to the
significant kinematic and spatial overlap of the MWTD with the thick-disk and
inner-halo components. Additional study (informed by more-detailed abundance
analyses of likely MWTD stars) must be given to this problem before its final
resolution can be obtained. Other sub-samples of stars from SDSS/SEGUE, beyond
the calibration stars discussed in our study, as well as stars from other
surveys such as RAVE, will help shed light on the nature of the Galactic
components we have described. It is evident that the final observational picture
has yet to be drawn; nuances continue to be revealed as our database of
information expands. Witness the ``highly flattened" inner-halo component
suggested by Morrison et al. (2009) that comprises stars with intermediate low
metallicities, and is confined to a region close to the Galactic plane, but (in
contrast to the MWTD) is not supported by rotation.

\acknowledgments

Funding for the SDSS and SDSS-II has been provided by the Alfred P. Sloan
Foundation, the Participating Institutions, the National Science Foundation, the
U.S. Department of Energy, the National Aeronautics and Space Administration,
the Japanese Monbukagakusho, the Max Planck Society, and the Higher Education
Funding Council for England. The SDSS Web Site is http://www.sdss.org/.

The SDSS is managed by the Astrophysical Research Consortium for the
Participating Institutions. The Participating Institutions are the American
Museum of Natural History, Astrophysical Institute Potsdam, University of Basel,
University of Cambridge, Case Western Reserve University, University of Chicago,
Drexel University, Fermilab, the Institute for Advanced Study, the Japan
Participation Group, Johns Hopkins University, the Joint Institute for Nuclear
Astrophysics, the Kavli Institute for Particle Astrophysics and Cosmology, the
Korean Scientist Group, the Chinese Academy of Sciences (LAMOST), Los Alamos
National Laboratory, the Max-Planck-Institute for Astronomy (MPIA), the
Max-Planck-Institute for Astrophysics (MPA), New Mexico State University, Ohio
State University, University of Pittsburgh, University of Portsmouth, Princeton
University, the United States Naval Observatory, and the University of
Washington.

D.C. acknowledges funding from RSAA ANU to pursue her research. She is
particularly grateful to W. E. Harris for useful discussions on the
mixture-modeling analysis and the R-Mix package, during his visit to Mount
Stromlo Observatory. She also is grateful for partial support from JINA, which
funded her multiple visits to MSU, and for the hospitality of its faculty,
staff, and students during her stays. T.C.B. and Y.S.L. acknowledge partial
funding of this work from grants PHY 02-16783 and PHY 08-22648: Physics Frontier
Center/Joint Institute for Nuclear Astrophysics (JINA), awarded by the U.S.
National Science Foundation. T.C.B. is grateful for the assistance and
hospitality of the faculty, staff, and students at Mount Stromlo Observatory
during a recent research leave. M.C. acknowledges support from a Grant-in-Aid
for Scientific Research (20340039) of the Ministry of Education, Culture,
Sports, Science and Technology in Japan. Studies at ANU of the most metal-poor
populations of the Milky Way are supported by Australian Reseach Council grants
DP0663562 and DP0984924. Z.I. acknowledges support from NSF grants AST 06-15991
and AAST-07 07901, as well as from grant AST 05-51161 to LSST for design and
development activities. 





{\it Facilities:} \facility{SDSS}.

{}

\clearpage

\begin{figure}
\figurenum{1}
\epsscale{1.0}
\plotone{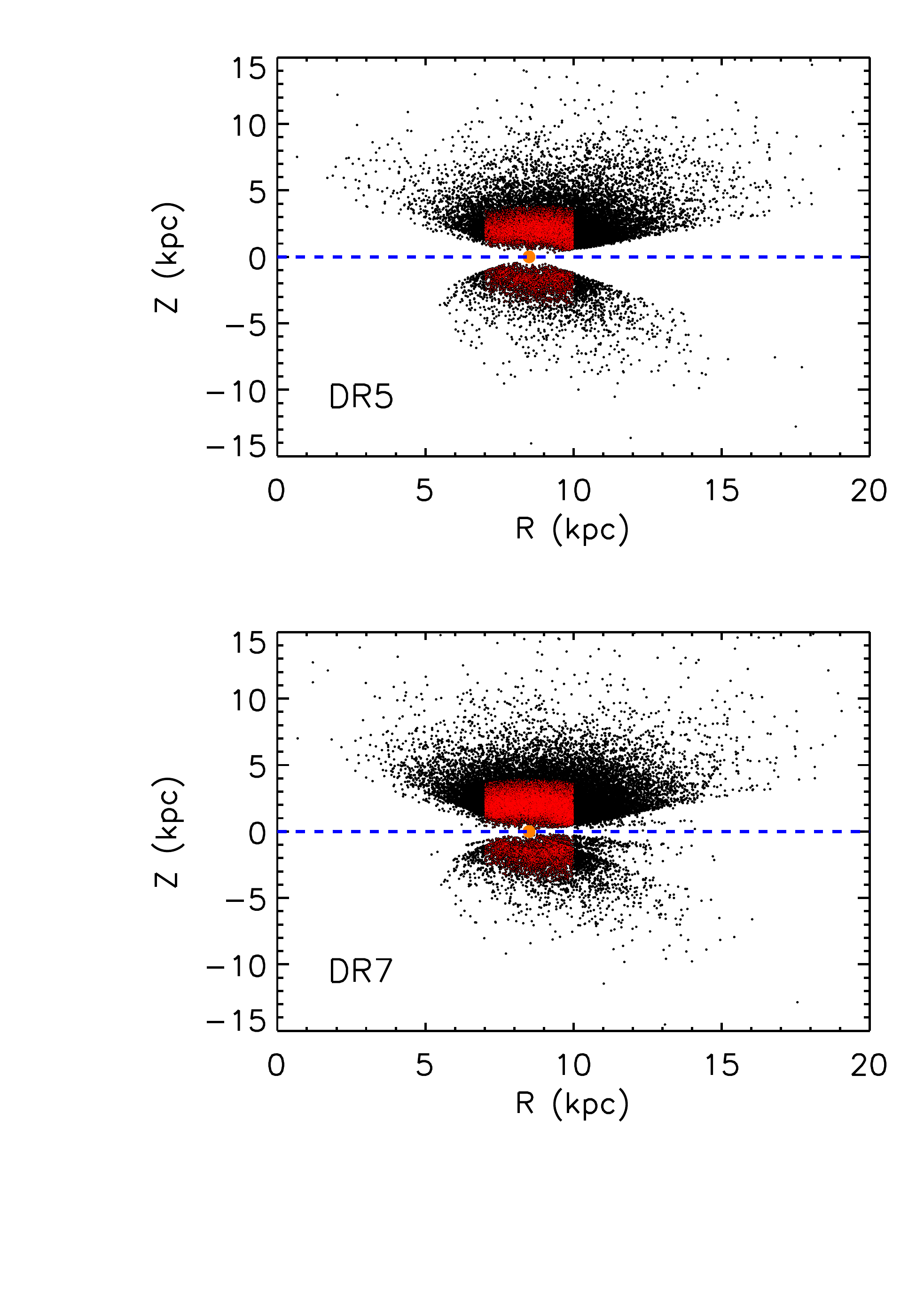}
\caption{Spatial distribution in the Z-R plane of the 20236 SDSS DR5 (upper panel),
and 32360 SDSS/SEGUE DR7 (lower panel) calibration stars.  The black
points represent the full sample, while the red points indicate the stars that
satisfy our criteria for a local sample. The dashed blue line represents the
Galactic plane, and the filled orange dot is the position of the Sun (at Z = 0
kpc; R = 8.5 kpc). In both panels, the wedge shape of the selection area is
the result of the limits of the SDSS footprint in Galactic latitude. The overall
increase in the numbers of calibration stars between DR5 and DR7 is clear, in
particular for regions closer to the Galactic plane, due to the lower
Galactic-latitude fields explored by SEGUE.}

\end{figure}
\clearpage

\begin{figure}[htp]
\figurenum{2}
\centering
\includegraphics[scale=0.7]{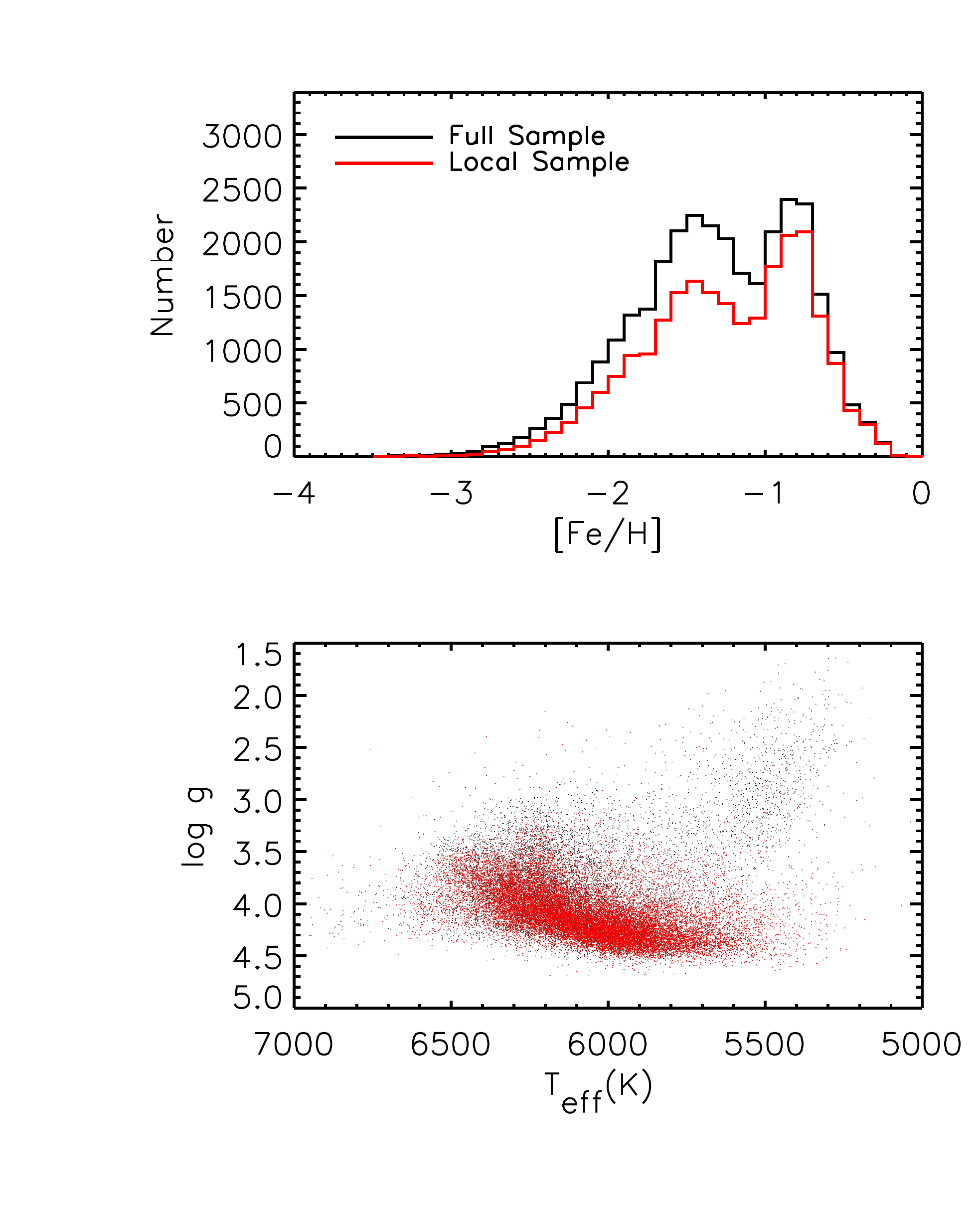}
\caption{Upper panel: Metallicity Distribution Function (MDF), {\it as observed}, for the full sample of
calibration stars (black histogram), and for the stars satisfying our criteria
for the local sample (red histogram).  Lower panel:  Distribution of surface
gravity, log g, versus effective temperature, T$_{\rm eff}$, for the full sample
(black dots) and local sample (red dots).  Note that the full sample contains
substantial numbers of main-sequence turnoff stars, subgiants, and giants,
while the local sample primarily comprises main-sequence turnoff stars and
dwarfs.}

\end{figure}
\clearpage

\begin{figure}[htp]
\figurenum{3}
\centering
\includegraphics[scale=0.7,angle=90]{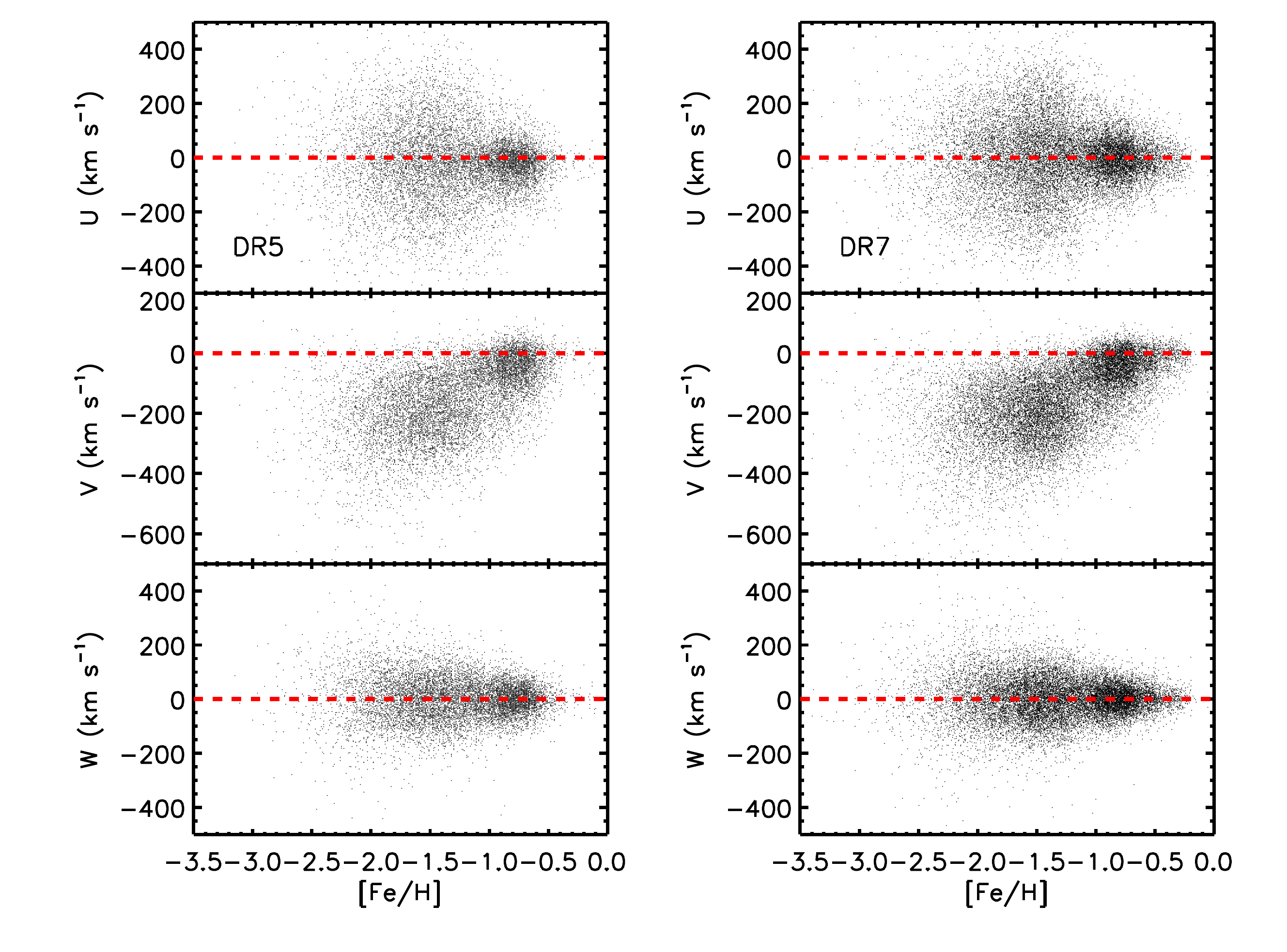}
\caption{Distribution of the velocity components (U,V,W) versus [Fe/H] for the
SDSS DR5 (left-hand panels) and the SDSS/SEGUE DR7 (right-hand panels)
calibration stars. In both sets of panels only the local sample has been
considered, comprising totals of 10120 and 16920 stars, respectively. The red
dashed line is the adopted LSR for stars in the solar neighborhood.  Note how
the primary components of the stellar populations represented in the local
volume, the thin disk, the thick disk, and the halo, are more clearly discerned in
the DR7 sample.  This is primarily due to refinements in the stellar parameter
estimates obtained by the SSPP.}

\end{figure}
\clearpage

\begin{figure}
\figurenum{4}
\epsscale{1.0}
\plotone{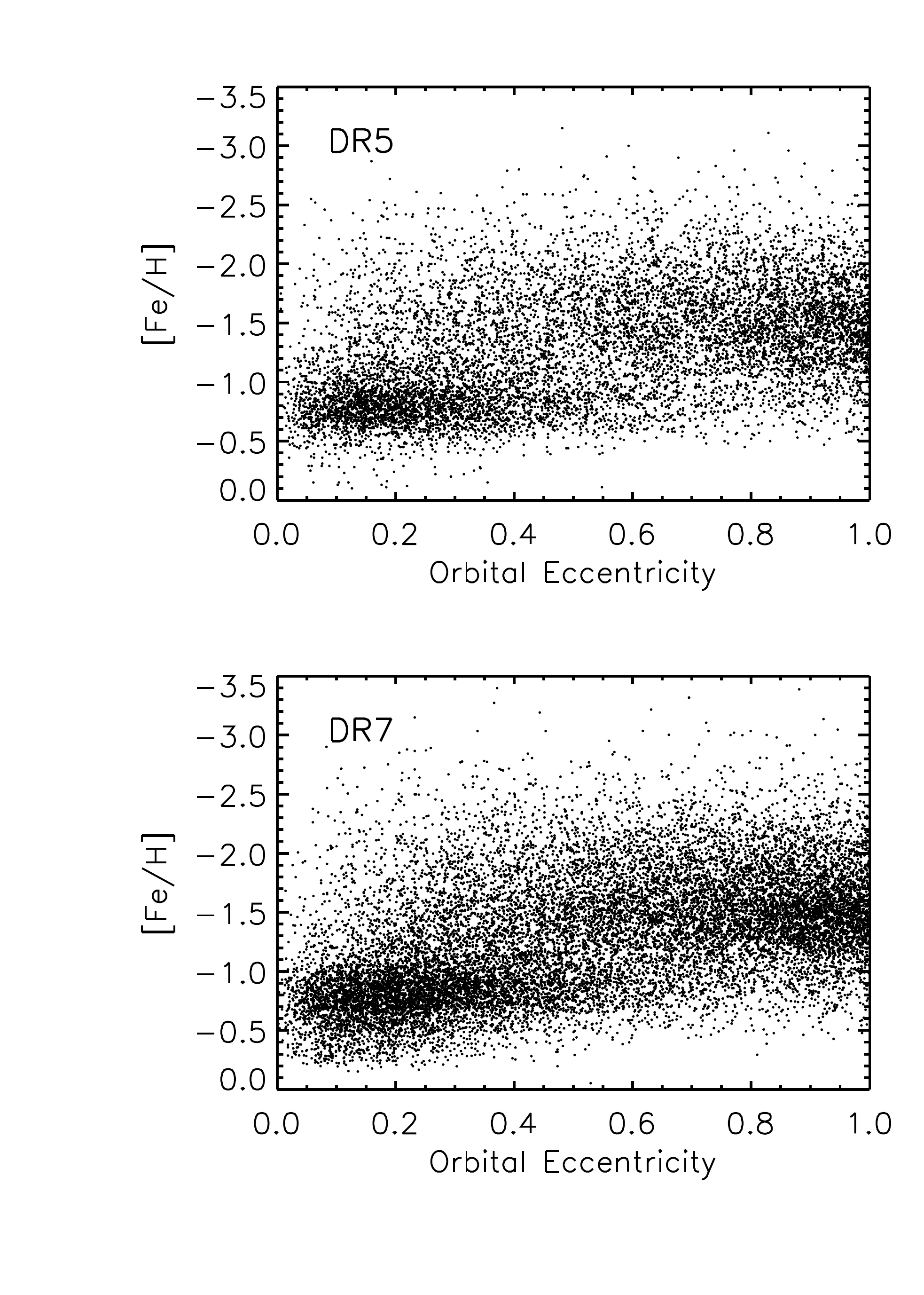}
\caption{[Fe/H] versus orbital eccentricity for the SDSS DR5
(upper panel) and the SDSS/SEGUE DR7 (lower panel) calibration stars
in the local volume.  Note the far greater numbers of stars from
DR7, as compared to DR5, in addition to the more densely populated disk-like
regions for the DR7 data, due to the
addition of the SEGUE fields.}

\end{figure}
\clearpage

\begin{figure}[htp]
\figurenum{5}
\centering
\includegraphics[scale=0.7]{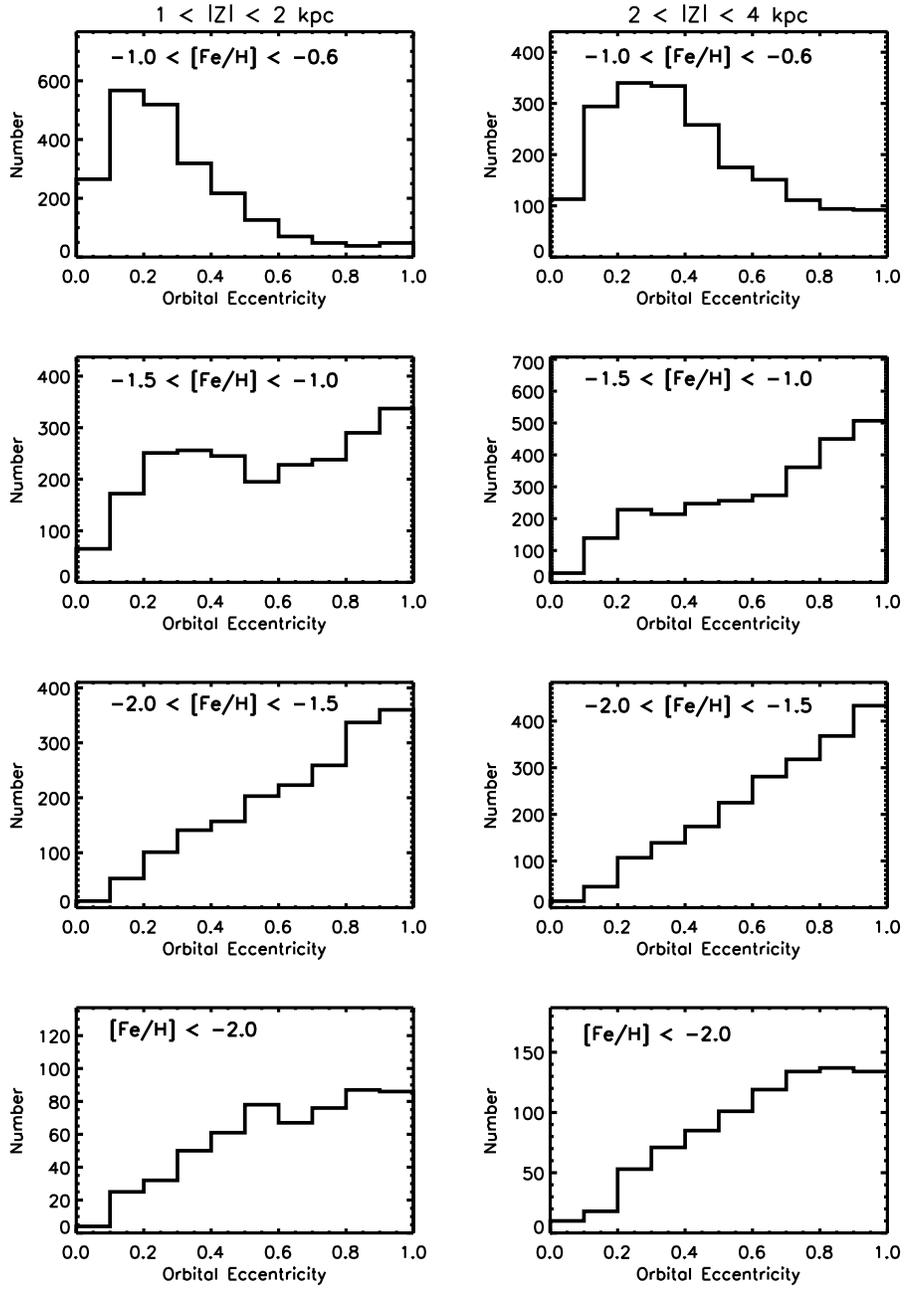}
\caption{Distribution of the derived orbital eccentricity for sub-samples of
stars selected in different metallicity ranges, for two different intervals on
vertical distance $|$Z$|$, close to (1 $<$ $|$Z$|$ $<$ 2 kpc; left-hand column of panels)
the Galactic plane, and farther from (2 $<$ $|$Z$|$ $<$ 4 kpc; right-hand column
of panels) the Galactic plane.}

\end{figure}
\clearpage

\begin{figure}[htp]
\figurenum{6}
\centering
\includegraphics[scale=0.6,angle=90]{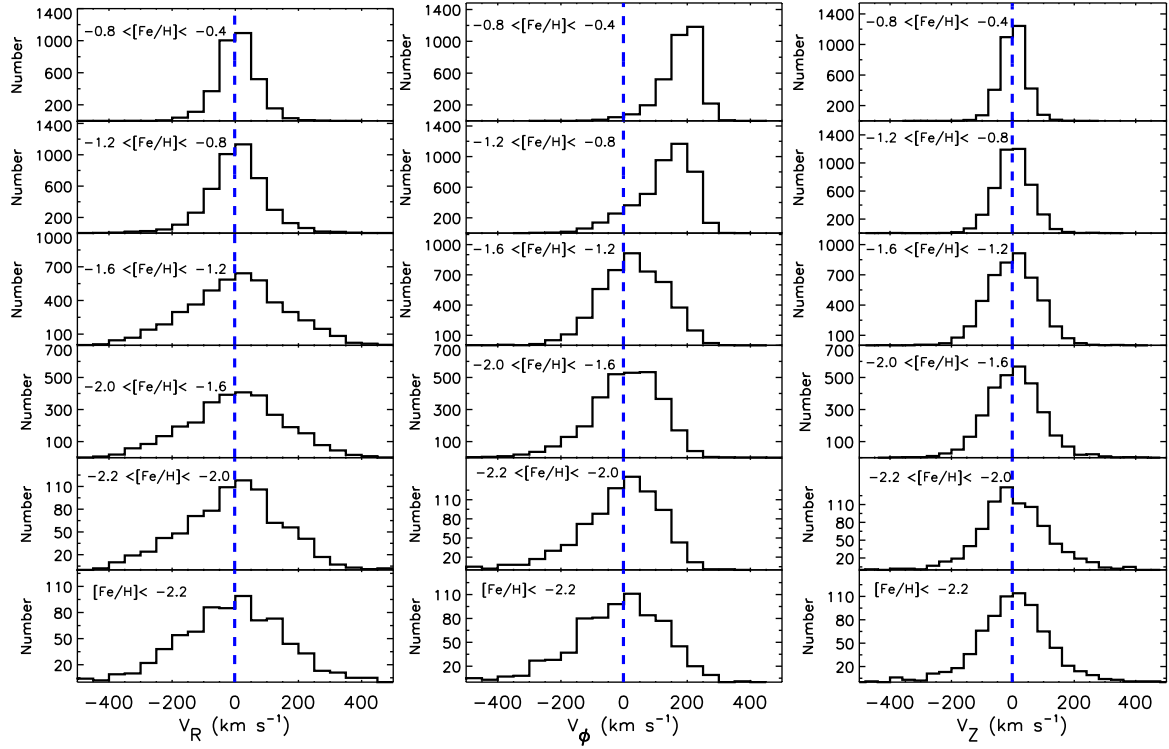}
\caption{Distribution of the derived velocity components in a cylindrical
Galactocentric frame,
(V$_{R}$,V$_{\phi}$,V$_{Z}$), for different ranges of [Fe/H]. The blue dashed line
is a reference line at zero for each component.
}

\end{figure}
\clearpage

\begin{figure}
\figurenum{7}
\epsscale{0.7}
\plotone{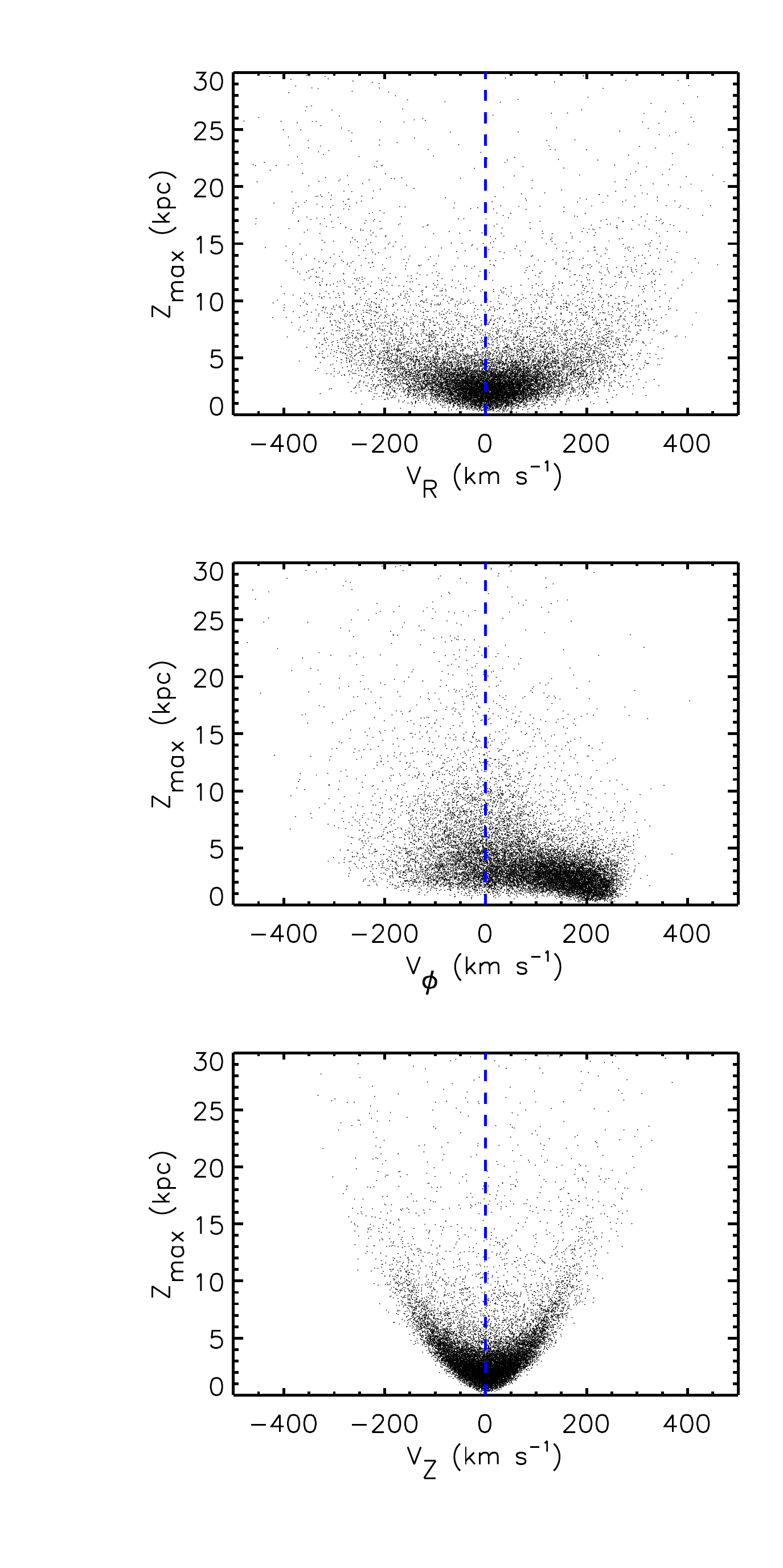}
\caption{The relation between Z$_{max}$ and the cylindrical velocity components
(V$_{R}$,V$_{\phi}$,V$_{Z}$) for the calibration stars in the local sample.
Note the strong correlation between Z$_{max}$ and the radial velocity
component shown in the upper panel, V$_{R}$, as well as with the vertical velocity component
shown in the lower panel, V$_{Z}$. The middle panel exhibits no strong
correlation between Z$_{max}$ and the rotational velocity component, V$_\phi$,
other than that expected from the presence of the thick-disk and halo
populations. The blue dashed line is a reference line at zero for each
component.  Note, in the middle panel, the clear excess of stars with
retrograde motions for Z$_{max}$ $>$ 15 kpc, which we associate with the
outer-halo component.} 

\end{figure}
\clearpage

\begin{figure}
\figurenum{8}
\epsscale{0.75}
\plotone{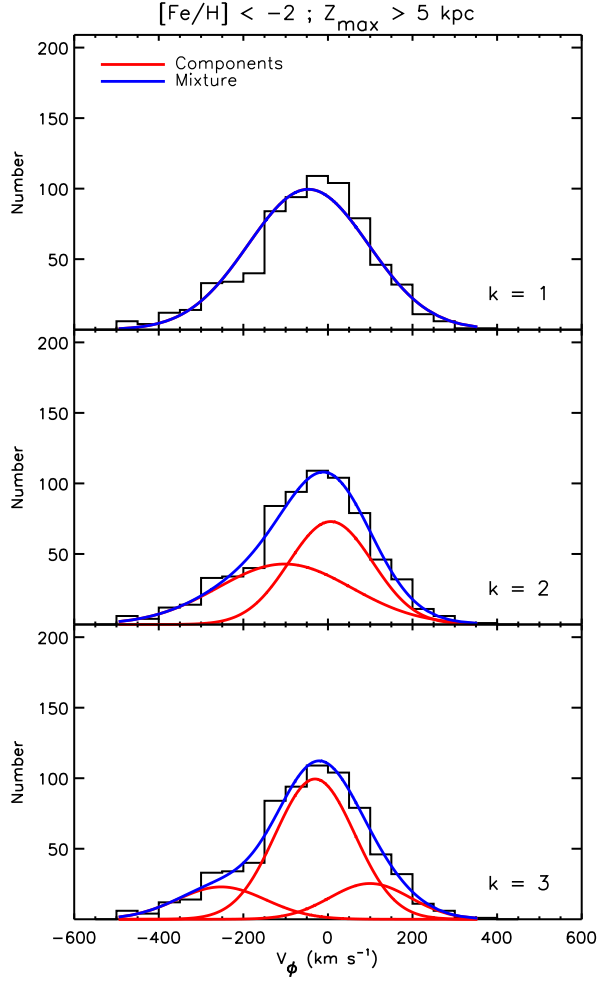}
\caption{R-Mix results for the low-metallicity ([Fe/H] $< -2.0$) sub-sample of local calibration stars
at large distances from the plane, Z$_{max}$ $> 5$ kpc, using as input guesses
the medoids and dispersions obtained with the clustering analysis (shown in
Table 1). The top panel
represents the fit for the case $k = 1$ (one component), while the second and
third panels show the fit for two and three components,
respectively. The blue lines in each panel denote the proposed mixture model for
the distribution of observed rotational velocities, while the red lines are the
individual Gaussian distributions included in the model.  As discussed in the
text, the one-component fit is strongly rejected, while both the two- and
three-component fits provide adequate descriptions of the data.}

\end{figure}
\clearpage

\begin{figure}
\figurenum{9}
\epsscale{1.0}
\plotone{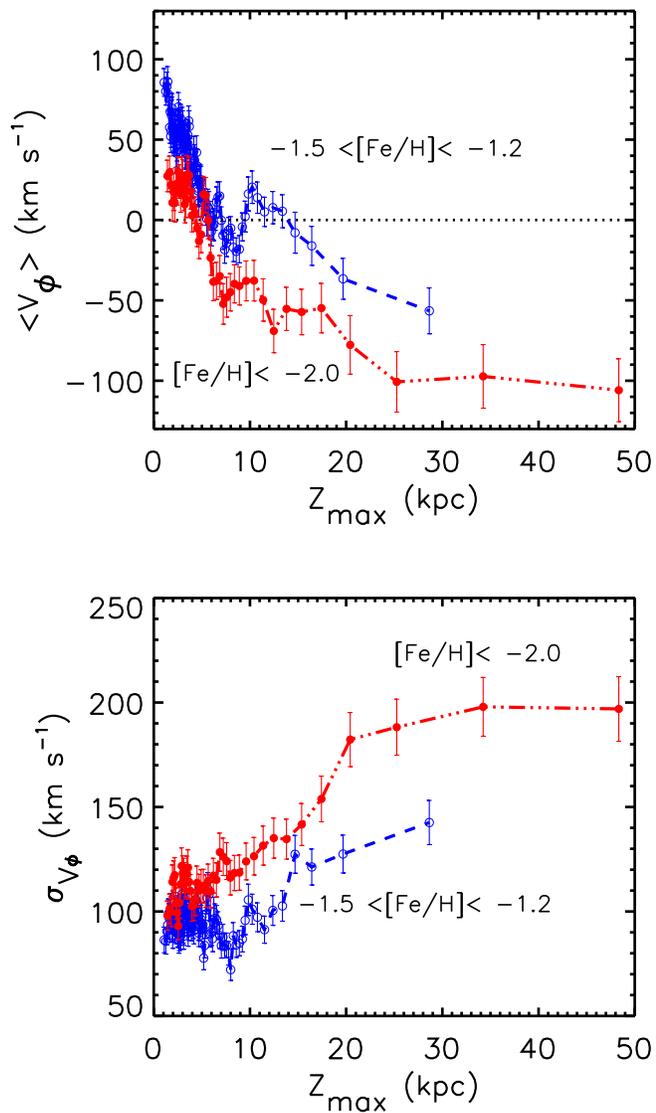}
\caption{Mean Galactocentric rotational velocity (upper panel),
and its dispersion (lower panel), as a function of Z$_{max}$, for
intermediate-metallicity stars ($-1.5 <$ [Fe/H] $< -1.2$; blue curves) and
low-metallicity stars ([Fe/H] $< -2.0$; red curves).  The values of these
quantities are obtained by passing a box of 100 stars, with an overlap of 70
stars per bin, through the data. The horizontal dotted line in the top panel
indicates a zero (non-rotating) velocity.  The values of the dispersion are
not corrected for observational errors.}

\end{figure}
\clearpage

\begin{figure}
\figurenum{10}
\epsscale{1.25}
\plotone{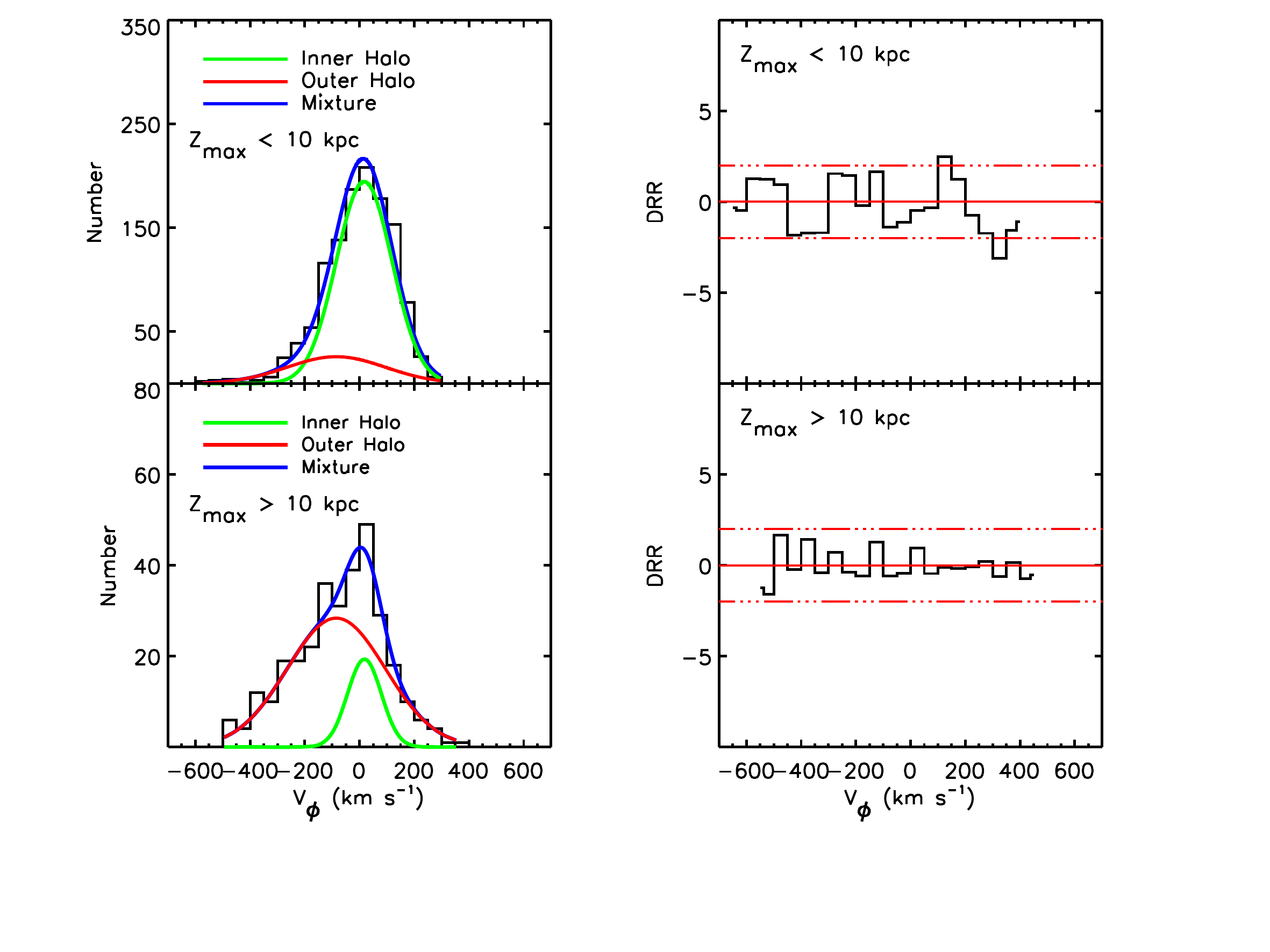}
\caption{Rotational properties for the low-metallicity sub-sample ([Fe/H] $<
-2.0$) of stars in the local calibration-star sample, divided into stars with
Z$_{max}$ above or below 10 kpc from the Galactic plane. The histograms in the
left-hand panels illustrate the observed distribution of V$_{\phi}$, while the
green (inner halo), and the red (outer halo) curves represent the results of the
ML analysis. The blue curves are the proposed mixture model for the distribution
of the rotational velocities for this sub-sample. The right-hand panels are the
Double Root Residual (DRRs) for these fits (see text for description). The
dot-dashed lines at $\pm 2$ indicate an approximate 95\% significance level.}

\end{figure}
\clearpage



\begin{figure}
\figurenum{11}
\epsscale{1.0}
\plotone{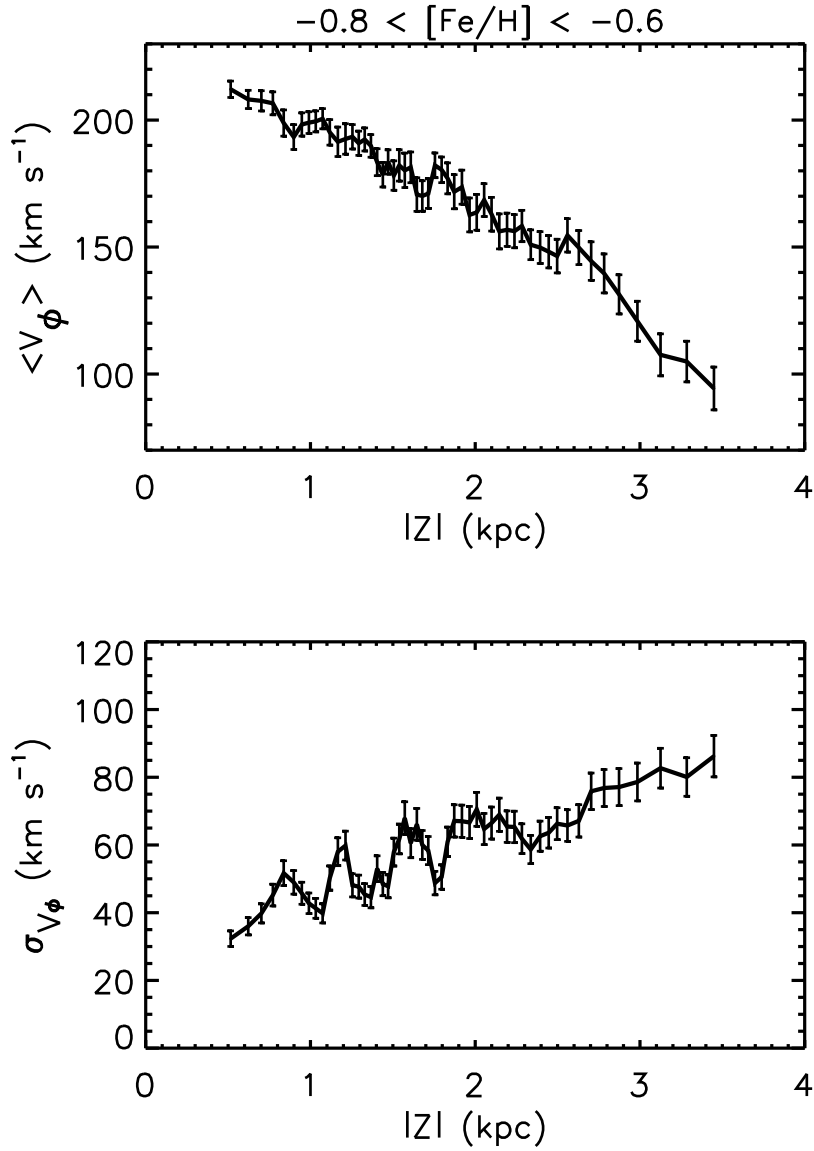}
\caption{Mean rotational velocity (upper panel) and dispersion (lower panel), as a function of the vertical
distance $|$Z$|$, for the high-metallicity sub-sample ($-0.8 <$ [Fe/H] $< -0.6$)
of stars in the local calibration-star sample. The values of these quantities
are obtained by passing a box of 100 stars, with an overlap of 70 stars per bin,
through the data. The values of the dispersions are
not corrected for observational errors.}

\end{figure}
\clearpage

\begin{figure}
\figurenum{12}
\epsscale{1.0}
\plotone{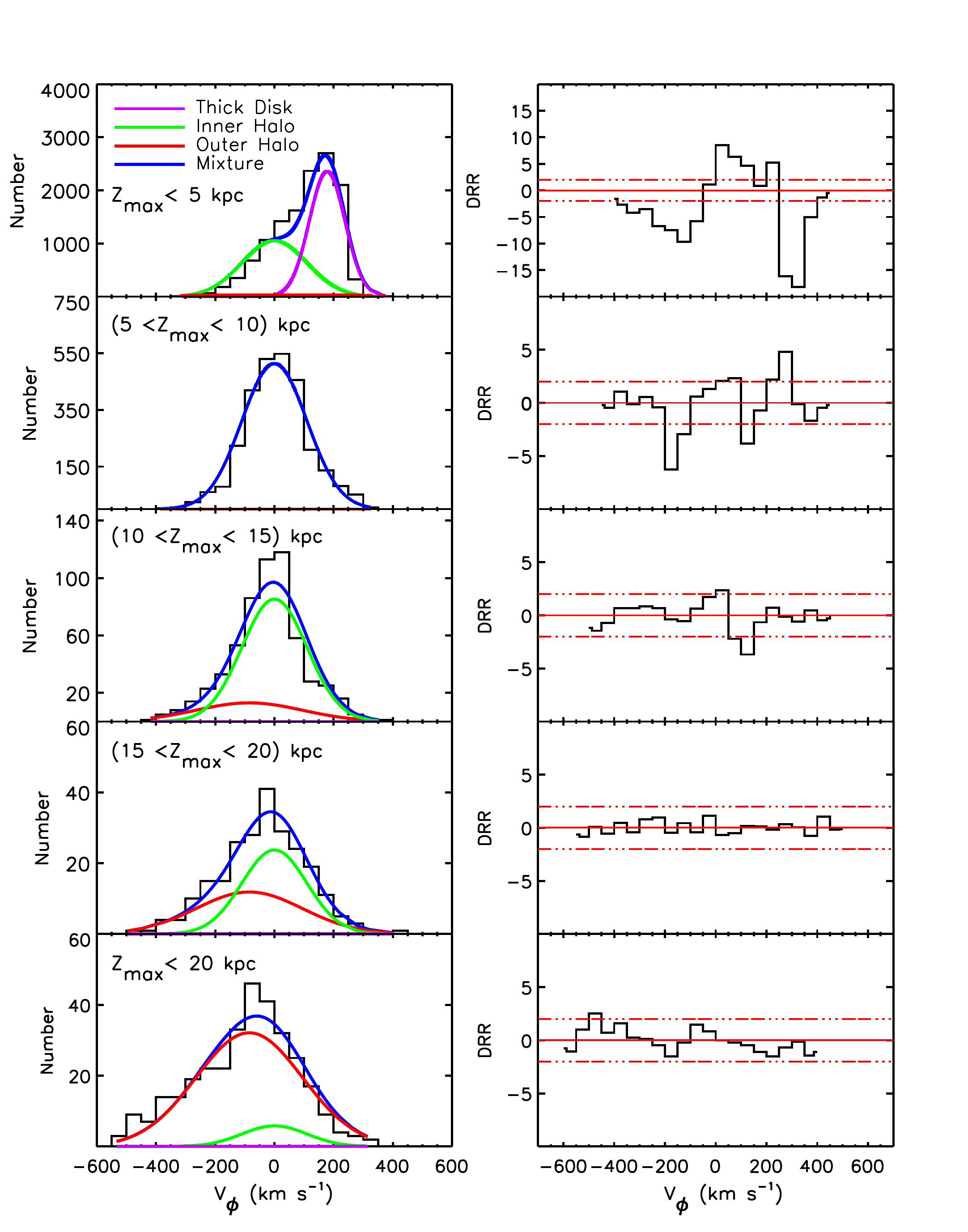}
\caption{Rotational properties for the full metallicity range
of stars in the local calibration-star sample. The histograms represent the observed
distribution of V$_{\phi}$, while the purple (thick disk),
green (inner halo), and red (outer halo) curves represent the results of the ML
analysis. The blue curves are the proposed mixture model for the
distribution of the rotational velocities for this sample. Each panel corresponds
to different cuts selected in intervals of Z$_{max}$, as indicated. The
right-hand panels are the Double Root Residuals (DRRs) for these fits (see text
for description). The dot-dashed lines at $\pm 2$ indicate an approximate 95\%
significance level.}

\end{figure}
\clearpage

\begin{figure}
\figurenum{13}
\epsscale{1.0}
\plotone{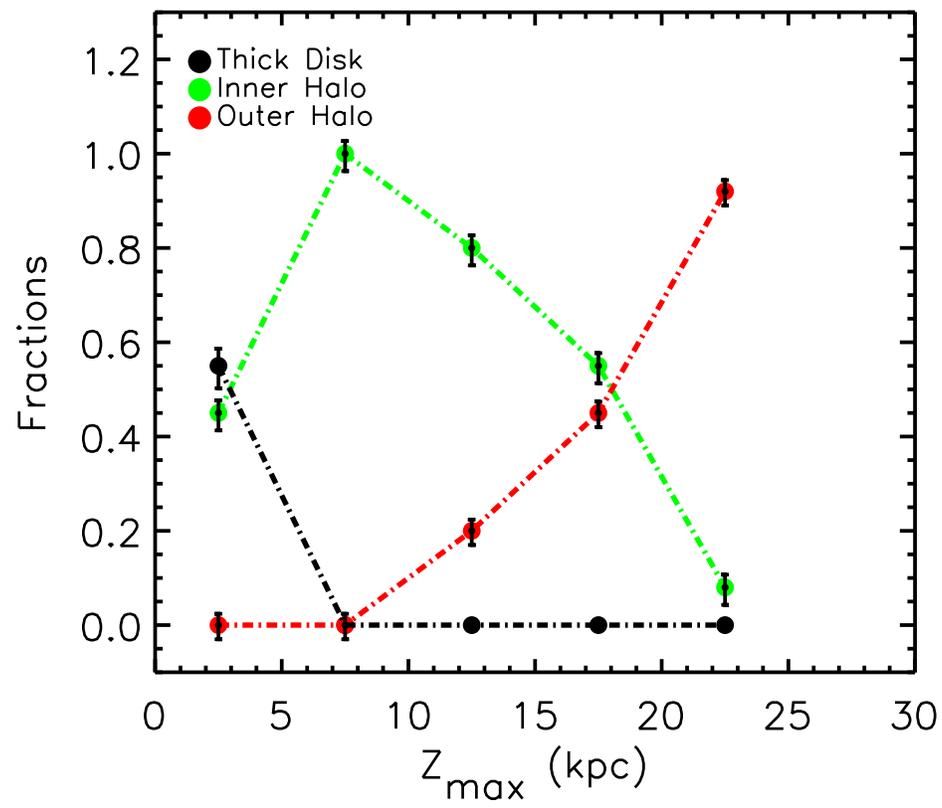}
\caption{Derived stellar fractions, as a function of Z$_{max}$,
for the thick disk (black dots), inner-halo (green dots), and outer-halo (red dots)
components. These fractions are evaluated with the ML analysis applied to the full
metallicity sample, with the input kinematic parameters for each component held
fixed.}

\end{figure}
\clearpage

\begin{figure}
\figurenum{14}
\epsscale{0.8}
\plotone{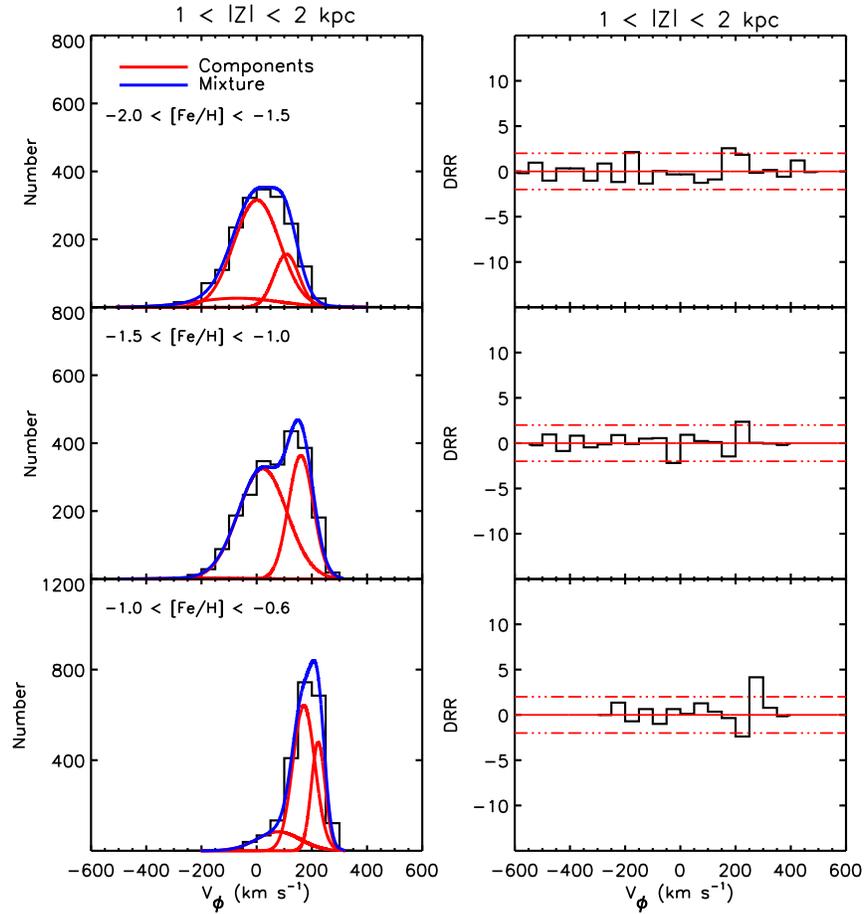}
\caption{R-Mix results for three sub-samples of local calibration stars covering
the range of metallicity $-$2.0 $<$ [Fe/H] $<$ $-$1.5 (upper panel), $-$1.5 $<$
[Fe/H] $<$ $-$1.0 (middle panel), and $-$1.0 $<$ [Fe/H] $<$ $-$0.6 (lower
panel), located close to the Galactic plane (1 $<$ $|$Z$|$ $<$ 2 kpc). The input
guesses are the medoids obtained with the CLARA clustering analysis. The histograms
represent the observed distribution of V$_{\phi}$, the blue lines in each
panel denote the proposed mixture model for the distribution of observed
rotational velocities, and the red lines are the individual Gaussian
distributions included in the model. The right-hand panels are the Double Root
Residuals (DRRs) for these fits (see text for description). The dot-dashed lines
at $\pm 2$ indicate an approximate 95\% significance level. The apparent lack of
fit for high velocity stars shown in the lower panel DRR plot is discussed in
the text.}

\end{figure}
\clearpage

\begin{figure}
\figurenum{15}
\epsscale{0.8}
\plotone{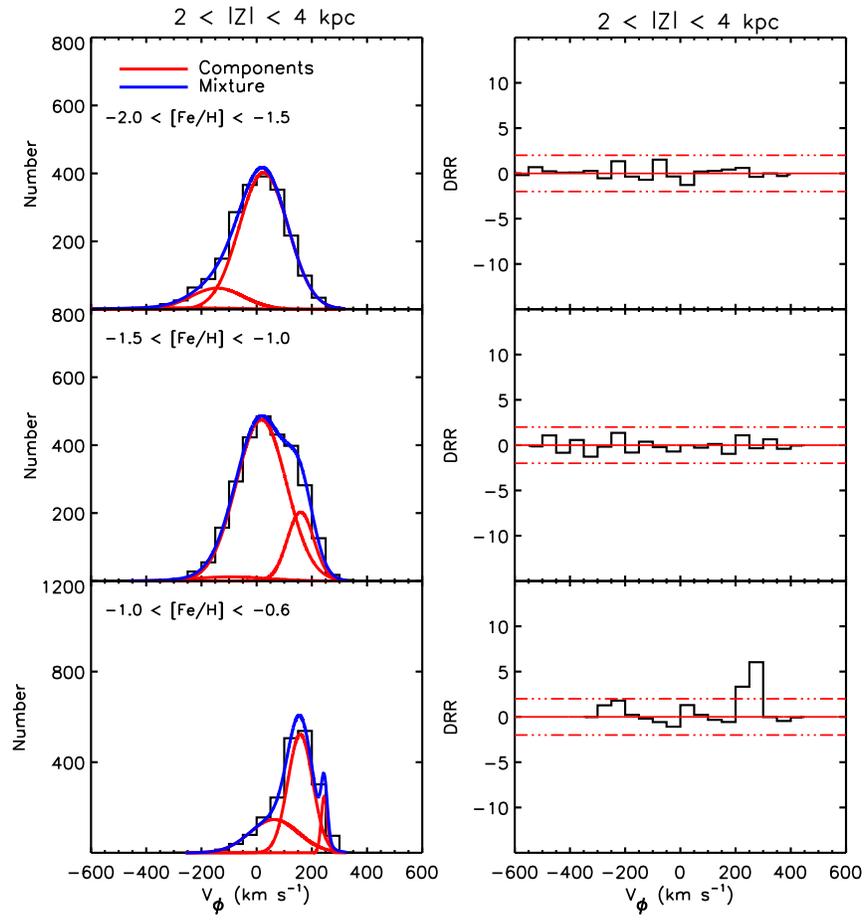}
\caption{The same as Figure 14, but for stars located farther from the Galactic
plane (2 $<$ $|$Z$|$ $<$ 4 kpc).  The apparent lack of
fit for high velocity stars shown in the lower panel DRR plot is discussed in
the text.}

\end{figure}
\clearpage

\begin{figure}
\figurenum{16}
\epsscale{1.0}
\plotone{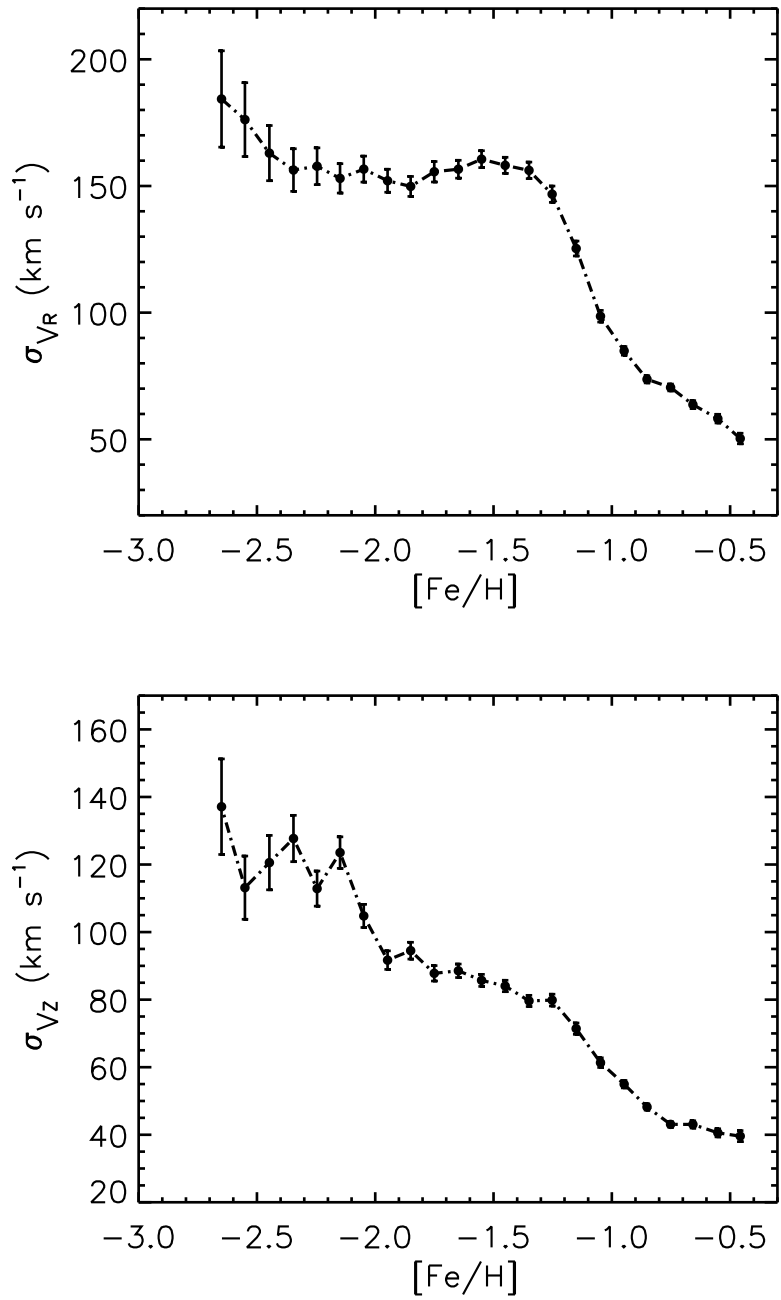}
\caption{Velocity dispersions (corrected for observational errors) for the
V$_{R}$ (upper panel) and V$_{Z}$ (lower panel) components, as a function of
metallicity. The plot employs stars covering the full [Fe/H] of the local
calibration-star sample. The bins represent a range of 0.1 dex in metallicity,
and are non-overlapping.}

\end{figure}
\clearpage

\begin{figure}
\figurenum{17}
\epsscale{1.0}
\plotone{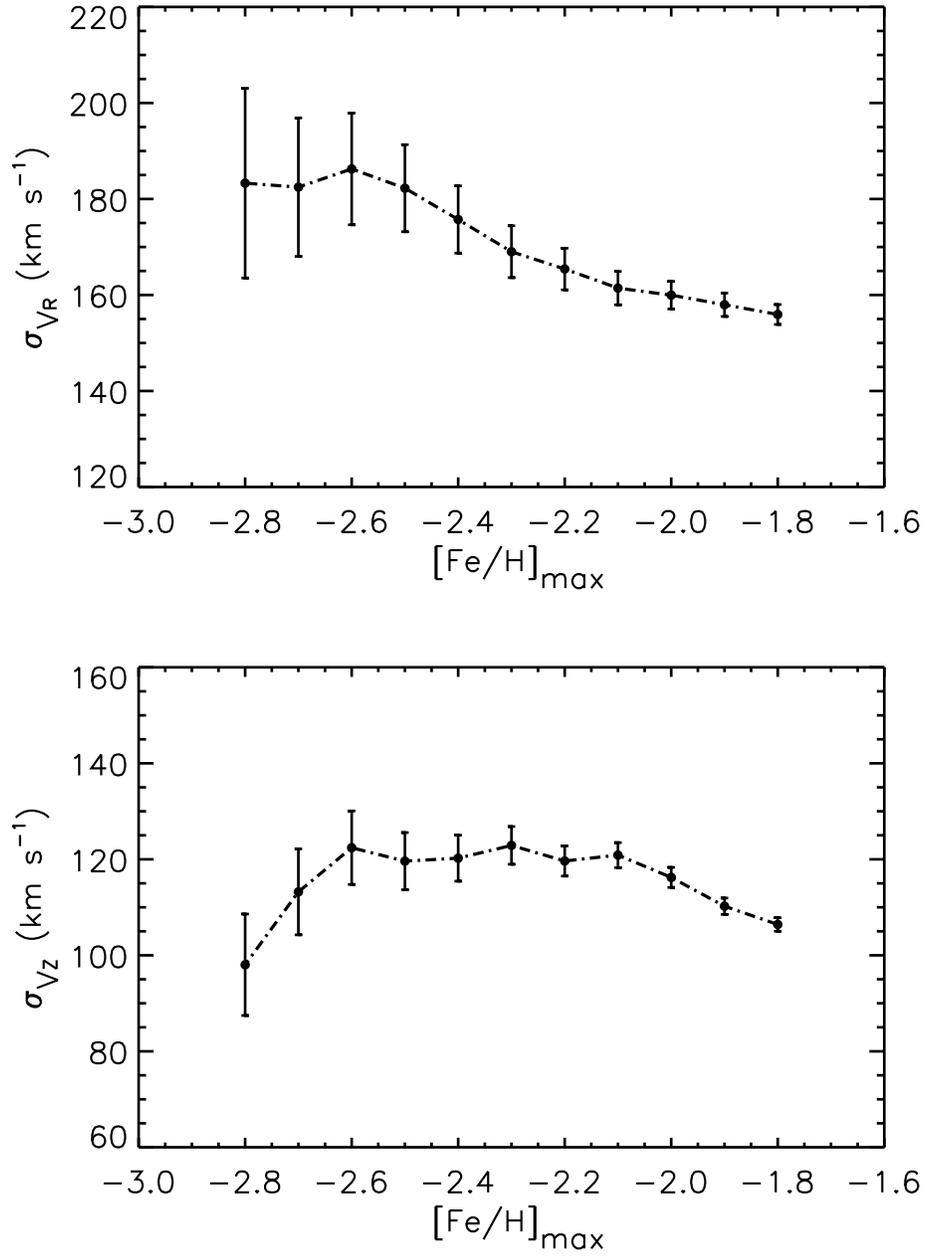}
\caption{Velocity dispersions (corrected for observational errors) for the
V$_{R}$ (upper panel) and V$_{Z}$ (lower panel) components, as a function of
metallicity, for metal-poor stars in the local sample. Each bin represents a
sub-sample of stars with [Fe/H] $<$ [Fe/H]$_{max}$.
}

\end{figure}
\clearpage

\begin{figure}
\figurenum{18}
\epsscale{1.0}
\plotone{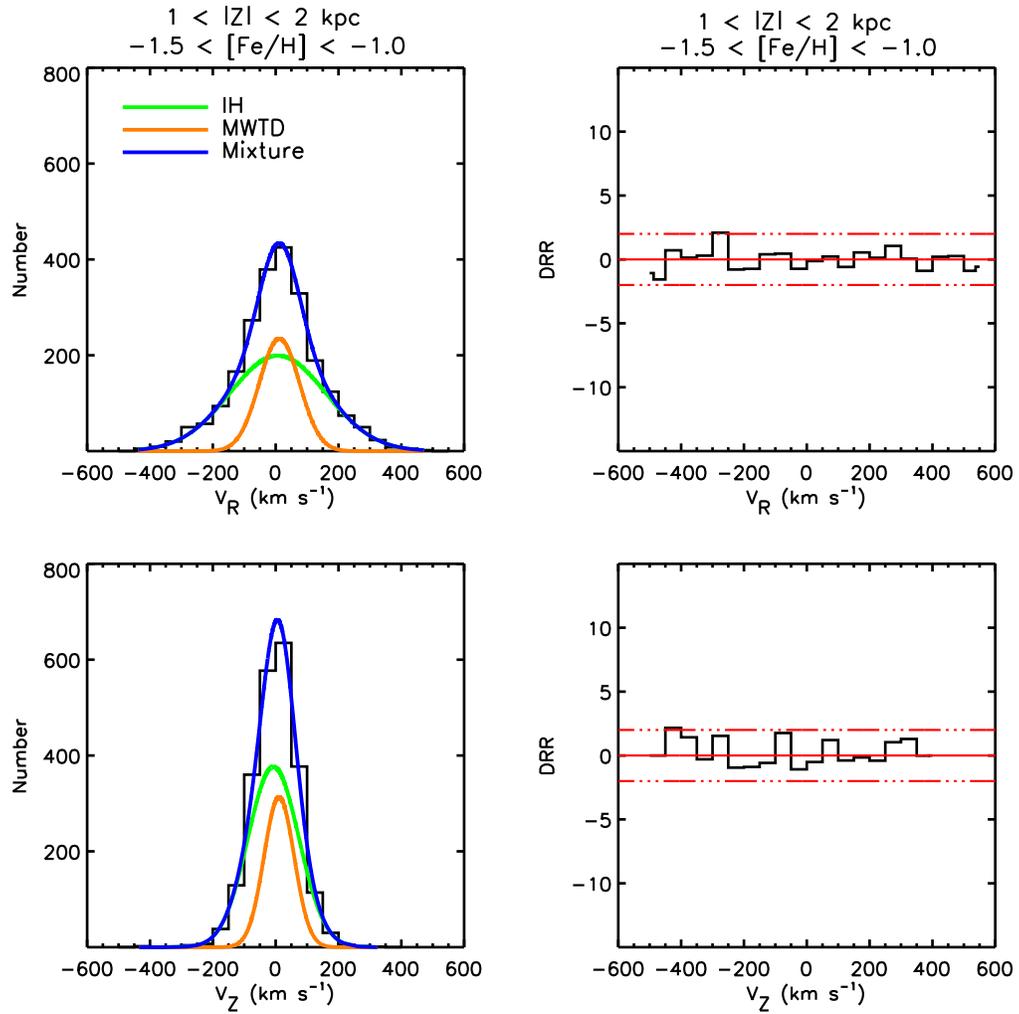}
\caption{R-Mix results for the V$_{R}$ (upper panel) and V$_{Z}$ (lower panel)
distributions. The selected range of metallicity is $-1.5 <$ [Fe/H] $<
-1.0$, and the stars are located close to the Galactic plane (1 $<$ $|$Z$|$ $<$
2 kpc). The histograms represent the observed distribution of V$_{R}$ and V$_Z$
in each panel, respectively. The blue lines in each panel denote the proposed
mixture model, while the orange and green lines are the individual Gaussians
included in the model, representing the MWTD and the inner halo,
respectively. The right-hand panels are the Double Root Residuals (DRRs) for
these fits (see text for description). The dot-dashed lines at $\pm 2$ indicate
an approximate 95\% significance level.}

\end{figure}
\clearpage

\begin{figure}
\figurenum{19}
\epsscale{1.0}
\plotone{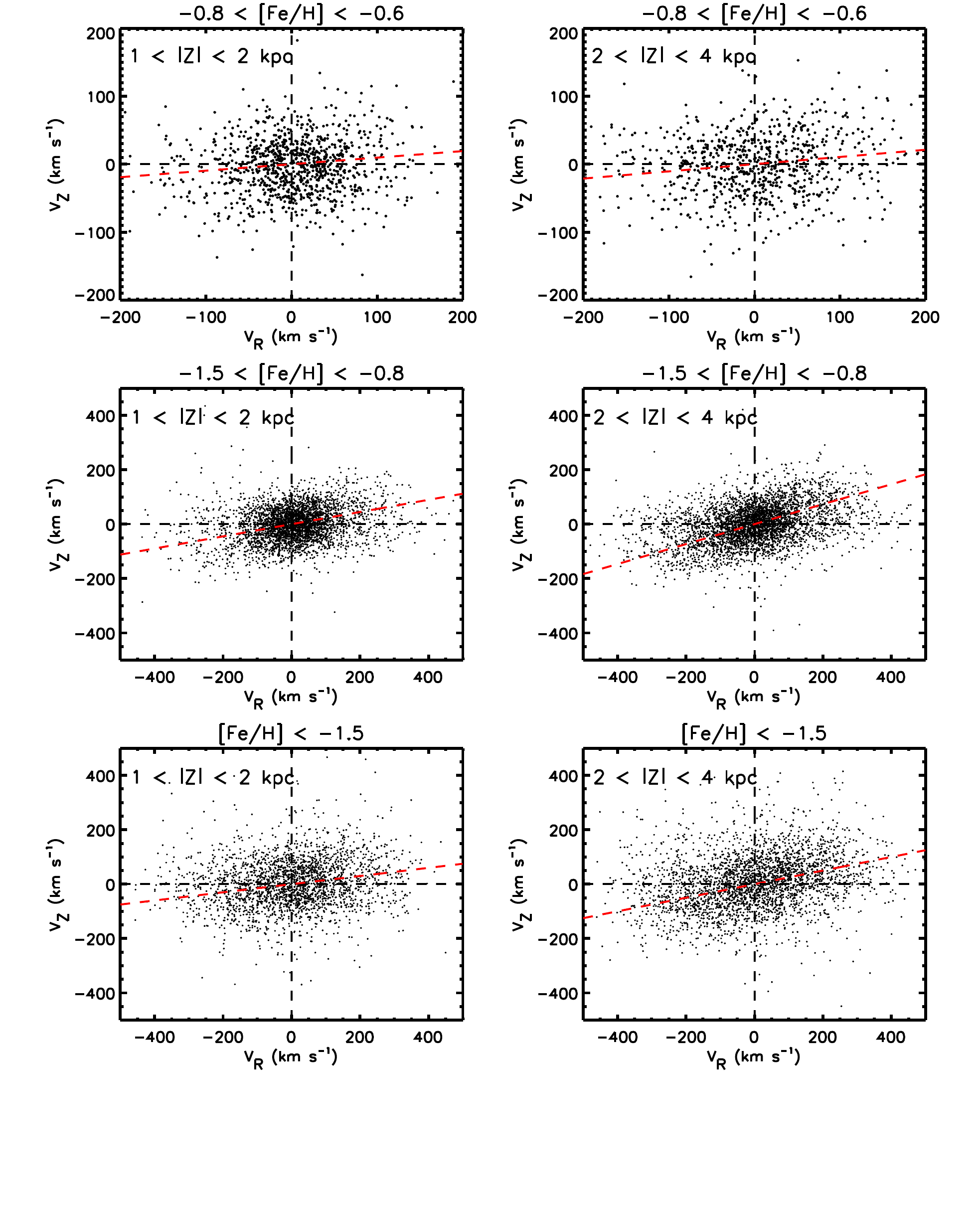}
\caption{Observed tilt of the velocity ellipsoids in the
(V$_{R}$,V$_{Z}$) plane for the metal-rich sub-sample with $-0.8 <$ [Fe/H] $<
-0.6$ (top panels), the intermediate-metallicity sub-sample with $-1.5 <$ [Fe/H]
$< -0.8$ (middle panels), and the low-metallicity sub-sample
with [Fe/H] $< -1.5$ (lower panels). The left column shows these
sub-samples close the Galactic plane, while the right column shows these
sub-samples selected at higher $|$Z$|$. In each panel, the black dots are the
distribution of the V$_{R}$ and V$_{Z}$ velocities , the black dashed lines
indicates the principal axes of the (cylindrical coordinate) velocity ellipsoids
in the meridional plane, and the red-dashed lines denote the derived
inclinations of the ellipsoids.}

\end{figure}
\clearpage

\begin{figure}
\figurenum{20}
\epsscale{1.0}
\plotone{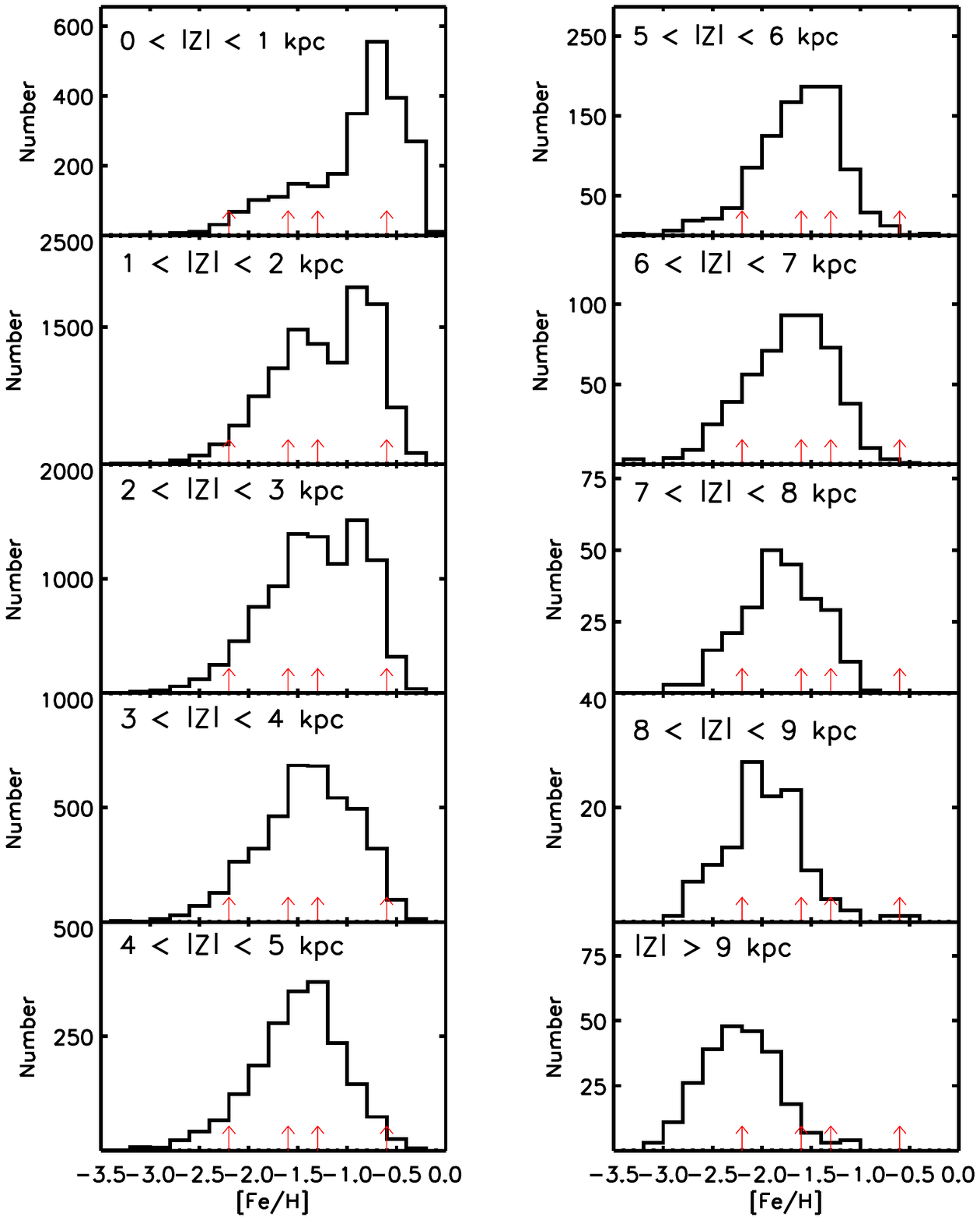}
\caption{Observed metallicity distribution functions (MDFs) for the full sample
of SDSS-SEGUE DR7 calibration stars as a function of vertical distance from the
Galactic plane. The black histograms represent the MDFs obtained at different
cuts of $|$Z$|$, while the red arrows denote the locations of the metallicity
peaks of the MDF for the thick disk ($-0.6$), the MWTD ($\sim -1.3$), the inner
halo ($-1.6$), and the outer halo ($-2.2$), respectively.
}

\end{figure}
\clearpage


\begin{thebibliography}{}

\bibitem[]{} Abadi, M.G., Navarro, J.F., Steinmetz, M., \& Eke, V.R. 2003, \apj, 597, 21

\bibitem[]{} Abazajian, K., et al. 2009, \apjs, 182, 543

\bibitem[]{} Adelman-McCarthy, J.K., et al. 2007, \apjs, 172, 634

\bibitem[]{} Adelman-McCarthy, J.K., et al. 2008, \apjs, 175, 297



\bibitem[]{} Aguerri, J.A.L., Balcells, M., \& Peletier, R.F. 2001, A\&A, 367, 428

\bibitem[]{} Allen, C., Schuster, W.J., \& Poveda, A. 1991, \aap, 244, 280

\bibitem[]{} Allende Prieto, C., et al., 2008, \aj, 136, 2070

\bibitem[]{} An, D., et al. 2009, \apj, 707, L64

\bibitem[]{} Batsleer, P., \& Dejonghe. H., 1994, A\&A, 287, 43

\bibitem[]{} Battaglia, G., Helmi, A., Tolstoy, E., Irwin, M., Hill, V., \&
Jablonka, P. 2008, \apj, 681, L13

\bibitem[]{} Beers, T.C., \& Sommer-Larsen, J. 1995, \apjs, 96, 175

\bibitem[]{} Beers, T.C., et al. 2000, \aj, 119, 2866

\bibitem[]{} Beers, T.C., et al. 2002, \aj, 124, 931

\bibitem[]{} Bell, E.F., et al. 2008, \apj, 680, 295

\bibitem[]{} Belokurov, V., et al. 2006, \apj, 642, L137

\bibitem[]{} Bienayme, O. et al. 2009, \aap, 500, 801

\bibitem[]{} Binney, J. \& Tremaine, S., 1988, Galactic Dynamics (Princeton: Princeton Univ. Press)

\bibitem[]{} Binney, J. \& Tremaine, S., 2008, Galactic Dynamics 2nd edition (Princeton: Princeton Univ. Press)

\bibitem[]{} Bond, N.A., et al. 2009, arXiv:0909.0013

\bibitem[]{} Brown, W.R., Geller, M.J., Kenyon, S.J., \& Diaferio, A. 2009, \aj, 139, 59

\bibitem[]{} Buser, R., et al. 1998, \aap, 331, 934

\bibitem[]{} Buser, R., Rong, J. \& Karaali, S. 1999, \aap, 348, 98

\bibitem[]{} Carney, B.W. 1984, \pasp, 96, 841

\bibitem[]{} Carney, B.W., Latham, D.W., \& Laird, J.B. 1990, \aj, 99, 572

\bibitem[]{} Carney, B.W., Laird, J.B., Latham, D.W., \& Aguilar, L.A. 1996, \aj, 112, 668

\bibitem[]{} Carollo, D., et al. 2007, Nature, 450, 1020

\bibitem[]{} Chapman, S.C., et al. 2006, \apj, 653, 255

\bibitem[]{} Chen, B., et al. 2001, \apj, 553, 184

\bibitem[]{} Chiba, M., \& Beers, T.C. 2000, \aj, 119, 2843

\bibitem[]{} Chiba, M., \& Beers, T. C., 2001, \apj, 549, 325

\bibitem[]{} Chiba, M., Yoshii, Y., \& Beers, T.C. 1999, in The Third Stromlo
Symposium: The Galactic Halo, ASP Conf. Ser. 165, eds. B.K. Gibson, T.S. Axelrod,
\& M.E. Putman, (San Francisco: ASP), p. 269

\bibitem[]{} Cooper, A.P, et al., 2009, arXiv:0910.3211

\bibitem[]{} Dejonghe, H., \& de Zeeuw, P. T., 1988, \apj, 333, 90

\bibitem[]{} De Lucia, G. \& Helmi, A., 2008, \apj, 391, 14

\bibitem[]{} Du et al. 2003, \aap, 407, 541

\bibitem[]{} Du et al. 2006, \mnras, 372, 1304

\bibitem[]{} Durrell, P.R., Harris, W.E., \& Pritchet, C.J. 2001, \aj, 121, 2557

\bibitem[]{} Eggen, O.J., Lynden-Bell, D., \& Sandage, A. R. 1962, \apj, 136, 748

\bibitem[]{} Frebel, A., Simon, J.D., Geha, M., \& Willman, B. 2009, \apj, 708, 560

\bibitem[]{} Freeman, K.C. 1987, \araa, 25, 603

\bibitem[]{} Ghez, A.M., et al., 2008, \apj, 689, 1044

\bibitem[]{} Gilmore, G., \& Reid, N. 1983, \mnras, 202, 1025

\bibitem[]{} Gilmore, G. \& Wyse, R.F.G. 1985, \aj, 90, 2015


\bibitem[]{} Gilmore, G., Wyse, R.F.G., \& Norris, J.E. 2002, \apj, 574, 39

\bibitem[]{} Girard, T. M., Korchagin, V.I., Casetti-Dinescu, D.I., van Altena,
W.F., Lopez, C.E., \& Monet, D.G. 2006, \aj, 132, 1768

\bibitem[]{} Grillmair, C.J., 2006, \apj, 651L, 29

\bibitem[]{} Grillmair, C.J., Carlin, J.L., \& Majewski, S.R. 2008, \apj, 689, L117


\bibitem[]{} Gunn, J.E., et al., 2006, \aj, 131, 2332

\bibitem[]{} Harris, W.E. 1976, \aj, 81, 1095

\bibitem[]{} Harris, W.E., Harris, G.L.H., Layden, A.C., \& Wehner, E.M.H. 2007, \apj, 666, 903

\bibitem[]{} Hartwick, F.D.A. 1987, in The Galaxy, NATO ESI Ser. 207,
eds. G. Gilmore \& B. Carswell, (Dordrecht: Reidel), p. 281

\bibitem[]{} Hayashi, H., \& Chiba, M., 2006, \pasj, 58, 835

\bibitem[]{} Helmi, A., et al., 2006, \apj, 651, L121

\bibitem[]{} Holmberg, J. \& Flynn, C. 2000, \mnras, 313, 209

\bibitem[]{} Ibata, R., Mouhcine, M., \& Rejkuba, M. 2009, \mnras, 395, 126

\bibitem[]{} Ishigaki, M., Chiba, M., \& Aoki, W. 2009, \pasj, in press (ArXiv:0912.0329)

\bibitem[]{} Ivezi{\'c}, {\v Z}., et al., 2000, \aj, 120, 963

\bibitem[]{} Ivezi{\'c}, {\v Z}., et al. 2008, \apj, 684, 287

\bibitem[]{} Jones, B.F., \& Walker M.F. 1988, \aj, 95, 1775

\bibitem[]{} Kalirai, J.S., et al. 2006a, \apj, 641, 268

\bibitem[]{} Kalirai, J.S., et al. 2006b, \apj, 648, 389

\bibitem[]{} Katz, D., et al. 1999, \apss, 265, 221

\bibitem[]{} Kaufman, L., \& Roesseeuw, P.J. 1990, Finding Groups in Data.  An
Introduction to Cluster Analysis, (New York: Wiley)

\bibitem[]{} Kazantzidis, S., Zentner, A.R., Kravtsov, A.V., Bullock, J.S., \& Debattista, V.P.
2009, \apj, 700, 1896

\bibitem[]{} Kerber, L.O., et al. 2001, \aap, 345, 424

\bibitem[]{} Kerr, F.J., \& Lynden-Bell, D. 1986, \mnras, 221, 1023

\bibitem[]{} Kinman, T.D., Suntzeff, N.B., \& Kraft, R.P. 1994, \aj, 108, 1722

\bibitem[]{} Kinman, T.D., Cacciari, C., Bragaglia, A., Buzzoni, A., \& Spagna,
A. 2007, \mnras, 371, 1381

\bibitem[]{} Kirby, E., et al., 2008, \aj, 685, L43

\bibitem[]{} Klement, R., et al. 2009, \apj, 698, 865

\bibitem[]{} Koch, A. 2009, Rev. Mod. Astron., 21, 9

\bibitem[]{} Koch, A., et al., 2008a, \apj, 689, 958

\bibitem[]{} Koch, A., et al., 2008b, \apj, 688, L13

\bibitem[]{} Koposov, S.E., Rix, H.-W., \& Hogg, D.W. 2009, \apj, submitted
(ArXiv:0907.1085)

\bibitem[]{} Kuijken, K. \& Gilmore, G. 1989, \mnras, 239, 605

\bibitem[]{} Kuijken, K. \& Gilmore, G. 1991, \apj, 367, L9

\bibitem[]{} Larsen, J.A. \& Humphreys, R.M 2003, \aj, 125, 1958

\bibitem[]{} Layden, A.C. 1995, \aj, 110, 2288


\bibitem[]{} Lee, Y.S., \& Beers, T.C. (2009), BAAS, 41, 227

\bibitem[]{} Lee, Y.S., et al. 2008a, \aj, 136, 2022

\bibitem[]{} Lee, Y.S., et al., 2008b, \aj, 136, 2050

\bibitem[]{} Lee, Y.W., Gim, H.B., \& Casetti-Dinescu, D.I. 2007, \aj, 661, L49

\bibitem[]{} Lupton, R.L., 1993, Statistics in Theory and Practice
(Princeton: Princeton Univ. Press)


\bibitem[]{} Majewski, S.R., 1992, \apjs, 78, 87

\bibitem[]{} Martin, J.C., \& Morrison, H.L. 1998, \aj, 116, 1724

\bibitem[]{} Miceli, A., et al. 2008, \apj, 678, 865

\bibitem[]{} Mihalas, D., \& Binney, J. 1981, Galactic Astronomy (San Francisco: Freeman)

\bibitem[]{} Mihos, J.C., Walker, I.R., Hernquist, L., Mendes de Oliveira, C.,
\& Bolte, M. 1995, \apj, L87

\bibitem[]{} Morrison, H.L., Flynn, C., \& Freeman, K.C. 1990, \aj, 100, 1191

\bibitem[]{} Morrison, H.L., et al. 2009, \apj, 694, 130

\bibitem[]{} Munn, J.A., et al. 2004, \aj, 127, 3034

\bibitem[]{} Munn, J.A., et al. 2008, \aj, 136, 895

\bibitem[]{} Ng, Y.K., et al. 1997, \aap, 324, 65

\bibitem[]{} Norris, J., Bessell, M.S., \& Pickles, A.J. 1985, \apjs, 58, 463

\bibitem[]{} Norris, J.E. 1994, \apj, 431, 635

\bibitem[]{} Norris, J.E. 1996, in Formation of the Galactic Halo, Inside and Out, ASP Conf.
Ser. 92, eds. M. Morrison \& A. Sarajedini, (San Francisco: ASP), p. 14

\bibitem[]{} Norris, J.E., et al. 2008, \apj, 689, L113

\bibitem[]{} Ojha, K. K., 2001, \mnras, 322, 426

\bibitem[]{} Press, W.G., Flannery, B.P., Teukolsky, S.A., \& Vetterling, W.T.
1992, Numerical Recipes in C: The Art of Scientific Computing
(Second Edition), (Cambridge: Cambridge Univ. Press)

\bibitem[]{} Preston, G.W., Shectman, S.A., \& Beers, T.C. 1991, \apj, 375, 121

\bibitem[]{} Quinn, P.J. \& Goodman, J. 1986, \apj, 309, 472

\bibitem[]{} Quinn, P.J., Hernquist, L., \& Fullagar, D.P. 1993, \apj, 403, 74

\bibitem[]{} Reddy, B.E., \& Lambert, D.L. 2008, \mnras, 391, 95

\bibitem[]{} Rejkuba, M., Greggio, L., Harris, W.E., Harris, G.L.H., \& Peng,
E.W. 2005, \apj 631, 262

\bibitem[]{} Rejkuba, M., Mouhcine, M., \& Ibata, R. 2009, \mnras, 396, 1231

\bibitem[]{} Reyl$\acute{e}$, C., \& Robin, A.C. 2001, A\&A, 373, 886

\bibitem[]{} Robin, A. \& Creze, M. 1986, \aap, 64, 53

\bibitem[]{} Robin, A.C., Haywood, M., Creze, M., Ojha, D.K., \& Bienayme, O.
1996, A\&A, 305, 125

\bibitem[]{} Rodgers, A.W., \& Roberts, W.H. 1993, \aj, 106, 6

\bibitem[]{} Roederer, I. 2009, \aj, 137, 272

\bibitem[]{} Ryan, S.G., \& Norris, J.E. 1991, \aj, 101, 1835

\bibitem[]{} Ryan, S.G., \& Lambert, D.L. 1995, \aj, 109, 2068

\bibitem[]{} Salvadori, S., Ferrara, A., Schneider, R., Scannnapieco, E., \&
Kawata, D. 2009, \mnras, 401, L5

\bibitem[]{} Sandage, A., \& Fouts, G. 1987, \aj, 93, 74

\bibitem[]{} Schlaufman, K.C., et al. 2009, \apj, 703, 2177

\bibitem[]{} Schlegel, D.J, Finkbeiner, D.P., \& Davis, M. 1998, \apj, 500, 525

\bibitem[]{} Schoenrich, R. \& Binney, J. 2009, \mnras, 396, 203

\bibitem[]{} Searle, L. \& Zinn, R. 1978, \apj, 225, 357

\bibitem[]{} Sesar, B., et al. 2009, \apj, 708, 717

\bibitem[]{} Sheffield A.A., et al., 2007, in IAU Symposium: A giant Step: from
Milli- to Micro-arcsecond Astrometry, IAU Conf. Ser. 248, eds. W.J. Jin, I.
Platais, \& M.A. Perryman, (Cambridge: ASP), p. 506

\bibitem[]{} Siebert, A., et al. 2008, \mnras, 391, 793

\bibitem[]{} Smith, M.C., et al. 2009, \mnras, 399, 1223

\bibitem[]{} Sommer-Larsen, J., \& Zhen, C. 1990, \mnras, 242, 10

\bibitem[]{} Sommer-Larsen, J., Beers, T.C., Flynn, C., Wilhelm, R.,
 \& Christensen, P.-R. 1997, \apj, 481, 775

\bibitem[]{} Soubiran, C., Bienayme, O., \& Siebert, A. 2003, \aap, 398, 141

\bibitem[]{} Spagna, A. et al. 1996, \aap, 311, 758

\bibitem[]{} Statler, T.S. 1988, \apj, 331, 71

\bibitem[]{} Tolstoy, E., Hill, V., \& Tosi, M. 2009, ARAA, 47, 371

\bibitem[]{} Twarog, B.A., \& Anthony-Twarog, B.J. 1996, \aj, 111, 220

\bibitem[]{} Vel$\acute{a}$zquez, H. \& White, S.D.M. 1999, \mnras, 304, 254

\bibitem[]{} Velleman, P., F. \& Hoaglin, D.C., 1981, Application, Basics,
and Computing of Exploratory Data Analysis (Boston: Duxbury Press)

\bibitem[]{} Villalobos, A., \& Hemli, A., 2008, \mnras, 391, 1806

\bibitem[]{} Vivas, A.K. \& Zinn, R. 2006, \aj, 132, 714

\bibitem[]{} Walker, I.R., Mihos, J.C., \& Hernquist, L. 1996, \apj, 460, 121

\bibitem[]{} Wall, J.,V. \& Jenkins, C., R., 2003, Practical Statistics for
Astronomers (Cambridge: Cambridge Univ. Press)

\bibitem[]{} Wilhelm, R. et al. 2005, in Cosmic Abundances as Records of Stellar
Evolution and Nucleosynthesis, ASP Conf. Ser. 336, eds. T.G. Barnes \& F.N.
Bash, (San Francisco: ASP), p. 371

\bibitem[]{} Xue, X.X. et al. 2008, \apj, 684, 1143

\bibitem[]{} Yanny, B., et al. 2003, \apj, 588, 824

\bibitem[]{} Yanny, B., et al. 2009, \aj, 137, 4377

\bibitem[]{} Yoachim, P., \& Dalcanton, J.J. 2008, \apj, 682, 1004

\bibitem[]{} York, D.G., et al., 2000, \aj, 120, 1579

\bibitem[]{} Zhang, L., et al. 2009, \apj, 706, 1095

\bibitem[]{} Zinn, R. 1985, \apj, 293, 424

\bibitem[]{} Zinn, R. 1993, in The Globular Clusters - Galaxy Connection,
ASP Conf. Ser. 48, eds. G.H. Smith, \& J.P. Brodie, (San Francisco: ASP), 38

\bibitem[]{} Zolotov, A., et al. 2009, \apj, 702, 1058

\end{thebibliography}
\end{document}